\documentclass[a4paper, USenglish, cleveref, autoref, thm-restate]{lipics-v2021}

\hideLIPIcs
\nolinenumbers

\usepackage{silence}
\usepackage{hyphenat}

\usepackage{amsmath,amssymb,amsfonts}
\usepackage{mathtools}
\usepackage{mathrsfs}
\usepackage[only,llbracket,rrbracket]{stmaryrd}

\usepackage{tikz}
\usepackage{tikz-cd}
\usetikzlibrary{positioning,arrows.meta,fit}

\usepackage{siunitx}
\usepackage{booktabs}
\usepackage{colortbl}
\usepackage{multirow}
\usepackage{rotating}

\usepackage{cite}

\usepackage{comment}

\usepackage{cleveref}

\usepackage[ruled,vlined]{algorithm2e}

\theoremstyle{definition}
\newtheorem{xtheorem}{Theorem}[section]
\newtheorem{xdefinition}[xtheorem]{Definition}
\newtheorem{xlemma}[xtheorem]{Lemma}
\newtheorem{xcorollary}[xtheorem]{Corollary}
\newtheorem{xproposition}[xtheorem]{Proposition}

\theoremstyle{remark}
\newtheorem{xremark}[xtheorem]{Remark}
\newtheorem{xexample}[xtheorem]{Example}

\crefname{section}{Section}{Sections}
\crefname{appendix}{Appendix}{Appendices}
\AddToHook{cmd/appendix/before}{%
    \crefalias{section}{appendix}%
    \crefalias{subsection}{appendix}
}

\crefname{xdefinition}{Definition}{Definitions}
\crefname{xtheorem}{Theorem}{Theorems}
\crefname{xlemma}{Lemma}{Lemmas}
\crefname{xcorollary}{Corollary}{Corollaries}
\crefname{xproposition}{Proposition}{Propositions}
\crefname{xfact}{Fact}{Facts}
\crefname{xexample}{Example}{Examples}
\crefname{xremark}{Remark}{Remarks}
\crefname{table}{Table}{Tables}
\crefname{listing}{Listing}{Listings}
\crefname{algorithm}{Algorithm}{Algorithms}

\makeatletter
\newcommand{\ie}{i.e.\@ifnextchar.{\!\@gobble}{}}
\newcommand{\eg}{e.g.\@ifnextchar.{\!\@gobble}{}}
\newcommand{\etc}{etc\@ifnextchar.{}{.\@}}
\makeatother

\DeclareMathOperator{\orb}{orb}
\DeclareMathOperator{\supp}{supp}
\DeclareMathOperator{\Sym}{Sym}
\DeclareMathOperator{\Aut}{Aut}
\DeclareMathOperator{\Inj}{Inj}
\DeclareMathOperator{\dom}{dom}
\DeclareMathOperator{\Fix}{Fix}
\DeclareMathOperator{\Stab}{Stab}
\DeclareMathOperator{\New}{New}
\renewcommand{\dim}{\operatorname{dim}}
\newcommand{\I}{\mathbb{I}}
\newcommand{\Nom}{\mathbf{Nom}}
\newcommand{\SNom}{\mathbf{SNom}}
\newcommand{\Set}{\mathbf{Set}}
\newcommand{\NSet}{\mathbf{NSet}}
\newcommand{\Fam}{\operatorname{Fam}}
\newcommand{\Proj}{\operatorname{Proj}}
\newcommand{\Kl}{\operatorname{Kl}}
\newcommand{\EM}{\operatorname{EM}}
\newcommand{\Sh}{\operatorname{Sh}}
\newcommand{\Eq}{\mathrm{Eq}}
\newcommand{\Ord}{\mathrm{Ord}}
\newcommand{\rowsim}{\sim_{{}_{\!\!\!\mathrm{row}}}}

\newcommand{\Namae}{\textsc{Namae}}
\newcommand{\DEQ}{\textrm{DEQ}}
\newcommand{\NLambda}{\textsc{N$\lambda$}}
\newcommand{\Lois}{\textsc{Lois}}
\newcommand{\Ons}{\textsc{Ons}}
\newcommand{\OnsHs}{\textsc{Ons-hs}}

\newcommand{\timeout}{$>10$\,min}

\newcommand{\para}[1]{\vspace*{-0.3em}\subparagraph{#1}}

\title{Efficient Nominal Automata Computation with Strong Nominal Sets}
\titlerunning{Efficient Nominal Automata Computation with Strong Nominal Sets}
\author{Hiroya Fujinami}{National Institute of Informatics, Tokyo, Japan \and The Graduate University for Advanced Studies (SOKENDAI), Kanagawa, Japan}{makenowjust@nii.ac.jp}{https://orcid.org/0009-0007-0794-5743}{}
\authorrunning{H. Fujinami}
\Copyright{Hiroya Fujinami}
\relatedversion{}

\EventEditors{}
\EventNoEds{0}
\EventLongTitle{}
\EventShortTitle{}
\EventAcronym{}
\EventYear{}
\EventDate{}
\EventLocation{}
\EventLogo{}
\SeriesVolume{0}
\ArticleNo{0}

\begin{document}
\hypersetup{pdfsubject={Copyright (C) Hiroya Fujinami; licensed under Creative Commons License CC-BY 4.0}}
\maketitle

\begin{abstract}
Nominal automata and register automata are expressively equivalent models of computation over infinite alphabets, yet an algorithmic gap separates them in practice: dedicated algorithms decide language equivalence of deterministic register automata in polynomial time by working orbit-wise, while general-purpose nominal-computation tools that rely on SMT solvers hide the orbit structure and degrade rapidly even on small instances.
We trace this gap to non-trivial \emph{local symmetries} in nominal sets and observe that the configuration spaces of register automata correspond to \emph{strong nominal sets}, whose elements carry only \emph{trivial} local symmetries.
We characterize strong nominal sets categorically, consolidating known and new equivalences, and derive a constraint-free array representation of orbit-finite strong nominal sets and equivariant maps; we further prove the existence of \emph{projective covers}, guaranteeing that every nominal set admits a strong approximation preserving both orbit count and dimensions, computable for quotients of strong nominal sets.
Building on these results, we develop concrete algorithms for language equivalence, minimization, and active learning, bridging the algorithmic gap for equivalence (in polynomial time for a fixed alphabet dimension) and extending the orbit-wise approach to minimization and learning.
Finally, our implementation \Namae{}, based on the array representation, outperforms existing nominal-computation tools and, at scale, stays within a small constant factor of a dedicated register-automaton equivalence tool.
\looseness=-1
\end{abstract}

\section{Introduction}\label{sec:Introduction}

\subparagraph{Register and nominal automata.}
\emph{Register automata}, introduced by Kaminski and Francez as \emph{finite-memory automata}~\cite{KaminskiF94}, are a natural model for systems that process data from infinite alphabets, such as data-manipulating programs and communication protocols.
They operate by storing data values in finitely many registers and testing inputs against the stored values for equality.
Algorithmically, deterministic register automata are well behaved: language equivalence~\cite{MurawskiRT18,MurawskiRT19} and minimization~\cite{BenediktLP10} are efficiently computable.
\looseness=-1

\begin{figure}[tb]
  \centering
  \scalebox{0.9}{
  \begin{tikzpicture}[>=latex, thick, font=\sffamily\small]
    \node[circle, draw, double, fill=black, text=white, minimum size=2.5em, inner sep=0pt] (q0) at (0,0) {$q_0$};
    \node[circle, draw, fill=black, text=white, minimum size=2.5em, inner sep=0pt] (q1) at (4,0) {$q_1$};
    \node[circle, draw, fill=black, text=white, minimum size=2.5em, inner sep=0pt] (q2) at (8,0) {$q_2$};

    \draw[->] (-1.2, 0) -- (q0);

    \draw[->] (q0) to[bend left=35] node[above, align=center] {$\mathsf{Open}(a)$ \\ $x_1 := a$} (q1);
    \draw[->] (q1) to[bend left=35] node[above, align=center] {$\mathsf{Open}(b) \mid x_1 \neq b$ \\ $x_2 := b$} (q2);

    \draw[->] (q1) to[bend left=35] node[below, align=center] {$\mathsf{Close}(a) \mid x_1 = a$} (q0);
    \draw[->] (q2) -- node[below, align=center] {$\mathsf{Close}(a) \mid x_2 = a$} (q1);
    \draw[->] (q2) to[bend left=35] node[below, align=center] {$\mathsf{Close}(a) \mid x_1 = a$ \\ $x_1 := x_2$} (q1);
  \end{tikzpicture}
  }
  \caption{
    The resource language $\mathrm{Res}(2)$ as a register automaton.
    The transition function is partial: if a transition is undefined, the automaton rejects the input.
  }
  \label{fig:Res2}
\end{figure}
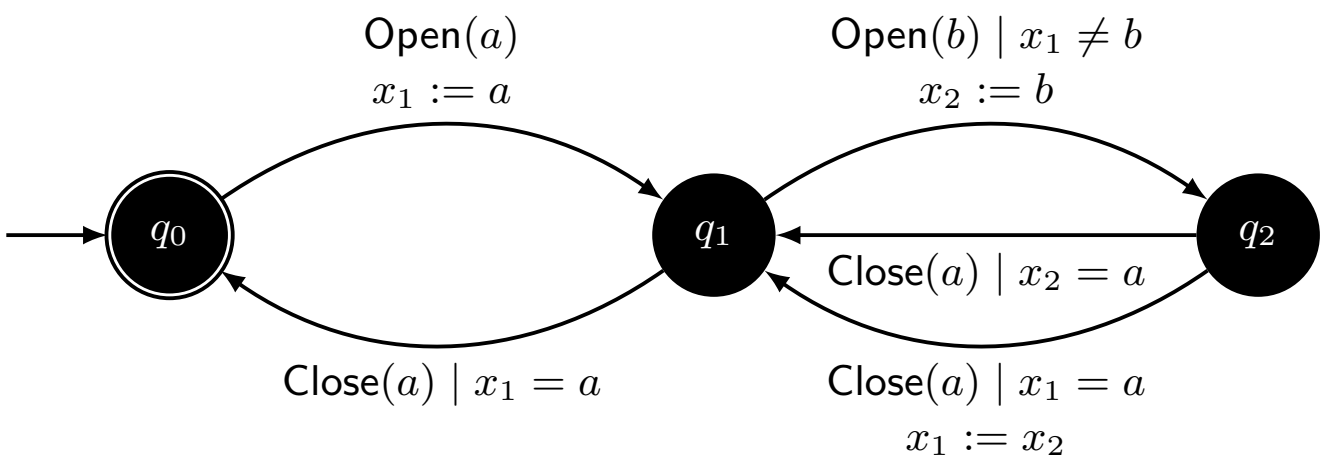

As a running example, \cref{fig:Res2} shows a register automaton for the resource language $\mathrm{Res}(2)$: at most two resources may be open simultaneously, and a word is accepted once every opened resource has been closed.
It uses three states $q_0, q_1, q_2$ and two registers $x_1, x_2$.
An $\mathsf{Open}(a)$ transition stores a new resource name $a$ in a vacant register, while a $\mathsf{Close}(a)$ transition releases the resource matching a stored value.
Since either open resource may be closed first, $q_2$ needs two $\mathsf{Close}$ transitions, one of which reassigns the registers ($x_1 := x_2$).
\looseness=-1

Nominal automata~\cite{BojanczykKL14} offer an alternative formulation based on \emph{nominal sets}~\cite{Pitts13}: sets equipped with an action of permutations of the data values, where each element has finite support.
States form an orbit-finite nominal set and transitions are equivariant maps.
The permitted permutations are a parameter called the \emph{data symmetry}~\cite{BojanczykKL14}: the \emph{equality symmetry} allows every permutation (data compared only for equality, as in \cref{fig:Res2}), while the \emph{total-order symmetry} allows only order-preserving permutations (ordered data).
Over the equality symmetry, nominal automata are expressively equivalent to register automata~\cite[Thm.~6.6]{BojanczykKL14}.
\looseness=-1

\para{The algorithmic gap and local symmetries.}
Despite their theoretical equivalence, an algorithmic gap separates the two models in practice.
For deterministic register automata, Murawski et al.~\cite{MurawskiRT18,MurawskiRT19} achieved a polynomial-time equivalence check using \emph{generating systems}, data structures that handle equivalence classes of configurations \emph{orbit-wise}.
In contrast, general-purpose nominal-computation tools, such as \NLambda{}~\cite{KlinS16} and \Lois{}~\cite{KopczynskiT17}, rely on SMT solvers,
hiding the orbit structure and leading to a performance bottleneck: ``direct custom-made register-automata algorithms are typically faster''~\cite{BalachanderFGT25}.
Over the total-order symmetry, \Ons{}~\cite{VenhoekMR22} already provides orbit-wise representations via constraint-free arrays.
The total-order symmetry can express every language over the equality symmetry, but each equality-symmetry orbit of dimension $d$ then splits into up to $d!$ orbits, as detailed in \cref{rem:LocalSymmetry}, so using the data symmetry that matches the language is crucial for efficiency~\cite{VenhoekMR22}.
\looseness=-1

We identify what blocks the \Ons{} recipe over the equality symmetry: non-trivial \emph{local symmetries}.
In $\mathrm{Res}(2)$, the Myhill--Nerode classes with two open resources form the orbit of \emph{unordered} pairs $\{a, b\}$; each such element is fixed by the non-trivial swap $(a\, b)$, \ie, carries a non-trivial local symmetry, so the orbit cannot be represented as a set of tuples of distinct atoms (the very format of a constraint-free array).
The register automaton of \cref{fig:Res2}, by contrast, avoids the symmetry by ordering its registers and reassigning them ($x_1 := x_2$).
\looseness=-1

\para{Strong nominal sets and the array representation.}
Our approach builds on \emph{strong nominal sets}~\cite{Tzevelekos07,MurawskiT16}, \ie, nominal sets with trivial local symmetries, as characterized in \cref{prop:LocalSymmetry}.
They coincide with the \emph{finitely named sets} underlying early history-dependent automata~\cite{MontanariP97}, a model that was soon enriched with permutation groups to enable canonical minimization~\cite{FerrariMP02}.
This paper moves in the \emph{opposite direction}: strong nominal sets correspond exactly to the configuration spaces of register automata, and the very absence of local symmetries is what makes them computationally lightweight.
Driven by this observation, our first contribution consolidates known and new
categorical equivalences of strong nominal sets into a single chain, shown in
\cref{thm:Characterization}, which grounds the rest of the paper:
\looseness=-1
\begin{equation*}
  \SNom \;\simeq\; \Kl(T)
         \;\simeq\; \Fam(\I^\mathrm{op})
         \;\simeq\; \Proj([\I, \Set])
         \;\simeq\; \NSet_\mathrm{fn},
\end{equation*}
where $\SNom$ denotes the category of strong nominal sets and equivariant maps, $\I$ is the category of finite subsets of $\mathbb{N}$ and injections, $T$ is the monad induced by the adjunction between the category $\Nom$ of nominal sets and the functor category $[|\I|, \Set]$~\cite{KurzPV10,Petrisan12}, $\Fam(\I^\mathrm{op})$ is the free coproduct completion of $\I^\mathrm{op}$, $\Proj([\I, \Set])$ is the category of projective objects of $[\I, \Set]$, and $\NSet_\mathrm{fn}$ is the category of finitely named sets~\cite{MontanariP97}.

Crucially, the equivalence $\SNom \simeq \Fam(\I^\mathrm{op})$ yields the concrete, constraint-free \emph{array representation} developed in \cref{ssec:DataRepresentations}.
In this representation, each orbit is determined solely by its dimension $d$, an element is a $d$-tuple of distinct atoms, and an equivariant map is an injective array of indices.
This representation allows orbits and equivariant maps to be manipulated directly, without relying on SMT solvers.

\para{Projective covers for quotient approximation.}
A challenge is that strong nominal sets are \emph{not} closed under quotients.
For instance, as \cref{ex:QuotientNotStrong} shows, taking the quotient of $\{(a,b) \mid a \neq b\}$ by $(a,b) \sim (b,a)$ yields $\{\{a,b\} \mid a \neq b\}$, thereby introducing local symmetries.
Consequently, the state space of a minimal nominal automaton
generally falls outside $\SNom$.
\looseness=-1

Beyond the equivalence $\SNom \simeq \Proj([\I, \Set])$, we show in \cref{thm:ProjectiveCover} that every nominal set $X$ admits a \emph{strong approximation} $X'$, obtained from a projective cover in $[\I, \Set]$: the induced equivariant surjection $p\colon X' \twoheadrightarrow X$ strips away local symmetries yet, by \cref{thm:OrbitCount}, merges no orbits and inflates no dimensions.
Since transitions lift along $p$ by projectivity, any nominal automaton over a strong alphabet admits a language-equivalent \emph{strong lift} with the same orbit count and dimensions by \cref{thm:StrongApproximation}.
Unlike the constructions assembled from known results in \cref{rem:PriorConstructions}, $X'$ is unique up to isomorphism and, by
\cref{lem:Quotient}, computable for quotients of strong nominal sets.
\looseness=-1

\para{Algorithms and implementation.}
Building on this foundation, we develop efficient algorithms for equivalence checking, minimization, and active learning on strong nominal automata.
By generalizing generating systems of register automata~\cite{MurawskiRT18,MurawskiRT19} to orbit-finite strong nominal sets, we obtain in \cref{thm:Equivalence} a language-equivalence algorithm that runs in polynomial time for a fixed alphabet dimension.
Algorithms based on quotients, such as Moore's minimization algorithm and active learning via $\nu L^\ast$~\cite{MoermanS0KS17}, are handled by strong approximations.
For a fixed alphabet dimension, minimization stays polynomial in the orbit count; for active learning, only the equivalence-query count and observation-table cells do, while the total membership-query count remains factorial, as shown in \cref{lem:Quotient,thm:Minimize,thm:Learn}.
\looseness=-1

We implemented the array representation and these algorithms in our library \Namae{}, and evaluated it in \cref{sec:ImplementationAndExperiments}.
We confirmed that \Namae{} outperforms the SMT-based tool \NLambda{} and remains within a small constant factor of the dedicated register-automaton tool \DEQ{}, demonstrating the practical scalability of the array representation.
\looseness=-1

\para{Contributions.}
We recast \emph{strong nominal sets}, for which canonical minimization is known
to fail, as an efficient foundation for nominal computation via the following
contributions:
\begin{enumerate}
  \item A categorical characterization of \emph{strong nominal sets}, consolidating known and new equivalences into a single theorem (\cref{ssec:CategoricalCharacterization}).
  \item An \emph{array representation} for orbit-finite strong nominal sets and equivariant maps between them, derived from the $\SNom \simeq \Fam(\I^\mathrm{op})$ equivalence (\cref{ssec:DataRepresentations}).
  \item An existence theorem for \emph{projective covers}, guaranteeing that every nominal set admits a strong approximation preserving its orbit count and dimensions (\cref{sec:ProjectiveCovers}).
  \item Concrete algorithms for language equivalence, minimization, and active learning on strong nominal automata, implemented in our library \Namae{} and evaluated (\cref{sec:AutomataAlgorithms,sec:ImplementationAndExperiments}).
\end{enumerate}

\para{Organization.}
\Cref{sec:Preliminaries} recalls preliminaries;
\cref{sec:StrongNominalSets} defines strong nominal sets, shows the four categorical equivalences, and develops the array representation;
\cref{sec:ProjectiveCovers} constructs projective covers, strong approximations, and strong lifts of automata;
\cref{sec:AutomataAlgorithms} presents concrete algorithms with complexity bounds;
\cref{sec:ImplementationAndExperiments} evaluates our approach;
\cref{sec:RelatedWork} discusses related work; and \cref{sec:Conclusion} concludes.
Omitted details appear in \crefrange{app:LocalSymmetry}{app:FullExperiments}.
\looseness=-1

\section{Preliminaries}\label{sec:Preliminaries}

We assume familiarity with group theory; our notation follows \cite{BojanczykKL14,VenhoekMR22}.
For a group $G$, we write $g_1 g_2$ for the multiplication (applying $g_2$ first), $\mathrm{id}$ for the unit, and $H \le G$ for a subgroup $H$.
For a set $X$, the \emph{symmetric group} $\Sym(X)$ consists of all permutations (bijections) of $X$.
We write $\mathbb{N}$ for the naturals and $[n] = \{1, \dots, n\}$, and abbreviate the symmetric group on $[n]$ as $S_n = \Sym([n])$.

\para{$G$-sets.}
A \emph{$G$-set} $X$ carries a group action ${\cdot}\colon G \times X \to X$.
A map $f\colon X \to Y$ between $G$-sets is \emph{equivariant} if $f(g \cdot x) = g \cdot f(x)$, and a subset $A \subseteq X$ is \emph{equivariant} if $g \cdot A = A$ for all $g \in G$.
The \emph{orbit} of $x \in X$ is $G \cdot x = \{g \cdot x \mid g \in G\}$.
We write $\orb(X)$ for the set of orbits, so $X = \bigsqcup_{O \in \orb(X)} O$.
A $G$-set $X$ is \emph{transitive} (or a \emph{single orbit}) if $\orb(X)$ is a singleton.

\para{Data symmetries and nominal sets.}
A \emph{data symmetry}~\cite{BojanczykKL14} is a pair $\mathcal{S} = (\mathbb{A}, G)$ of a countably infinite set $\mathbb{A}$ of \emph{atoms} and a subgroup $G \le \Sym(\mathbb{A})$.
We focus on two symmetries: the \emph{equality symmetry} $\Eq = (\mathbb{N}, \Sym(\mathbb{N}))$ and the \emph{total-order symmetry} $\Ord = (\mathbb{Q}, \Aut(\mathbb{Q}, <))$, where $\Aut(\mathbb{Q}, <)$ comprises all order-preserving permutations.

For a finite set $C \subseteq \mathbb{A}$, its \emph{pointwise stabilizer} is $\Fix_G(C) = \{g \in G \mid \forall a \in C.\, g(a) = a\}$.
We say $C$ \emph{supports} an element $x$ of a $G$-set if $\Fix_G(C) \subseteq \Stab_G(x)$, where $\Stab_G(x) = \{g \in G \mid g \cdot x = x\}$ is the \emph{stabilizer} of $x$.
A \emph{nominal set} over $\mathcal{S}$ is a $G$-set where every element has a finite support.
We denote by $\Nom_{\mathcal{S}}$ the category of nominal sets and equivariant maps, and by $\Nom_{\mathcal{S}}^1$ its full subcategory of single-orbit nominal sets.
A nominal set $X$ is \emph{orbit-finite} if $\orb(X)$ is finite; we write $N(X) = |\orb(X)|$ for its number of orbits.
Over $\Eq$ and $\Ord$, orbit-finite nominal sets play the role of ``finite'' sets in nominal computation: they admit finite descriptions and are closed under equivariant subsets and finite products~\cite{BojanczykKL14}.

\para{Support and dimension.}
A data symmetry \emph{admits least supports} if every element $x$ in any nominal set has a unique least support $\supp(x)$.
Both $\Eq$ and $\Ord$ have this property~\cite{BojanczykKL14,VenhoekMR22}.
The \emph{dimension} of $x$ is $\dim(x) = |\supp(x)|$.
Since $\supp(g \cdot x) = g \cdot \supp(x)$, all elements of an orbit $O$ share the same dimension, written $\dim(O)$; for an orbit-finite set $X$ we define $\dim(X) = \max_{O \in \orb(X)} \dim(O)$.
The \emph{local symmetry} of $x$ is the restriction $\Stab_G(x)|_{\supp(x)}$ of its stabilizer to its least support; it is a subgroup of $\Sym(\supp(x))$, since every $g \in \Stab_G(x)$ satisfies $g \cdot \supp(x) = \supp(g \cdot x) = \supp(x)$.
When least supports exist, every single-orbit nominal set is determined up to isomorphism by a \emph{support representation} $(C, S)$ of a finite support $C \subseteq \mathbb{A}$ and a local symmetry group $S \le \Sym(C)$~\cite[Prop.~9.15]{BojanczykKL14}.

\begin{xexample}\label{ex:NominalSets}
  We illustrate these notions over $\Eq$:
  \begin{itemize}
    \item The set of atoms $\mathbb{N}$ with action $g \cdot a = g(a)$ is a single orbit (\ie, $N(\mathbb{N}) = 1$).
      Since $\supp(a) = \{a\}$, it has dimension $1$ and trivial local symmetry.
    \item A product $X \times Y$ of nominal sets carries the pointwise action $g \cdot (x, y) = (g \cdot x, g \cdot y)$.
      Thus $\mathbb{N}^2$ splits into two orbits: the diagonal $\{(a, a) \mid a \in \mathbb{N} \}$ of dimension $1$, and the off-diagonal $\{(a, b) \mid a \neq b\}$ of dimension $2$.
      Both have trivial local symmetries.
    \item The powerset of a nominal set has the action $g \cdot A = \{g \cdot a \mid a \in A\}$.
      The set of two-element subsets $\{\{a, b\} \mid a \neq b\}$ forms a single orbit of dimension $2$, with local symmetry $\Sym(\{a, b\}) \cong S_2$ swapping the atoms.
    \item Any set $X$ with trivial action $g \cdot x = x$ is a nominal set with dimension $0$.
  \end{itemize}
\end{xexample}

\begin{xremark}[Local symmetry over $\Eq$ and $\Ord$]\label{rem:LocalSymmetry}
  Under $\Eq$, a dimension-$d$ element can carry any local symmetry $S \le \Sym([d])$; under $\Ord$, local symmetries are always trivial~\cite[Lem.~5.1]{VenhoekMR22}, so each orbit is determined solely by its dimension.
  This triviality comes at the cost of more orbits: the set of $d$-tuples of distinct atoms, a single $\Eq$-orbit of dimension $d$ with trivial local symmetry, splits into $d!$ $\Ord$-orbits when taken over the atoms $\mathbb{Q}$ of $\Ord$.
  For example, the off-diagonal $\{(a, b) \mid a \neq b\}$, a single $\Eq$-orbit, splits into $\{(a, b) \mid a < b\}$ and $\{(a, b) \mid a > b\}$ over $\Ord$.
\end{xremark}

\para{Nominal automata.}
Fix an orbit-finite nominal set $\Sigma$ as the input alphabet.
The set $\Sigma^\ast$ of finite words carries the pointwise action $g \cdot (a_1 \cdots a_n) = (g \cdot a_1) \cdots (g \cdot a_n)$; an \emph{equivariant language} is an equivariant subset $L \subseteq \Sigma^\ast$.

\begin{xdefinition}[nominal automaton]\label{def:NominalAutomaton}
  A \emph{(deterministic, orbit-finite) nominal automaton} is a tuple $\mathcal{A} = (\Sigma, Q, q_0, \delta, F)$ where $Q$ is an orbit-finite nominal set of states, $q_0 \in Q$ is an initial state with $\supp(q_0) = \emptyset$, $\delta\colon Q \times \Sigma \to Q$ is an equivariant transition map, and $F \subseteq Q$ is an equivariant set of accepting states.
\end{xdefinition}
The transition map extends to words by $\delta^\ast(q, \varepsilon) = q$ and $\delta^\ast(q, wa) = \delta(\delta^\ast(q, w), a)$, and the language of $\mathcal{A}$ is $L(\mathcal{A}) = \{w \in \Sigma^\ast \mid \delta^\ast(q_0, w) \in F\}$, which is an equivariant language.
We call $\mathcal{A}$ \emph{reachable} if every state is of the form $\delta^\ast(q_0, w)$ for some $w \in \Sigma^\ast$.

\para{Nominal Myhill--Nerode theorem and minimal nominal automata.}
For an equivariant language $L \subseteq \Sigma^\ast$, the \emph{Myhill--Nerode equivalence} is $u \sim_L v \iff \forall w \in \Sigma^\ast.\, (uw \in L \iff vw \in L)$.
As $\sim_L$ is equivariant, the quotient $\Sigma^\ast / {\sim_L}$ is again a nominal set.
The nominal Myhill--Nerode theorem~\cite{BojanczykKL14} states that $L$ is recognizable by a nominal automaton iff $\Sigma^\ast / {\sim_L}$ is orbit-finite.
In this case, the \emph{minimal nominal automaton} $\mathcal{A}_{\min}$ takes $\Sigma^\ast / {\sim_L}$ as its states, with initial state $[\varepsilon]_L$, transitions $\delta_{\min}([u]_L, a) = [ua]_L$, and accepting states $\{[u]_L \mid u \in L\}$.
Among all nominal automata recognizing $L$, $\mathcal{A}_{\min}$ simultaneously has the fewest state orbits and the smallest state dimension, as the reachable part of any such automaton maps onto it by an equivariant surjection, but its states may carry non-trivial local symmetries~\cite[Sec.~6.2]{BojanczykKL14}.

\para{Categories and functors.}
We assume the basic vocabulary of category theory (see standard texts, \eg, \cite{MacLane98}) and use $e\colon X \twoheadrightarrow Y$ for epimorphisms, $m\colon Z \hookrightarrow X$ for monomorphisms, $[\mathscr{C}, \mathscr{D}]$ for functor categories, and $|\mathscr{C}|$ for the discrete category on the objects of $\mathscr{C}$.
A functor is \emph{fully faithful} if it is bijective on hom-sets, and an \emph{equivalence} $\mathscr{C} \simeq \mathscr{D}$ is a fully faithful, essentially surjective functor.
An adjunction $F \dashv U\colon \mathscr{C} \leftrightarrows \mathscr{D}$ induces a monad $T = UF$ on $\mathscr{D}$; we use its Eilenberg--Moore category $\EM(T)$ of $T$-algebras and its Kleisli category $\Kl(T)$.

\para{Free completions and projective objects.}
The \emph{free coproduct completion} $\Fam(\mathscr{C})$ has objects $(A_i)_{i \in I}$ (set-indexed families of $\mathscr{C}$-objects) and morphisms $(\rho, (f_i)_{i \in I})$ with $\rho\colon I \to J$ and $f_i\colon A_i \to B_{\rho(i)}$.
An object $P$ is \emph{projective} if morphisms $P \to Y$ lift along epimorphisms $X \twoheadrightarrow Y$, and $\Proj(\mathscr{C})$ is the full subcategory of projective objects.
A \emph{projective cover} of $X$ is a pair $(P, e)$ of a projective object $P$ and a superfluous epimorphism $e\colon P \twoheadrightarrow X$, meaning $e$ restricted to any proper subobject of $P$ is no longer an epimorphism.

\para{Nominal sets and the functor category $[\I, \Set]$.}
Let $\I$ denote the category of finite subsets of $\mathbb{N}$ and injections.
It is established~\cite{KurzPV10,Petrisan12} that there is an adjunction ${F \dashv U}\colon \Nom_{\Eq} \leftrightarrows [|\I|, \Set]$ with $U(X)(C) = \{x \in X \mid \supp(x) \subseteq C\}$ and $F(Y) = \coprod_C Y(C) \times \Inj(C, \mathbb{N})$.
The induced monad $T = UF$ has $\mathrm{EM}(T) \simeq [\I, \Set]$.
The comparison functor $J\colon \Nom_{\Eq} \to [\I, \Set]$, $J(X)(C) = \{x \in X \mid \supp(x) \subseteq C\}$, is fully faithful but not essentially surjective~\cite{KurzPV10,Petrisan12}, embedding $\Nom_{\Eq}$ as a full subcategory of the functor category $[\I, \Set]$.

\section{Strong Nominal Sets}\label{sec:StrongNominalSets}

This section introduces \emph{strong nominal sets}: nominal sets with trivial local symmetries, hence computationally lightweight.
We fix a data symmetry $\mathcal{S} = (\mathbb{A}, G)$ admitting least supports.

\begin{xdefinition}[strong support and strong nominal sets]\label{def:StrongNominalSet}
  Let $x$ be an element of a nominal set $X$ with least support $C = \supp(x)$.
  We say $C$ \emph{strongly supports} $x$ if $\Stab_G(x) \subseteq \Fix_G(C)$, \ie, $\Stab_G(x) = \Fix_G(C)$, as $\Fix_G(C) \subseteq \Stab_G(x)$ holds because $C$ supports $x$.
  A nominal set $X$ is \emph{strong} if every element is strongly supported by its least support.
\end{xdefinition}
We write $\SNom_{\mathcal{S}}$ for the full subcategory of $\Nom_{\mathcal{S}}$ on the strong nominal sets, and $\SNom_{\mathcal{S}}^1$ for its full subcategory of single-orbit strong nominal sets.
Strong support was originally introduced in nominal game semantics to ensure well-defined strategy composition~\cite{Tzevelekos07,Tzevelekos08,MurawskiT16}, there defined for an arbitrary finite supporting set.
A few of its categorical properties were briefly noted in the literature~\cite{KurzPV10,Petrisan12}.

\begin{xexample}\label{ex:StrongNominalSets}
  Over $\Eq$, the atoms $\mathbb{N}$ and the product $\mathbb{N}^2$ from \cref{ex:NominalSets} are strong.
  The set $\{\{a, b\} \mid a \neq b\}$ is \emph{not} strong: the swap $(a\,b)$ satisfies $(a\,b) \cdot \{a, b\} = \{a, b\}$.
  Its ordered counterpart $\{(a, b) \mid a \neq b\}$ has the same dimension and orbit count, but is strong.
  More generally, every \emph{polynomial orbit-finite (pof) set} $\mathbb{N}^{n_1} + \cdots + \mathbb{N}^{n_k}$~\cite[Def.~1.1]{BojanczykSIS} is strong, but not conversely: up to isomorphism, the orbit-finite strong nominal sets are exactly the equivariant subsets of pof sets.
  For instance, the off-diagonal $\{(a, b) \mid a \neq b\} \subsetneq \mathbb{N}^2$ is strong but is a proper equivariant subset, not a pof set itself.
\end{xexample}

As suggested above, strong support is inherently tied to the triviality of local symmetries.

\begin{restatable}[local-symmetry characterization]{xproposition}{LocalSymmetry}\label{prop:LocalSymmetry}
  A nominal set $X$ is strong if and only if every element $x \in X$ has trivial local symmetry, \ie, $\Stab_G(x)|_{\supp(x)} = \{\mathrm{id}\}$.
\end{restatable}

Recall that a single-orbit nominal set is determined up to isomorphism by a support representation $(C, S)$. For strong nominal sets, $S = \{\mathrm{id}\}$, meaning orbits are uniquely determined by their dimension $|C|$.
Over $\Ord$, local symmetries are always trivial~\cite[Lem.~5.1]{VenhoekMR22}, hence $\SNom_{\Ord} = \Nom_{\Ord}$.

\subsection{Categorical Characterization}\label{ssec:CategoricalCharacterization}

From now on, we fix $\mathcal{S} = \Eq$ and drop it from the notation.
Recall that $\Nom$ is equivalent to the Schanuel topos $\Sh(\I^\mathrm{op})$ (sheaves for the atomic topology) and to the category $\NSet_\I$ of \emph{named sets with symmetries}, where elements carry permutation groups~\cite{FerrariMP02,GadducciMM06,Pitts13}; the notation $\NSet_\I$ follows~\cite[Sec.~5.1]{Staton07}.
The subcategory $\SNom$ admits an analogous but simpler description where all local symmetry groups are trivial.
\looseness=-1

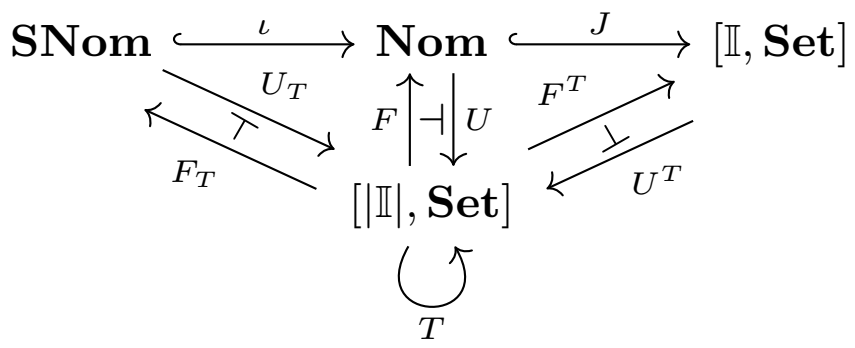
\begin{figure}[bt]
  \centering
  \footnotesize
  \begin{tikzcd}[column sep=3.4em, row sep=1.8em]
    \SNom
      \arrow[r, hook, "\iota"]
      \arrow[dr, "U_T", shift left=1, shorten <=3pt, shorten >=3pt, ""{name=UTL}]
    &
    \Nom
      \arrow[r, hook, "J"]
      \arrow[d, "U", shift left=2, ""{name=UNom}]
    &
    {[\I, \Set]}
      \arrow[dl, "U^T", shift left=3, shorten <=3pt, shorten >=3pt, ""{name=UTR}]
    \\
    &
    {[|\I|, \Set]}
      \arrow[u, "F", shift left=2, ""{name=FNom}]
      \arrow[ul, "F_T", shift left=3, shorten <=3pt, shorten >=3pt, ""{name=FTL}]
      \arrow[ur, "F^T", shift left=1, shorten <=3pt, shorten >=3pt, ""{name=FTR}]
      \arrow["T"', loop, out=-120, in=-60, distance=2em]
    &
    \arrow[from=UTL, to=FTL, phantom, "{\dashv}"{rotate=60}]
    \arrow[from=FNom, to=UNom, phantom, "\dashv"]
    \arrow[from=UTR, to=FTR, phantom, "{\dashv}"{rotate=-60}]
  \end{tikzcd}
  \caption{The categorical relationship between $\SNom$, $\Nom$, and $[\I, \Set]$.}
  \label{fig:CategoricalRelationship}
\end{figure}

\begin{xtheorem}[categorical characterization of strong nominal sets]\label{thm:Characterization}
  Let $T = U F$ be the monad on $[|\I|, \Set]$ induced by the adjunction $F \dashv U\colon \Nom \leftrightarrows [|\I|, \Set]$ of \cref{sec:Preliminaries}.
  Then, there are equivalences of categories
  \begin{equation*}
    \SNom
      \;\simeq\; \Kl(T)
      \;\simeq\; \Fam(\I^\mathrm{op})
      \;\simeq\; \Proj([\I, \Set])
      \;\simeq\; \NSet_\mathrm{fn},
  \end{equation*}
  where $\NSet_\mathrm{fn}$ is the category of \emph{finitely named sets}~\cite{MontanariP97}: sets whose elements each carry a finite name set, related by injective renamings, as detailed in \cref{app:NSet}.
\end{xtheorem}
We prove these in \cref{app:Characterization}; the sequel builds on $\Fam(\I^\mathrm{op})$ and $\Proj([\I, \Set])$, while the other two equivalences relate $\SNom$ to prior frameworks:
\begin{itemize}
  \item $\SNom \simeq \Kl(T)$ identifies $\SNom$ with the \emph{free} $T$-algebras~\cite{KurzPV10,Petrisan12}, as depicted in \cref{fig:CategoricalRelationship}.
  \item $\SNom \simeq \Fam(\I^\mathrm{op})$ underlies the array representation of \cref{ssec:DataRepresentations} via $\SNom^1 \simeq \I^\mathrm{op}$.
  \item $\SNom \simeq \Proj([\I, \Set])$ provides the correct setting for the projective covers of \cref{sec:ProjectiveCovers}\footnote{%
    \cite[Exercise~116]{Bojanczyk19} poses that an equivariant orbit-finite set is \emph{straight}~\cite[Def.~6.7]{Bojanczyk19}, equivalently strong~\cite[Footnote~3]{Bojanczyk19}, iff it is projective in $\Nom$; its solution is left blank, and the new draft~\cite{BojanczykSIS} omits the exercise.
    As posed this fails: only $\emptyset$ is projective in $\Nom$, as \cref{app:Proj} shows.
    The lifting argument of \cite[Claim~6.10]{Bojanczyk19} makes straight sets projective relative to \emph{support-reflecting} surjections; \cref{thm:Characterization} removes this side condition by moving to $[\I, \Set]$.%
  }.
  \item $\SNom \simeq \NSet_\mathrm{fn}$ links $\SNom$ to the early named-set development~\cite{MontanariP97}.
\end{itemize}

\subsection{Data Representations}\label{ssec:DataRepresentations}

From the equivalence $\SNom \simeq \Fam(\I^\mathrm{op})$ of \cref{thm:Characterization}, we derive a concrete, constraint-free representation of data and equivariant maps, the basis of our implementation in \cref{sec:ImplementationAndExperiments}.
It is the $\Eq$ counterpart of the sorted array representation for $\Nom_\Ord$~\cite{VenhoekMR22}: since $\SNom_\Ord = \Nom_\Ord$, theirs covers all of $\Nom_\Ord$, whereas ours captures precisely the strong nominal sets.
\looseness=-1

\para{Array representation.}
Under $\SNom^1 \simeq \I^\mathrm{op}$, a single-orbit strong nominal set is isomorphic to $\Inj(C, \mathbb{N})$ for a finite set $C$. Choosing $C = [d]$, we identify the orbit with its \emph{dimension} $d$, and an element with an injection $[d] \hookrightarrow \mathbb{N}$, \ie, a \emph{tuple of $d$ distinct atoms} $\langle a_1, \dots, a_d \rangle$.
By $\SNom \simeq \Fam(\I^\mathrm{op})$, any orbit-finite strong nominal set $X \cong \coprod_{i \in I} \Inj([d_i], \mathbb{N})$ is determined by a finite list of dimensions $(d_i)_{i \in I}$, with elements represented as pairs $(i, \langle a_1, \dots, a_{d_i} \rangle)$.
We call this the \emph{array representation}.
Unlike over $\Ord$, where elements are uniquely determined by their least supports as \emph{sorted} arrays~\cite{VenhoekMR22}, an element over $\Eq$ is an arbitrary array of $d$ distinct atoms: its least support is the set of its entries, and the array additionally fixes their order.

Equivariant maps are similarly concrete.
An equivariant map $\Inj([m], \mathbb{N}) \to \Inj([n], \mathbb{N})$ is precomposition with a unique injection $h\colon [n] \hookrightarrow [m]$, mapping elements via $\langle a_1, \dots, a_m \rangle \mapsto \langle a_{h(1)}, \dots, a_{h(n)} \rangle$.
Generally, an equivariant map is represented by assigning each source orbit a target orbit and an index array $\langle h(1), \dots, h(n) \rangle$.
Consequently, both elements and functions are represented as finite integer arrays.
\looseness=-1

\para{Connection to register automata.}
This array representation concretely matches the configuration spaces of register automata with injective (possibly partial) register assignments~\cite{KaminskiF94,BojanczykKL14}.
A configuration consisting of a state $q$ and an assignment of $d_q$ distinct atoms naturally forms an element $(q, \langle a_1, \dots, a_{d_q} \rangle) \in \coprod_{q \in Q} \Inj([d_q], \mathbb{N})$, which is an element of an orbit-finite strong nominal set.
For instance, the configuration space of \cref{fig:Res2} forms a strong nominal set with three orbits: no open resources $(q_0, \langle \rangle)$, one $(q_1, \langle a \rangle)$, and two $(q_2, \langle a, b \rangle)$.
Thus, register-automaton configurations are not merely nominal, but \emph{strong} nominal sets, naturally justifying their array representation; cf.\ the closely related observation for pDRAs~\cite[Rem.~6]{BalachanderFGT25}.
\looseness=-1

\para{Products.}
Products have a less transparent orbit structure, as they mix atoms from two factors.
For single orbits $X$ and $Y$ of dimensions $m$ and $n$, a pair $(\langle a_1, \dots, a_m \rangle, \langle b_1, \dots, b_n \rangle)$ is classified up to the $G$-action by its shared atoms, captured by the relation $(i, j) \in \pi \iff a_i = b_j$; distinctness on each side makes $\pi$ a partial bijection $\pi\colon [m] \rightharpoonup [n]$.
Thus, the orbits of $X \times Y$ correspond bijectively to the partial bijections $[m] \rightharpoonup [n]$, where the orbit for $\pi$ has dimension $m + n - |\dom(\pi)|$; this specializes the product representations of~\cite[Sec.~10.3]{BojanczykKL14}~\cite[Sec.~5]{CianciaKM10} to strong nominal sets, with~\cite[Sec.~5.3]{VenhoekMR22} as the $\Ord$ analogue.
Each such orbit is itself strong, so strong nominal sets are closed under finite products.
Counting these partial bijections yields the number of product orbits:
\looseness=-1

\begin{xproposition}[number of product orbits]\label{prop:Product}
  Let $X$ and $Y$ be single strong orbits of dimensions $m$ and $n$.
  The number of orbits of $X \times Y$ equals the number of partial bijections, that is,
  \looseness=-1
  \begin{equation*}
    N(X \times Y) = \sum_{k=0}^{\min(m, n)} \binom{m}{k} \binom{n}{k} \, k!.
  \end{equation*}
\end{xproposition}
This count grows factorially in $\min(m, n)$, a combinatorial blow-up that dominates the costs of the algorithms of \cref{sec:AutomataAlgorithms}, which enumerate these orbits.

\para{Effective operations.}
The array representation renders standard nominal operations effective.
In practice, we manipulate \emph{orbit-finite} strong nominal sets, typically presented as equivariant subsets.
They are closed under the following operations:
\begin{itemize}
  \item union $X \cup Y$ and intersection $X \cap Y$ within a common strong nominal set;
  \item coproduct $X \sqcup Y$ and finite product $X \times Y$;
  \item image $\{ f(x) \mid x \in X \}$ along an equivariant map $f\colon X \to Y$;
  \item filtering $\{x \in X \mid p(x)\}$ by an equivariant predicate $p\colon X \to \{0, 1\}$.
\end{itemize}
Furthermore, the representation supports basic decision and inspection primitives:
\begin{itemize}
  \item checking membership $x \in X$, inclusion $Y \subseteq X$, and emptiness $X = \emptyset$;
  \item enumerating orbits and selecting representative elements.
\end{itemize}
\Cref{app:Operations} details the worst-case costs of all these operations.

\section{Projective Covers}\label{sec:ProjectiveCovers}

Unlike the operations in \cref{ssec:DataRepresentations}, quotients \emph{fall outside} the class of strong nominal sets.

\begin{xexample}[strong nominal sets are not closed under quotients]\label{ex:QuotientNotStrong}
  Recall from \cref{ex:StrongNominalSets} that the off-diagonal $X = \{(a, b) \mid a \neq b\}$ is a strong single orbit.
  Consider the equivalence relation
  \looseness=-1
  \begin{equation*}
    R = \{((a, b), (a, b)) \mid a \neq b\} \cup \{((a, b), (b, a)) \mid a \neq b\},
  \end{equation*}
  which identifies $(a, b)$ with its swap $(b, a)$, so that its classes are the unordered pairs $\{a, b\}$.
  As an equivariant subset of $X \times X$, the relation $R$ is strong.
  However, as noted in \cref{ex:StrongNominalSets}, the quotient $X / R \cong \{\{a, b\} \mid a \neq b\}$ is \emph{not} strong.
\end{xexample}

Quotients are unavoidable in automata theory: the minimal automaton's state space $\Sigma^\ast / {\sim_L}$ typically falls outside $\SNom$; for the running example $\mathrm{Res}(2)$, the Myhill--Nerode classes with two open resources are exactly these unordered pairs, as observed in \cref{sec:Introduction}.
To compute with such objects in the array representation, we approximate them using \emph{projective covers}.
We develop this notion in \cref{ssec:ProjectiveCoversAsStrongApproximation} and apply it to automata in \cref{ssec:MinimalStrongNominalAutomata}.
\looseness=-1

\subsection{Projective Covers as Strong Approximation}\label{ssec:ProjectiveCoversAsStrongApproximation}

By \cref{thm:Characterization}, the functor $J' = J \circ \iota \colon \SNom \to [\I, \Set]$ identifies $\SNom$ with $\Proj([\I, \Set])$.
This projectivity already makes strong nominal sets convenient domains, as maps from them lift along epimorphisms.
Moreover, projective covers always exist:

\begin{restatable}[projective cover existence]{xtheorem}{ProjectiveCover}\label{thm:ProjectiveCover}
  Every object $\mathscr{F} \in [\I, \Set]$ admits a projective cover $(J'(X'), e)$ with a strong nominal set $X'$ and a superfluous epimorphism $e\colon J'(X') \twoheadrightarrow \mathscr{F}$.
\end{restatable}

We provide an explicit construction in \cref{app:ProjectiveCovers}.
Being projective with a superfluous epimorphism, $J'(X')$ is a retract of every projective object mapping onto $\mathscr{F}$, so it approximates $\mathscr{F}$ optimally.
When applied to $J(X)$ for $X \in \Nom$, full faithfulness of $J$ yields a unique equivariant map $p\colon X' \to X$ with $J(p) = e$, and since $e$ is a pointwise epimorphism, $p$ is an equivariant surjection.
This $p$ induces a surjective map $\orb(p)\colon \orb(X') \twoheadrightarrow \orb(X)$ on orbits.
Crucially, the projective cover merges no orbits and inflates no dimensions:

\begin{restatable}[orbit-count preservation]{xtheorem}{OrbitCount}\label{thm:OrbitCount}
  Let $X \in \Nom$ and let $p\colon X' \twoheadrightarrow X$ be the equivariant surjection corresponding to the projective cover
  of $J(X)$, as above.
  Then, $\orb(p)$ is a dimension-preserving bijection, \ie, $\dim(O') = \dim(\orb(p)(O'))$ for every orbit $O'$ of $X'$.
  \looseness=-1
\end{restatable}

By \cref{app:ProjectiveCovers}, the projective cover is unique up to isomorphism, so we call $X'$ \emph{the strong approximation} of $X$.
Intuitively, $p$ strips local symmetries from $X$ without merging orbits or inflating dimensions; the discarded symmetry is recorded in the kernel pair of $p$.
For instance, the non-strong orbit $X = \{\{a, b\} \mid a \neq b\}$ has the strong approximation $X' = \{(a, b) \mid a \neq b\}$ with $p(a, b) = \{a, b\}$.
Since $\orb(p)$ is bijective, $X$ is orbit-finite iff $X'$ is, with $N(X') = N(X)$; the existence of such a surjection is essentially \cite[Cor.~4.14]{BojanczykKL14}, applied orbit by orbit.

\subsection{Strong Lifts of the Minimal Nominal Automaton}\label{ssec:MinimalStrongNominalAutomata}

By combining projectivity with the orbit-count preservation of \cref{thm:OrbitCount}, we transform any nominal automaton into a language-equivalent strong automaton with the same orbit count and dimensions, which we call a \emph{strong lift}.
We say a nominal automaton is \emph{strong} if its alphabet and state space are both strong nominal sets.
Throughout, we assume the alphabet $\Sigma$ is strong, \ie, a strong nominal set, and call it a \emph{strong alphabet}; this holds for typical data alphabets such as tuples of pairwise distinct atoms.

\begin{restatable}[strong lifts of automata]{xtheorem}{StrongApproximation}\label{thm:StrongApproximation}
  Let $\mathcal{A} = (\Sigma, Q, q_0, \delta, F)$ be a nominal automaton over a strong alphabet $\Sigma$, and let $p\colon Q' \twoheadrightarrow Q$ be the equivariant surjection corresponding to the projective cover of $J(Q)$.
  Then, there exists a strong nominal automaton $\mathcal{A}' = (\Sigma, Q', q_0', \delta', F')$ for which $p$ is an automaton homomorphism, that is, $p(q_0') = q_0$, $F' = p^{-1}(F)$, and $p(\delta'(q', \sigma)) = \delta(p(q'), \sigma)$ for all $q' \in Q'$ and $\sigma \in \Sigma$; in particular, $L(\mathcal{A}') = L(\mathcal{A})$.
  \looseness=-1
\end{restatable}

\begin{proof}[Proof sketch]
  Since \cref{thm:ProjectiveCover} ensures $Q'$ is strong and $\Sigma$ is strong by assumption, $J'(Q' \times \Sigma)$ is projective by \cref{thm:Characterization}, so the map $\delta \circ (p \times \mathrm{id}_\Sigma)$ lifts along $p$ via projectivity; furthermore, \cref{thm:OrbitCount} dictates that $q_0'$ is the unique dimension-0 element in the pre-image of $q_0$, and we set $F' = p^{-1}(F)$.
  Full details appear in \cref{app:MinimalStrongNominalAutomata}.
\end{proof}

Consequently, by \cref{thm:OrbitCount}, $\orb(p)$ is a dimension-preserving bijection, so $Q'$ inherits the orbit structure of $Q$.
Specializing to the minimal automaton yields our main result:

\begin{xcorollary}[strong lift of the minimal nominal automaton]\label{cor:MinimalStrong}
  For a strong alphabet $\Sigma$, every equivariant language $L \subseteq \Sigma^\ast$ recognized by a nominal automaton is also recognized by a strong nominal automaton whose state set has the same orbit count and dimensions as the minimal nominal automaton $\mathcal{A}_{\min}$ of $L$.
  \looseness=-1
\end{xcorollary}

\begin{xremark}[relation to register automata]\label{rem:PriorConstructions}
  \Cref{cor:MinimalStrong} can be assembled from known results: combining the least-support-preserving surjections of~\cite[Cor.~4.14]{BojanczykKL14} with the transition-lifting lemma~\cite[Lem.~6.7]{BojanczykKL14} yields what we call a strong lift of the minimal automaton (a similar statement and observation appeared in~\cite[Lem.~6.9]{Bojanczyk19} and~\cite{BalachanderFGT25}).
  Our contribution is to identify the structure behind these constructions as the projective cover: the resulting state space is exactly the strong approximation of the minimal one, unique up to isomorphism and, crucially for \cref{sec:AutomataAlgorithms}, computable for quotients of strong nominal sets by \cref{lem:Quotient}.
  \looseness=-1
\end{xremark}

\begin{figure}[tb]
  \centering\small
  \tikzset{
    snaut/.style={baseline={(current bounding box.center)}, >={Stealth[]}, on grid, semithick},
  }
  \scalebox{0.8}{
  \begin{tabular}{c@{\hspace{2em}}c}
    \begin{tikzpicture}[snaut, node distance=1.5cm]
      \node (s) {$\{a, b\}$};
      \node (t) [right=of s] {$\{a, b\}$};
      \draw[->] (s) -- node[above] {$a, b$} (t);
    \end{tikzpicture}
    &
    \begin{tabular}{c@{\quad}c}
      \begin{tikzpicture}[snaut, node distance=1.7cm]
        \node (s) {$(a, b)$};
        \node (t) [right=of s] {$(a, b)$};
        \draw[->] (s) to[bend left=20] node[above] {$a$} (t);
        \draw[->] (s) to[bend right=20] node[below] {$b$} (t);
      \end{tikzpicture}
      &
      \begin{tikzpicture}[snaut, node distance=1.7cm]
        \node (s) {$(a, b)$};
        \node (t) [right=of s] {$(b, a)$};
        \draw[->] (s) to[bend left=20] node[above] {$a$} (t);
        \draw[->] (s) to[bend right=20] node[below] {$b$} (t);
      \end{tikzpicture}
      \\[0.3ex]
      \begin{tikzpicture}[snaut]
        \node (s) {$(a, b)$};
        \node (t1) [above right=0.5cm and 1.6cm of s] {$(a, b)$};
        \node (t2) [below right=0.5cm and 1.6cm of s] {$(b, a)$};
        \node[draw, dotted, rounded corners, inner sep=3pt, fit=(t1)(t2)] {};
        \draw[->] (s) -- node[above left=-2pt] {$a$} (t1);
        \draw[->] (s) -- node[below left=-2pt] {$b$} (t2);
      \end{tikzpicture}
      &
      \begin{tikzpicture}[snaut]
        \node (s) {$(a, b)$};
        \node (t1) [above right=0.5cm and 1.6cm of s] {$(b, a)$};
        \node (t2) [below right=0.5cm and 1.6cm of s] {$(a, b)$};
        \node[draw, dotted, rounded corners, inner sep=3pt, fit=(t1)(t2)] {};
        \draw[->] (s) -- node[above left=-2pt] {$a$} (t1);
        \draw[->] (s) -- node[below left=-2pt] {$b$} (t2);
      \end{tikzpicture}
    \end{tabular}
    \\[0.3ex]
    {\small non-strong transition} & {\small the four projective-cover lifts}
  \end{tabular}
  }
  \caption{%
    Non-uniqueness of strong lifts.
    The non-strong transition (left) has four projective-cover lifts (right): two uniform (top) and two mixed (bottom), whose targets share an orbit (dotted).%
  }
  \label{fig:NonUniqueLift}
\end{figure}
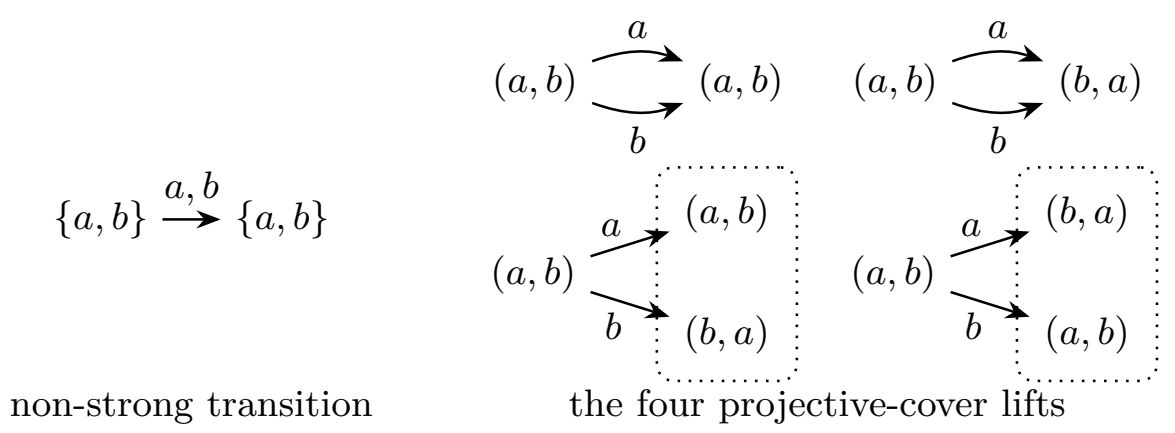

However, strong lifts are \emph{not} canonical: the lifted transition $\delta'$ is generally non-unique (per domain transition orbit, the valid choices of $\delta'$ into a target orbit with local symmetry group $H$ are in bijection with $H$).
A further cost is that the domain $Q' \times \Sigma$ may contain more orbits than $Q \times \Sigma$, as local symmetries that previously fused transition orbits are removed (\eg, the two $\mathsf{Close}$ transitions from $q_2$ in \cref{fig:Res2} form a single orbit in the minimal automaton).
\Cref{fig:NonUniqueLift} illustrates both: a non-strong self-loop on input $\{a, b\}$ with local symmetry $S_2$ (left) yields \emph{four} possible projective-cover lifts (right).
Stripping $S_2$ splits the single transition orbit into two, each with two valid target choices.
This non-uniqueness was already observed in~\cite{BojanczykKL14,Bojanczyk17}, where transitions are lifted via per-orbit representative choices; in our formulation, the valid lifts differ exactly by the discarded local symmetries, recorded in the kernel pair of $p$.

This orbit inflation is the deliberate price of enforcing trivial local symmetries; as our experiments show in \cref{sec:ImplementationAndExperiments}, the computational savings of the constraint-free array representation of \cref{ssec:DataRepresentations} far outweigh this overhead.

\section{Automata Algorithms}\label{sec:AutomataAlgorithms}

We translate the developments of \cref{sec:ProjectiveCovers} into algorithms.
\Cref{ssec:LanguageEquivalence} presents a language-equivalence check for strong nominal automata that runs in polynomial time for a fixed alphabet dimension, generalizing the register-automaton equivalence algorithm~\cite{MurawskiRT18}.
\Cref{ssec:MinimizationLearning} addresses minimization and learning via a \emph{quotient operation} that computes the strong approximation of the quotient by the Myhill--Nerode equivalence; this operation is polynomial in the orbit count but, by \cref{prop:Product}, factorial in the dimension.
All algorithms are implemented in our library \Namae{} and evaluated in \cref{sec:ImplementationAndExperiments}.
\looseness=-1

\para{Computational model.}
All inputs (automata and alphabets) are presented in the array representation of \cref{ssec:DataRepresentations}, hence strong, and we measure algorithmic cost by the number of orbits $N(X)$ and the dimension $\dim(X)$; this loses no generality, as \cref{thm:StrongApproximation} provides language-equivalent strong lifts preserving both measures.
We assume orbit-level primitives (mapping an element to its orbit, comparing two orbits, selecting a representative) take constant time, as in~\cite[Sec.~8]{VenhoekMR22}; \cref{app:Operations} details the costs of the effective operations under this assumption.
Two extra quantities appear in the algorithmic costs:
$\mu(d)$ bounds the cost of Schreier--Sims operations on a subgroup of $S_d$ (polynomial in $d$)~\cite{FurstHL80}; and $\beta(m, n)$ is the number of partial bijections $[m] \rightharpoonup [n]$, \ie, the product-orbit count of \cref{prop:Product}.
For a fixed $n$, $\beta(m, n) = \Theta(m^n)$ is polynomial in $m$, whereas $\beta(d, d) = O(2^d d!)$ grows factorially.
\looseness=-1

\subsection{Language Equivalence}\label{ssec:LanguageEquivalence}

Let $\mathcal{A}_1$ and $\mathcal{A}_2$ be strong nominal automata with state spaces $Q_1$ and $Q_2$, and suppose we wish to decide whether $L(\mathcal{A}_1) = L(\mathcal{A}_2)$.
Since the automata are deterministic, language equivalence coincides with bisimilarity. This can be decided by a breadth-first search over $Q_1 \times Q_2$ starting from the initial states.
Naively, this requires exploring all orbits of $Q_1 \times Q_2$ in the worst case; by \cref{prop:Product}, this number grows factorially with the dimension (\ie, $\beta(d, d) = O(2^d d!)$).
\looseness=-1

To overcome this, we employ a \emph{generating system} $\mathcal{G}$, the core data structure behind the polynomial-time equivalence check for register automata~\cite{MurawskiRT18}, which originates in the group-theoretic representation of bisimulations~\cite{MurawskiRT15,MurawskiRT25} and is implemented in \DEQ{}~\cite{MurawskiRT19}.
Since register-automaton configurations form strong nominal sets, as \cref{ssec:DataRepresentations} shows, this data structure naturally generalizes to our setting.\footnote{%
  The generalization is along the alphabet axis: we do not cover the \emph{globally fresh} transitions of fresh-register automata~\cite{Tzevelekos11,MurawskiRT18}, which test an input for freshness against all data seen in the run.
  Recognizing such languages requires state supports that grow with the run history, which no orbit-finite state space provides.\looseness=-1%
}
The generating system maintains an equivariant equivalence relation ${\sim}$ on a strong nominal set $X$ (here, $X = Q_1 \sqcup Q_2$), using time and space polynomial in both the orbit count and the dimension.
Its efficiency comes from how each equivalence class is stored: by a single orbit together with generators of its symmetry group (a subgroup of $S_d$), rather than explicitly listing all related pairs in $X \times X$.
It supports four operations, summarized in \cref{tab:GeneratingSystem}; \cref{app:GeneratingSystem} gives the aggregate analysis.

\begin{table}[t]
  \caption{Operations of the generating system $\mathcal{G}$ and their worst-case per-call costs.}\label{tab:GeneratingSystem}
  \centering
  \footnotesize
  \begin{tabular}{@{}lp{0.46\linewidth}l@{}}
    \toprule
    Operation & Description & Cost \\
    \midrule
    $\mathcal{G} \leftarrow \textsf{create}(X)$ & Initializes each orbit of $X$ as a singleton class. & $O(N(X) \cdot d)$ \\
    $\mathcal{G}.\textsf{contains}(x, y)$ & Checks whether $x \sim y$ holds. & $O(d + \mu(d))$ \\
     $\mathcal{G}.\textsf{merge}(x, y)$ & Records $x \sim y$. & $O(N(X) \cdot d + d \cdot \mu(d))$ \\
    $\mathcal{G}.\textsf{classes}$ & Returns the equivalence classes and the equivariant quotient lift from $X$. & $O(N(X) \cdot d + N(X) \log N(X))$ \\
    \bottomrule
  \end{tabular}
\end{table}

This generalization is correct: $\mathcal{G}$ always represents an equivariant equivalence relation, after any sequence of $\textsf{merge}$s it maintains the finest one relating the merged pairs, and $\textsf{classes}$ returns exactly the strong approximation of the quotient together with its lift.
The complexity is given by the following theorem (correctness, pseudocode, and proofs in \cref{app:GeneratingSystem}).

\begin{restatable}[language equivalence]{xtheorem}{Equivalence}\label{thm:Equivalence}
  Let $\mathcal{A}_1$ and $\mathcal{A}_2$ be strong nominal automata with state spaces $Q_1$ and $Q_2$ over a common strong alphabet $\Sigma$, with $N = N(Q_1) + N(Q_2)$, $d = \dim(Q_1 \sqcup Q_2)$, and $e = \dim(\Sigma)$.
  Their language equivalence is decidable in time
  \begin{equation*}
    O\!\left( N \cdot d^2 \cdot N(\Sigma) \cdot \beta(2d, e) \cdot \mu(d) + N^2 d \right),
  \end{equation*}
  which, for a fixed alphabet dimension $e$, is polynomial in $N$, $N(\Sigma)$, and $d$.
\end{restatable}

This generalizes the polynomial-time result for register automata~\cite{MurawskiRT18} to strong nominal automata, and gives a more precise complexity bound in terms of $N$ and $d$.

\subsection{Minimization and Learning}\label{ssec:MinimizationLearning}

Both minimization and learning rely on quotienting by the Myhill--Nerode equivalence.
Since \cref{ex:QuotientNotStrong} demonstrates that strong nominal sets are not closed under quotients, we instead compute the strong approximation of the quotient via the generating system introduced in \cref{ssec:LanguageEquivalence}, formalized by the following lemma (pseudocode and proof in \cref{app:MinimizationProofs}).
\looseness=-1

\begin{restatable}[quotient]{xlemma}{Quotient}\label{lem:Quotient}
  Let $X$ be an orbit-finite strong nominal set with $d = \dim(X)$, and let ${\equiv}$ be an equivariant equivalence relation on $X$ decidable in time $C$.
  Then, one can compute the strong approximation $P$ of $X / {\equiv}$ together with a lift $h\colon X \twoheadrightarrow P$ of the quotient map $q\colon X \to X / {\equiv}$, \ie, an equivariant surjection satisfying $p \circ h = q$ for the surjection $p\colon P \twoheadrightarrow X / {\equiv}$ of the strong approximation.
  This computation runs in time
  \begin{equation*}
    O\!\left( N(X)^2 \cdot \beta(d, d) \cdot (C + \mu(d)) \right),
  \end{equation*}
  polynomial in $N(X)$ but, since $\beta(d, d) = O(2^d d!)$, factorial in $d$.
\end{restatable}

The factorial cost is inherent to this construction, as the algorithm
must enumerate all $\beta(d, d)$ partial bijections per orbit pair.
By \cref{lem:Quotient}, minimization (Moore's algorithm~\cite{Moore56}) and
learning ($\nu L^\ast$~\cite{MoermanS0KS17}) yield the following bounds
(pseudocode and proofs in \cref{app:MinimizationProofs}).
\looseness=-1

\begin{restatable}[minimization]{xtheorem}{Minimize}\label{thm:Minimize}
  Let $\mathcal{A}$ be a reachable strong nominal automaton over a strong alphabet $\Sigma$, with $N(Q)$ state orbits, $d = \dim(Q)$, and $e = \dim(\Sigma)$.
  Moore's algorithm computes a strong lift of the minimal automaton of $L(\mathcal{A})$ in time
  \begin{equation*}
    O\!\left( N(Q)^3 \cdot d \cdot \beta(d, d) \cdot N(\Sigma)\,\beta(2d, e) \cdot \mu(d) \right),
  \end{equation*}
  which, for a fixed $e$, is polynomial in $N(Q)$ and $N(\Sigma)$ but factorial in $d$.
\end{restatable}

\begin{restatable}[learning]{xtheorem}{Learn}\label{thm:Learn}
  Let $\mathcal{A}_{\min}$ be the minimal nominal automaton of an equivariant language, with $n$ state orbits, state dimension $d$, and strong alphabet $\Sigma$ of dimension $e = \dim(\Sigma)$; let $m$ be the length of the longest counterexample and $\ell$ the maximum column length of the observation table (the proof shows $\ell = m + O(n d)$).
  Then, $\nu L^\ast$ learns a strong lift of $\mathcal{A}_{\min}$ using $O(n d)$ equivalence queries and
  \begin{equation*}
    O\!\left( n^2 d \cdot (m + 1) \cdot N(\Sigma) \cdot \beta(en, e) \cdot \beta(en, e\ell) \right)
  \end{equation*}
  membership queries.
  Each hypothesis is constructed by \cref{lem:Quotient} in time factorial in the observation dimensions, at most $en$ (rows) and $e\ell$ (columns).
\end{restatable}

The equivalence-query count is polynomial in all parameters; the table has polynomially many cells for a fixed alphabet dimension~\cite{MoermanS0KS17}, but filling each cell costs one membership query per product orbit of \cref{prop:Product}, whose number is factorial in $en$ and $e\ell$.
\looseness=-1

\section{Implementation and Experiments}\label{sec:ImplementationAndExperiments}

We evaluate the algorithms of \cref{sec:AutomataAlgorithms} via our library \Namae{}.
We address two questions:
\begin{description}
  \item[\textbf{(Q1)}] Language equivalence (\cref{thm:Equivalence}): does the algorithm remain polynomial in practice, and does \Namae{} outperform SMT-based tools such as \NLambda{}?
  \item[\textbf{(Q2)}] Minimization and learning (\cref{thm:Minimize,thm:Learn}): do the algorithms show the expected factorial scaling in practice, and does \Namae{} outperform existing tools?
\end{description}
All experiments were run on Apple M1 Pro (8 cores, 32\,GB) under macOS~26.5 with Java~SE~25.0.3, Scala~3.8.4, GHC~9.4.8, Apple clang~21.0.0, and Z3~4.16.0; \Namae{} used \texttt{-Xmx16G} with 50 JIT warm-up iterations on fixed small instances where timed in-process, and each benchmark was run 3~times, with the median time reported and a 10-minute timeout.\footnote{Experimental details: \cref{app:FullExperiments}.}

\subsection{Implementation}\label{ssec:Implementation}

\Namae{} is a Scala (JVM) library that implements the three algorithms of \cref{sec:AutomataAlgorithms} using the \emph{array representation} of \cref{ssec:DataRepresentations}: for language equivalence, the generating-system bisimulation search of \cref{ssec:LanguageEquivalence}, backed by a Schreier--Sims implementation; for minimization and learning, Moore's algorithm and $\nu L^\ast$ of \cref{ssec:MinimizationLearning}.
The latter two rely on the quotient operation of \cref{lem:Quotient}: minimization uses it to reduce the state space, and learning to construct each hypothesis.
Although this paper focuses on $\Eq$, \Namae{} also implements $\Ord$~\cite{VenhoekMR22} within the same codebase; its supplementary $\Ord$ benchmarks are available in \cref{app:FullResults}.
\looseness=-1

\subsection{Language Equivalence}\label{ssec:EquivalenceExperiments}

We evaluate the equivalence-check algorithm of \cref{thm:Equivalence} along two axes: \textbf{(1)} against the SMT-based tool \NLambda{}~\cite{KlinS16} to expose the computational advantage of the array representation, and \textbf{(2)} against \DEQ{}~\cite{MurawskiRT19} to quantify the overhead of generalizing from register automata to strong nominal automata.
We follow \DEQ{}'s benchmark suite~\cite{MurawskiRT19}: \emph{stacks} (a stack of capacity $n$) for $n = 10, 20, \dots, 500$, and \emph{CPT} (``partial permutations represented compactly''~\cite{MurawskiRT19}, stressing bisimulations with many partial permutations) for $n = 10, 20, \dots, 100$.
For the comparison against \NLambda{}, which degrades rapidly, we additionally use small stacks $n = 1, 2, \dots, 15$ and restrict \NLambda{} to this range, following~\cite{MurawskiRT19}.
\looseness=-1

\begin{figure}[t]
  \centering
  \includegraphics[width=0.66\linewidth]{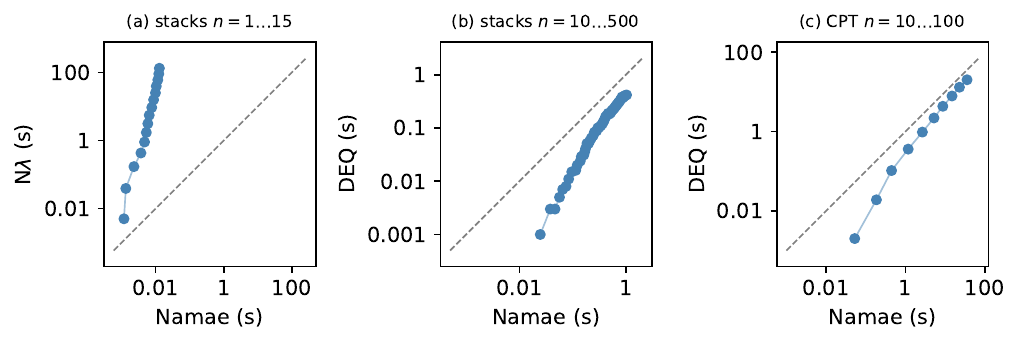}
  \caption{Language equivalence runtimes (log-log). (a) \Namae{} vs.\ \NLambda{} on stacks $n=1,\dots,15$. (b) \Namae{} vs.\ \DEQ{} on stacks $n=10,\dots,500$. (c) \Namae{} vs.\ \DEQ{} on CPT $n=10,\dots,100$.}\label{fig:EquivMain}
\end{figure}

\Cref{fig:EquivMain} reports three log-log scatter plots.
Panel~(a) contrasts \Namae{} with \NLambda{} on stacks: \Namae{} takes $0.013$\,s at $n=15$, while \NLambda{} needs $133$\,s and scales steeply.
Panels~(b) and~(c) compare \Namae{} with \DEQ{}: both scale polynomially on two structurally different benchmarks.
Although \DEQ{} is faster, the gap narrows to $1.8$--$2.5\times$ at the largest sizes; this residual overhead likely stems from the generalization from unary data alphabets to orbit-finite strong alphabets and from runtime constants (JVM vs.\ native code).
\looseness=-1

This answers \textbf{(Q1)}: \Namae{} stays within a small constant factor of the polynomial-time \DEQ{}, consistent with \cref{thm:Equivalence}, and outperforms \NLambda{}.
\looseness=-1

\subsection{Minimization and Learning}\label{ssec:MinimizationLearningExperiments}

We evaluate the quotient, minimization, and learning algorithms of \cref{lem:Quotient,thm:Minimize,thm:Learn} along two axes: \textbf{(1)} against \NLambda{} (under $\Eq$) to show the efficiency gains over symbolic constraint solving, and \textbf{(2)} against \OnsHs{} to expose the orbit-splitting penalty of using $\Ord$.
We use three benchmark languages: $\mathrm{Res}(n)$ (the running example's language with capacity $n$), $\mathrm{Res}_{\textsc{FIFO}}(n)$ (FIFO variant; trivial local symmetry), and $\mathrm{MRes}(m)$ (a variant of $\mathrm{Res}(1)$ whose input alphabet has dimension $m$; minimization only).
\looseness=-1

\begin{table}[t]
  \caption{Minimization results. $N_{\mathrm{in}}$, $N_{\min}$: state orbit counts before/after minimization; $\Delta_{\min}$: minimized transition-function size; \timeout{}: timeout; ---: unavailable. Full data: \cref{app:FullResults}.}\label{tab:MinimizeMain}
  \centering
  \small
  \scalebox{0.7}{\begin{tabular}{lrrrrrrrrrr}
  \toprule
  & \multicolumn{6}{c}{Equality Symmetry $\Eq$} & \multicolumn{4}{c}{Total Order Symmetry $\Ord$} \\
  \cmidrule(lr){2-7}\cmidrule(lr){8-11}
  & \multicolumn{2}{c}{size} & \multicolumn{2}{c}{\Namae{}} & \multicolumn{2}{c}{\NLambda{}}
  & \multicolumn{2}{c}{size} & \multicolumn{2}{c}{\OnsHs{}} \\
  \cmidrule(lr){2-3}\cmidrule(lr){4-5}\cmidrule(lr){6-7}
  \cmidrule(lr){8-9}\cmidrule(lr){10-11}
  Target & $N_\mathrm{in}$ & $N_{\min}$ & time (s) & $\Delta_{\min}$ & time (s) & $\Delta_{\min}$
         & $N_\mathrm{in}$ & $N_{\min}$ & time (s) & $\Delta_{\min}$ \\
  \midrule
  \rowcolor{gray!20}$\mathrm{Res}(3)$ & 5 & 5 & 0.06 & 22 & 6.99 & 15 & 11 & 5 & 0.04 & 34 \\
  $\mathrm{Res}(4)$ & 6 & 6 & 0.10 & 32 & 461.26 & 19 & 35 & 6 & 0.62 & 52 \\
  \rowcolor{gray!20}$\mathrm{Res}(5)$ & 7 & 7 & 0.26 & 44 & \timeout{} & --- & 155 & 7 & 21.59 & 74 \\
  \midrule
  $\mathrm{Res}_\textsc{FIFO}(3)$ & 11 & 5 & 0.07 & 22 & 80.53 & 51 & 34 & 11 & 0.38 & 114 \\
  \rowcolor{gray!20}$\mathrm{Res}_\textsc{FIFO}(4)$ & 16 & 6 & 0.20 & 32 & \timeout{} & --- & 154 & 35 & 52.10 & 546 \\
  $\mathrm{Res}_\textsc{FIFO}(5)$ & 22 & 7 & 0.84 & 44 & --- & --- & 874 & --- & \timeout{} & --- \\
  \midrule
  \rowcolor{gray!20}$\mathrm{MRes}(2)$ & 6 & 6 & 0.05 & 102 & 7.62 & 58 & 7 & 7 & 0.04 & 298 \\
  $\mathrm{MRes}(3)$ & 16 & 16 & 0.26 & 1752 & \timeout{} & --- & 27 & 27 & 19.77 & 18730 \\
  \rowcolor{gray!20}$\mathrm{MRes}(4)$ & 53 & 53 & 38.37 & 42292 & --- & --- & 151 & --- & \timeout{} & --- \\
  \bottomrule
\end{tabular}
}
\end{table}

\Cref{tab:MinimizeMain} reports representative minimization results.
\Namae{} scales significantly better than \NLambda{} under $\Eq$, minimizing instances where \NLambda{} times out; this gap stems from computing the quotient orbit-wise on constraint-free arrays instead of by symbolic constraint solving.
The outputs differ accordingly: \NLambda{} returns the minimal nominal automaton itself, whereas \Namae{} returns a strong lift of it, which by \cref{cor:MinimalStrong} attains the minimal orbit count and dimensions on the state space but not on the transition function, whose orbits may split as in \cref{fig:NonUniqueLift}; trading this minimality for trivial local symmetries is exactly what enables the efficiency observed here.
Under $\Ord$, \OnsHs{} suffers from orbit splitting: for $\mathrm{Res}_{\textsc{FIFO}}$, $\Ord$ distinguishes atom orderings the language leaves unordered, splitting each $\Eq$-orbit into up to $d!$ $\Ord$-orbits, as predicted by \cref{rem:LocalSymmetry}.
This penalty is a property of the symmetry, not the implementations: indeed, \Namae{} and \OnsHs{} perform comparably on the $\Ord$ inputs in \cref{app:FullResults}, with the remaining differences attributable to implementation details, showing that our speedup is due to using $\Eq$.
Conversely, $\Ord$ remains the only option for genuinely order-based languages~\cite{VenhoekMR22} that cannot be expressed over $\Eq$.
The $\mathrm{MRes}(m)$ family instead isolates the effect of the alphabet dimension: as $m$ grows, \Namae{}'s time rises sharply, consistent with \cref{thm:Minimize}.
\looseness=-1

\begin{table}[t]
  \caption{$\nu L^\ast$ learning results. $N$: state orbit count of the target; $\Delta$: learned transition-function size; MQs: membership queries; \timeout{}: timeout; ---: unavailable. Full data: \cref{app:FullResults}.}\label{tab:LearningMain}
  \centering
  \small
  \scalebox{0.7}{\begin{tabular}{lrrrrrrrrrrr}
  \toprule
  & \multicolumn{7}{c}{Equality Symmetry $\Eq$} & \multicolumn{4}{c}{Total Order Symmetry $\Ord$} \\
  \cmidrule(lr){2-8}\cmidrule(lr){9-12}
  & & \multicolumn{3}{c}{\Namae{}} & \multicolumn{3}{c}{\NLambda{}}
  & & \multicolumn{3}{c}{\OnsHs{}} \\
  \cmidrule(lr){3-5}\cmidrule(lr){6-8}\cmidrule(lr){10-12}
  Target & $N$ & time (s) & MQs & $\Delta$ & time (s) & MQs & $\Delta$
         & $N$ & time (s) & MQs & $\Delta$ \\
  \midrule
  \rowcolor{gray!20}$\mathrm{Res}(3)$ & 5 & 0.94 & 2638 & 22 & 525.50 & 2699 & 10 & 5 & 29.67 & 10671 & 34 \\
  $\mathrm{Res}(4)$ & 6 & 11.09 & 24259 & 32 & \timeout{} & --- & --- & 6 & \timeout{} & --- & --- \\
  \midrule
  \rowcolor{gray!20}$\mathrm{Res}_\textsc{FIFO}(3)$ & 5 & 0.91 & 2638 & 22 & 498.29 & 2698 & 10 & 11 & \timeout{} & --- & --- \\
  \bottomrule
\end{tabular}
}
\end{table}

\Cref{tab:LearningMain} reports representative $\nu L^\ast$ learning results.
\Namae{} learns these languages in seconds, while \NLambda{} times out on larger instances.
The measured membership-query counts remain modest, far below the worst-case bound of \cref{thm:Learn}; \OnsHs{} suffers from the same orbit-splitting blowup as in minimization.
\looseness=-1

This answers \textbf{(Q2)}: the expected factorial scaling is indeed observed in practice, and \Namae{} nonetheless outperforms \NLambda{} and \OnsHs{}.

\section{Related Work}\label{sec:RelatedWork}

\subparagraph{Nominal-computation implementations.}
\NLambda{}~\cite{KlinS16} and \Lois{}~\cite{KopczynskiT17} encode nominal sets as first-order formulas for SMT solvers. \NLambda{}'s predecessor~\cite{BojanczykBKL12} explored an explicit orbit representation but abandoned it: sizes are proportional to orbit counts, and its hull construction hinders extension to other symmetries such as the partial order~\cite[Sec.~5]{KlinS16}; it is, moreover, not canonical.
Over $\Eq$, our constraint-free arrays accept the orbit-count cost and provide a canonical representation of each orbit.
Our experiments show that \Namae{}'s array representation outperforms \NLambda{} and scales to minimization and learning instances where \NLambda{} times out.
However, the SMT-based approach natively handles non-trivial local symmetries without approximations, suggesting the two paradigms could be integrated complementarily.
For $\Ord$, \Ons{} and \OnsHs{}~\cite{VenhoekMR22} implement the sorted array representation.
Because $\Ord$ admits only trivial local symmetries, their underlying data model coincides with the $\Ord$-variant of our theory.
\Namae{} handles both $\Eq$ and $\Ord$ within a unified codebase; extending it to other Fra\"iss\'e symmetries~\cite{BojanczykBKL12}, such as the timed symmetry~\cite{BojanczykL12,Ochremiak14}, is an interesting future direction.
\looseness=-1

\para{Register automata.}
As \cref{ssec:DataRepresentations} shows, register-automaton configurations form strong nominal sets, which lets the generating system of \DEQ{}~\cite{MurawskiRT18} generalize to strong nominal automata in \cref{sec:AutomataAlgorithms}.
Recently, Balachander et al.\ introduced pDRA~\cite{BalachanderFGT25}, a register automaton model with transition permutations.
Over a single-atom alphabet, their injective register assignments and transition permutations mirror our array representation for states and equivariant maps.
While they observed that pDRAs correspond to nominal automata with straight state sets~\cite[Rem.~6]{BalachanderFGT25}, we identify the structure behind their observations as the projective cover: their minimal pDRA matches a strong lift of \cref{cor:MinimalStrong}, and their register-permutation isomorphism~\cite[Rem.~20]{BalachanderFGT25} reflects the freedom of projective-cover lifts discussed in \cref{ssec:MinimalStrongNominalAutomata}.
Generalizing the memorable-query minimization of Benedikt et al.~\cite{BenediktLP10}, they also proposed $\Pi$-DRA~\cite{BalachanderFGT25}, which fixes a global permutation policy $\Pi$ at the cost of a factorial blow-up in states and admits polynomial-time active learning using memorable queries~\cite[Thm.~34]{BalachanderFGT25}.
Since their memorability query ($\mathrm{mem}_L(u) = \supp(u^{-1}L)$) is simply a support query in nominal terms, this offers a promising foundation for learning strong nominal automata.
Tree-based register-automaton learning ($SL^\ast$~\cite{CasselHJS14} and $SL^\lambda$~\cite{DierlFHJST24}) is another natural target for strong nominal adaptation.
\looseness=-1

\para{Other adjunctions.}
Beyond the adjunction $F \dashv U\colon \Nom \leftrightarrows [|\I|, \Set]$ underlying our theory in \cref{sec:StrongNominalSets}, two other categorical frameworks offer relevant perspectives.
First, Wi\ss{}mann~\cite{Wissmann23} proves that $\Nom_{\mathcal{S}}$ is monadic over supported sets $\mathrm{Supp}(\mathbb{A})$ for various symmetries.
This monadicity provides an alternative algebraic foundation where a register automaton is a finite supported coalgebra whose generalized powerset yields the nominal configuration automaton, structurally complementing our view of strong nominal sets as register-automaton configurations.
Second, Moerman and Rot~\cite{MoermanR20} define an adjunction between nominal sets with substitution and standard nominal sets (\ie, $\Nom_{\Eq}$).
Its left adjoint maps the \emph{separated product} $X \mathop{\ast} Y$ (pairs with disjoint supports) to the Cartesian product.
This leads to \emph{separated nominal automata} with restricted transition domains $Q \mathop{\ast} \Sigma$, which can be exponentially more concise for substitution-closed languages.
As support disjointness is efficiently computable in our array representation, integrating the separated product is a promising extension.

\section{Conclusion}\label{sec:Conclusion}

This paper recasts \emph{strong nominal sets}, for which canonical minimization is known to fail, as an efficient foundation for nominal computation.
Specifically, through \cref{thm:Characterization}, the absence of local symmetries enables the constraint-free \emph{array representation}, and \cref{thm:ProjectiveCover,thm:OrbitCount} ensure that projective covers provide \emph{strong approximations} of nominal sets preserving orbit count and dimensions.
This shift in perspective bridges the algorithmic gap for equivalence, which \cref{thm:Equivalence} decides in polynomial time for a fixed alphabet dimension, and extends the orbit-wise approach to minimization and learning in \cref{thm:Minimize,thm:Learn}.
In experiments, \Namae{} outperforms existing nominal-computation tools and, at scale, stays within a small constant factor of the dedicated register-automaton tool \DEQ{}.
\looseness=-1

Several promising directions remain.
Extending the categorical characterization beyond the equality symmetry to other Fra\"iss\'e symmetries, such as the timed symmetry~\cite{BojanczykL12,Ochremiak14}, is a natural next step.
Furthermore, integrating concepts such as memorable queries~\cite{BalachanderFGT25} and separated products~\cite{MoermanR20} into strong nominal sets could yield further algorithmic improvements.
\looseness=-1

\bibliographystyle{plainurl}
\bibliography{main}

\appendix

\section{Proof of \texorpdfstring{\cref{prop:LocalSymmetry}}{the Local-Symmetry Characterization}}\label{app:LocalSymmetry}

This appendix proves the local-symmetry characterization of \cref{sec:StrongNominalSets}; as there, $\mathcal{S}$ is an arbitrary data symmetry admitting least supports.

\LocalSymmetry*

\begin{proof}
  The statement follows by unfolding the definitions:
  \begin{align*}
    &X \text{ is a strong nominal set} \\
    &\iff \forall x \in X.\, \supp(x) \text{ strongly supports } x
      && \text{(def. of strong nominal set)} \\
    &\iff \forall x \in X, g \in G.\, g \cdot x = x \implies g \in \Fix_G(\supp(x))
      && \text{(def. of strong support)} \\
    &\iff \forall x \in X, g \in \Stab_G(x), a \in \supp(x).\, g(a) = a
      && \text{(def. of $\Stab$, $\Fix$)} \\
    &\iff \forall x \in X.\, \Stab_G(x)|_{\supp(x)} = \{\mathrm{id}\}
      && \text{(def. of local symmetry)}.\qedhere
  \end{align*}
\end{proof}

\section{Proofs for \texorpdfstring{\cref{ssec:CategoricalCharacterization}}{the Categorical Characterization}}\label{app:Characterization}

This appendix proves \cref{thm:Characterization} by establishing its four equivalences separately, as \cref{lem:Kleisli,lem:Fam,lem:Proj,lem:NSet}.
Throughout, $\mathcal{S} = \Eq$ and we drop it from the notation; unlike \cref{app:LocalSymmetry}, which works over any data symmetry admitting least supports, the equality symmetry is essential here, as the swap arguments below use arbitrary transpositions.

\subsection{The Kleisli Characterization}\label{app:Kleisli}

\begin{xlemma}[$\SNom \simeq \Kl(T)$]\label{lem:Kleisli}
  The comparison functor $K\colon \Kl(T) \to \Nom$, $K(Y) = F(Y)$, restricts to an equivalence $\Kl(T) \simeq \SNom$.
\end{xlemma}

\begin{proof}
  For any adjunction $F \dashv U$, the Kleisli category is equivalent to the full subcategory of free algebras, and $K$ is the induced inclusion into $\Nom$; hence $K$ is fully faithful and its essential image is the image of $F$.
  By~\cite[Lem.~2.4.2]{Petrisan12}, $X \in \Nom$ is strong iff $X \cong F(Y)$ for some $Y \in [|\I|, \Set]$, so the essential image of $K$ is exactly $\SNom$.
  Therefore $K$ restricts to an equivalence $\Kl(T) \simeq \SNom$.
\end{proof}

The fact that $F$ lands in the strong nominal sets was already observed in~\cite[Footnote~1]{KurzPV10}; \cref{lem:Kleisli} records the resulting equivalence and identifies $\SNom$ with the free part of $\Nom$.
Our formula $F(Y) = \coprod_C Y(C) \times \Inj(C, \mathbb{N})$ from \cref{ssec:CategoricalCharacterization} is the tuple presentation of the left adjoint of~\cite{KurzPV10,Petrisan12}, to which it is naturally isomorphic.

\subsection{The Free-Coproduct-Completion Characterization}\label{app:Fam}

\begin{xlemma}[$\SNom^1 \simeq \I^\mathrm{op}$]\label{lem:SNom1}
  The functor $\Phi\colon \I^\mathrm{op} \to \SNom^1$ defined by $\Phi(C) = \Inj(C, \mathbb{N})$ and $\Phi(h)(e) = e \circ h$ for an injection $h\colon D \hookrightarrow C$ in $\I$ is an equivalence of categories.
\end{xlemma}

\begin{proof}
  First, $\Phi$ is well defined: each $\Inj(C, \mathbb{N})$ is a transitive $G$-set and is strong, since $g \cdot e = g \circ e = e$ forces $g$ to fix $e(C) = \supp(e)$ pointwise, and $\Phi(h)$ is equivariant.
  It remains to show $\Phi$ is fully faithful and essentially surjective.
  \begin{description}
    \item[Fully faithful.]
      Let $f\colon \Inj(C, \mathbb{N}) \to \Inj(D, \mathbb{N})$ be equivariant; fix $e_0$ and put $e_0' = f(e_0)$.
      By transitivity, every $e$ has the form $g \circ e_0$, so $f(e) = g \circ e_0'$ and $f$ is determined by $e_0'$.
      If some $d$ had $e_0'(d) = a$ and $a \notin e_0(C)$, choose $b \notin e_0(C) \cup e_0'(D)$; then $(a\,b) \circ e_0 = e_0$ while $(a\,b) \circ e_0' \neq e_0'$, contradicting equivariance.
      Hence $e_0'(D) \subseteq e_0(C)$, and $h = e_0^{-1} \circ e_0'\colon D \hookrightarrow C$ satisfies $f = \Phi(h)$; injectivity of $e_0$ makes $h$ unique.
    \item[Essentially surjective.]
      For $X \in \SNom^1$ pick $x_0$ with $C = \supp(x_0)$ and let $e_0\colon C \hookrightarrow \mathbb{N}$ be the inclusion.
      Strongness gives $\Stab_G(x_0) = \Fix_G(C) = \Stab_G(e_0)$, so by the orbit--stabilizer theorem $X \cong G / \Fix_G(C) \cong \Inj(C, \mathbb{N}) = \Phi(C)$.\qedhere
  \end{description}
\end{proof}

\begin{xlemma}[$\SNom \simeq \Fam(\I^\mathrm{op})$]\label{lem:Fam}
  $\SNom$ is equivalent to the free coproduct completion $\Fam(\I^\mathrm{op})$.
\end{xlemma}

\begin{proof}
  Every strong nominal set is the coproduct of its orbits, and by transitivity of the $G$-action, every equivariant map from a single orbit into a coproduct factors through a unique summand.
  Hence $\SNom$ is the free coproduct completion of its full subcategory $\SNom^1$ of single orbits, and \cref{lem:SNom1} gives $\SNom \simeq \Fam(\SNom^1) \simeq \Fam(\I^\mathrm{op})$.
\end{proof}

Under $\Phi$, a single orbit corresponds to a finite set $C$, hence to its dimension $|C|$, and an element to an injection $C \hookrightarrow \mathbb{N}$, \ie, a tuple of $|C|$ distinct atoms.
This is the source of the array representation of \cref{ssec:DataRepresentations}.

Family-style representations are classical for arbitrary orbit-finite nominal sets, with a local symmetry group attached to each orbit: the single-orbit case over Fra\"iss\'e symmetries is given in~\cite[Thm.~10.9]{BojanczykKL14}, and extended to orbit-finite families over algebraic atoms in~\cite[Thm.~4.19]{Ochremiak14}.
\Cref{lem:Fam} complements these by identifying the strong fragment as precisely a \emph{free coproduct completion}.

A free-coproduct-completion presentation also appears in the named-set literature:
the bijection-based category $\NSet_\mathbb{B}$ of named sets with symmetries arises as a free coproduct completion of the coset actions of the symmetric groups, with $\NSet_\mathbb{B} \simeq [\mathbb{B}, \Set]$ for the groupoid $\mathbb{B}$ of finite sets and bijections~\cite[Sec.~5.1]{Staton07};
this is generalized to arbitrary small categories $C$ of monos in~\cite{CianciaKM10}: pullback-preserving presheaves over $C$ are equivalent to $\Fam(\mathrm{Sym}(C)^\mathrm{op})$, where $\mathrm{Sym}(C)$ is the category of automorphism groups over $C$.
Over $C = \I$, this equivalence gives $\Nom \simeq \Fam(\mathrm{Sym}(\I)^\mathrm{op})$; \cref{lem:Fam} identifies the trivial-symmetry fragment $\Fam(\I^\mathrm{op})$ as exactly $\SNom$.

\subsection{The Projective Characterization}\label{app:Proj}

A \emph{section} of an epimorphism $e\colon X \twoheadrightarrow Y$ is a map $s\colon Y \to X$ with $e \circ s = \mathrm{id}_Y$; in particular, if $Y$ is projective in the sense of \cref{sec:Preliminaries}, then lifting $\mathrm{id}_Y$ along $e$ shows that \emph{every epimorphism onto $Y$ admits a section}.
More generally, $B$ is a \emph{retract} of $A$ if there are $i\colon B \to A$ and $r\colon A \to B$ with $r \circ i = \mathrm{id}_B$, \ie, $i$ is a section of $r$; retracts of projective objects are again projective (lift along $r$, then compose with $i$).
A category is \emph{idempotent complete} if every idempotent $e\colon A \to A$ (\ie, $e \circ e = e$) splits as $e = i \circ r$ for some retract pair $i, r$ as above.

\begin{xlemma}[$\Fam(\mathscr{C})$ inherits idempotent completeness]\label{lem:FamIdempotent}
  If $\mathscr{C}$ is idempotent complete, so is $\Fam(\mathscr{C})$.
\end{xlemma}
\begin{proof}
  An idempotent of a family $(A_i)_{i \in I} \in \Fam(\mathscr{C})$ is a morphism $(\rho, (f_i)_{i \in I})$ with $\rho\colon I \to I$ and $f_i\colon A_i \to A_{\rho(i)}$ satisfying $\rho \circ \rho = \rho$ and $f_{\rho(i)} \circ f_i = f_i$ for all $i \in I$; in particular, every $\rho(i)$ is a fixed point of $\rho$, and $f_i$ is an idempotent endomorphism for each fixed point $i$.
  Write $I' \subseteq I$ for the set of fixed points of $\rho$ and, using the idempotent completeness of $\mathscr{C}$, split $f_i = m_i \circ r_i$ through an object $B_i$ for each $i \in I'$.
  Then the family $(B_i)_{i \in I'}$ splits the idempotent: the morphisms $(\rho, (r_{\rho(i)} \circ f_i)_{i \in I})\colon (A_i)_{i \in I} \to (B_i)_{i \in I'}$ and $(\iota, (m_i)_{i \in I'})$ in the opposite direction, where $\iota\colon I' \hookrightarrow I$ is the inclusion, compose to $\mathrm{id}$ on $(B_i)_{i \in I'}$ and to $(\rho, (f_i)_{i \in I})$ on $(A_i)_{i \in I}$, using $r_i \circ m_i = \mathrm{id}_{B_i}$ and $f_{\rho(i)} \circ f_i = f_i$.
\end{proof}

The only idempotents in $\I$ are identities, so $\I$ (and hence $\I^\mathrm{op}$) is idempotent complete.
By \cref{lem:FamIdempotent}, $\Fam(\I^\mathrm{op})$ is idempotent complete, and therefore so is $\SNom$.

We record the action of the comparison functor $J$ on morphisms, which \cref{ssec:CategoricalCharacterization} leaves implicit: for a morphism $i\colon C \to D$ of $\I$ and $x \in J(X)(C)$, we set $J(X)(i)(x) = \hat{i} \cdot x$, where $\hat{i} \in G$ is any finite permutation extending $i$.
This is well defined, since two extensions of $i$ differ by a permutation fixing $C$ pointwise, which fixes $x$ as $\supp(x) \subseteq C$; moreover $\supp(\hat{i} \cdot x) = \hat{i}(\supp(x)) \subseteq D$ places $\hat{i} \cdot x$ in $J(X)(D)$, and functoriality follows directly.

\begin{xlemma}[{$\SNom \simeq \Proj([\I, \Set])$}]\label{lem:Proj}
  The restricted comparison functor $J' = J \circ \iota\colon \SNom \to [\I, \Set]$ (where $\iota\colon \SNom \hookrightarrow \Nom$ is the inclusion, see \cref{fig:CategoricalRelationship}) is an equivalence onto the full subcategory $\Proj([\I, \Set])$ of projective objects.
\end{xlemma}

\begin{proof}
  On a single orbit, $J'(\Inj(C, \mathbb{N})) \cong \I(C, -)$, the representable functor at $C$.
  By the Yoneda lemma representables are projective, and coproducts of projectives are projective, so the essential image of $J'$ is contained in $\Proj([\I, \Set])$.
  Conversely, every projective object of $[\I, \Set]$ is a retract of a coproduct of representables, hence a retract of some $J'(X)$ with $X \in \SNom$.
  Since $\SNom$ is idempotent complete and $J'$ is fully faithful, the essential image of $J'$ is closed under retracts.
  So every such projective is isomorphic to some $J'(X)$.
  Thus $J'$ is essentially surjective onto $\Proj([\I, \Set])$.
\end{proof}

Projectivity here is relative to $[\I, \Set]$.
In $\Nom$ itself, only the empty set $\emptyset$ is projective: for any non-empty nominal set $X$, the equivariant surjection $\textsf{fst}\colon \{(x, a) \in X \times \mathbb{N} \mid a \notin \supp(x)\} \twoheadrightarrow X$, defined by $\textsf{fst}(x, a) = x$, has no equivariant section.
Indeed, any section $s$ of $\textsf{fst}$ must have the form $s(x) = (x, \alpha(x))$ for some function $\alpha\colon X \to \mathbb{N}$, and since $s(x)$ lies in the domain of $\textsf{fst}$, $\alpha(x) \notin \supp(x)$ for all $x$.
The domain of $\textsf{fst}$ is orbit-finite whenever $X$ is, so the failure persists among orbit-finite nominal sets.
Equivariance of $s$ forces $\alpha$ to be equivariant as well, so $\supp(\alpha(x)) \subseteq \supp(x)$; since $\alpha(x)$ is itself an atom, $\supp(\alpha(x)) = \{\alpha(x)\}$, giving $\alpha(x) \in \supp(x)$, which contradicts $\alpha(x) \notin \supp(x)$.
Therefore, any non-empty nominal set $X$ is not projective in $\Nom$.
The strong nominal sets acquire projectivity only once transported along $J'$ into the functor category.

\subsection{The Named-Set Characterization}\label{app:NSet}

We first recall the terminology of named sets from the theory of history-dependent automata.
We follow the early definition of~\cite{MontanariP97}, which does not yet involve permutation groups (\ie, local symmetries): a \emph{named set} is a set $I$ of elements equipped with a \emph{name set} $C_i$ for each $i \in I$, and a \emph{named function} from $(C_i)_{i \in I}$ to $(D_j)_{j \in J}$ is a pair $(\rho, (h_i)_{i \in I})$ of a function $\rho\colon I \to J$ and \emph{name embeddings}, \ie, injections $h_i\colon D_{\rho(i)} \hookrightarrow C_i$.
A named set is \emph{finitely named} if every name set $C_i$ is finite; without loss of generality, we take the name sets of finitely named sets to be finite subsets of $\mathbb{N}$.
We write $\NSet_\mathrm{fn}$ for the category of finitely named sets and named functions.

\begin{xlemma}[$\SNom \simeq \NSet_\mathrm{fn}$]\label{lem:NSet}
  $\SNom$ is equivalent to the category $\NSet_\mathrm{fn}$ of finitely named sets.
\end{xlemma}

\begin{proof}
  An object of $\Fam(\I^\mathrm{op})$ is a family $(C_i)_{i \in I}$ of finite subsets of $\mathbb{N}$, which is exactly a finitely named set, and a morphism $(\rho, (h_i)_{i \in I})$ of $\Fam(\I^\mathrm{op})$ is exactly a named function between finitely named sets.
  This identification is the identity on objects and morphisms, so $\Fam(\I^\mathrm{op}) = \NSet_\mathrm{fn}$, and \cref{lem:Fam} gives $\SNom \simeq \NSet_\mathrm{fn}$.
\end{proof}

The contemporary definition of named sets~\cite[Sec.~3.1]{FerrariMP02}~\cite[Defs.~22--23]{GadducciMM06} additionally equips each element with a permutation group and defines named functions up to the action of these groups: a morphism assigns to each element a \emph{non-empty} set of injections, saturated under the permutation groups of its source and target.
This yields the category $\NSet_\I$ of named sets with permutation groups, as recalled in \cref{ssec:CategoricalCharacterization}; the notation follows~\cite[Sec.~5.1]{Staton07}, which distinguishes these injection-based morphisms from the bijection-based variant $\NSet_\mathbb{B} \simeq [\mathbb{B}, \Set]$ discussed in \cref{app:Fam}, proves $\NSet_\I$ equivalent to the Schanuel topos, and identifies it as ``essentially'' the category of~\cite[Sec.~3.1]{FerrariMP02}, the differences (a total order on the elements, and possibly empty sets of injections there) being immaterial.
Under $\Nom \simeq \NSet_\I$, strong nominal sets correspond, by \cref{prop:LocalSymmetry}, to those named sets whose permutation groups are all trivial; the saturation condition then collapses each set of injections to a singleton, so between such objects the morphisms of $\NSet_\I$ are exactly the named functions of \cref{lem:NSet}.
This full subcategory is precisely $\NSet_\mathrm{fn}$, formally linking strong nominal sets to the original formulation of history-dependent automata.

Finally, assembling these four categorical equivalences completes the full picture of strong nominal sets.

\begin{proof}[Proof of \cref{thm:Characterization}]
  \Cref{lem:Kleisli,lem:Fam,lem:Proj,lem:NSet} together establish the theorem.
\end{proof}

\section{Proofs for \texorpdfstring{\cref{ssec:DataRepresentations}}{Counting Product Orbits}}\label{app:Product}

We prove \cref{prop:Product} by separating its three claims: each partial bijection determines one orbit per \cref{lem:ProductOrbit}, these orbits exhaust the product according to \cref{lem:ProductPartition}, and the partial bijections are counted by the stated sum as verified in \cref{lem:CountPBij}.

By the array representation, an element of $X \times Y$ is a pair $(\vec{a}, \vec{b})$ of tuples $\vec{a} = \langle a_1, \dots, a_m \rangle$ and $\vec{b} = \langle b_1, \dots, b_n \rangle$, each consisting of pairwise distinct atoms.
Its \emph{coincidence pattern} is the relation $\pi_{(\vec{a}, \vec{b})} \subseteq [m] \times [n]$ defined by $\pi_{(\vec{a}, \vec{b})}(i) = j$ iff $a_i = b_j$.
Since the $a_i$ are pairwise distinct and so are the $b_j$, each index $i$ corresponds to at most one index $j$, and vice versa, so $\pi_{(\vec{a}, \vec{b})}$ is a partial bijection $[m] \rightharpoonup [n]$.
For a partial bijection $\pi\colon [m] \rightharpoonup [n]$, we write
\begin{equation*}
  \llbracket \pi \rrbracket = \{ (\vec{a}, \vec{b}) \in X \times Y \mid \pi_{(\vec{a}, \vec{b})} = \pi \}
\end{equation*}
for the set of pairs realizing exactly the coincidence pattern $\pi$.

\begin{xlemma}[partial bijections determine orbits]\label{lem:ProductOrbit}
  For every partial bijection $\pi\colon [m] \rightharpoonup [n]$, the set $\llbracket \pi \rrbracket$ is a single \emph{strong} orbit of $X \times Y$ of dimension $m + n - |{\dom(\pi)}|$.
\end{xlemma}

\begin{proof}
  First, $\llbracket \pi \rrbracket$ is equivariant: every $g \in G$ acts by relabeling atoms and hence preserves coincidences, so $\pi_{g \cdot (\vec{a}, \vec{b})} = \pi_{(\vec{a}, \vec{b})}$, whence $g \cdot \llbracket \pi \rrbracket = \llbracket \pi \rrbracket$.
  It is nonempty, as $\pi$ is realized by choosing atoms that coincide exactly as $\pi$ prescribes (which is always possible since the supply of atoms is infinite).
  It is a single orbit: given $(\vec{a}, \vec{b}), (\vec{a}', \vec{b}') \in \llbracket \pi \rrbracket$, the map sending each atom of $(\vec{a}, \vec{b})$ to the atom in the corresponding position of $(\vec{a}', \vec{b}')$ is well defined and injective precisely because both share the coincidence pattern $\pi$; since $G = \Sym(\mathbb{N})$, as a finite partial injection of $\mathbb{N}$ it extends to some $g \in G$ with $g \cdot (\vec{a}, \vec{b}) = (\vec{a}', \vec{b}')$.
  The least support of $(\vec{a}, \vec{b})$ is the set of atoms occurring in the pair; the two tuples contribute $m + n$ atoms with exactly $|{\dom(\pi)}|$ coincidences identified, so $\dim(\llbracket \pi \rrbracket) = m + n - |{\dom(\pi)}|$.
  Finally, $\llbracket \pi \rrbracket$ is strong: if $g \cdot (\vec{a}, \vec{b}) = (\vec{a}, \vec{b})$, then $g$ fixes every atom occurring in the pair, hence all of $\supp(\vec{a}, \vec{b})$.
\end{proof}

\begin{xlemma}[the orbits exhaust the product]\label{lem:ProductPartition}
  $X \times Y = \coprod_{\pi} \llbracket \pi \rrbracket$, where $\pi$ ranges over all partial bijections $[m] \rightharpoonup [n]$.
\end{xlemma}

\begin{proof}
  Each pair $(\vec{a}, \vec{b}) \in X \times Y$ has a unique coincidence pattern $\pi_{(\vec{a}, \vec{b})}$, which is a partial bijection, so it belongs to exactly one $\llbracket \pi \rrbracket$.
  Thus the sets $\llbracket \pi \rrbracket$ are pairwise disjoint and cover $X \times Y$.
\end{proof}

\begin{xlemma}[number of partial bijections]\label{lem:CountPBij}
  The number of partial bijections $[m] \rightharpoonup [n]$ is
  \begin{equation*}
    \sum_{k=0}^{\min(m, n)} \binom{m}{k} \binom{n}{k} \, k!.
  \end{equation*}
\end{xlemma}

\begin{proof}
  A partial bijection with domain of size $k$ is determined by a $k$-subset of $[m]$, a $k$-subset of $[n]$, and a bijection between them, giving $\binom{m}{k} \binom{n}{k} \, k!$ choices.
  Summing over the possible sizes $0 \le k \le \min(m, n)$ gives the claim.
\end{proof}

\begin{proof}[Proof of \cref{prop:Product}]
  By \cref{lem:ProductOrbit,lem:ProductPartition}, the orbits of $X \times Y$ are exactly the sets $\llbracket \pi \rrbracket$ for partial bijections $\pi\colon [m] \rightharpoonup [n]$, with $\dim(\llbracket \pi \rrbracket) = m + n - |{\dom(\pi)}|$.
  Their number is counted by \cref{lem:CountPBij}.
\end{proof}

The decomposition into the orbits $\llbracket \pi \rrbracket$ specializes the
product representation of~\cite[Sec.~10.3]{BojanczykKL14}, which describes
products of single orbits over general Fra\"iss\'e symmetries as coproducts
indexed by partial isomorphisms between the underlying structures; over $\Eq$
with strong orbits, these partial isomorphisms are exactly the partial bijections
$[m] \rightharpoonup [n]$, and the absence of local symmetries enables the
explicit count of \cref{prop:Product}.
At the family level, the same decomposition instantiates the product in
$\Fam(C^\mathrm{op})$ constructed via multi-coproducts of
interfaces~\cite[Sec.~5]{CianciaKM10}: over $C = \I$, the canonical cospans
between $[m]$ and $[n]$ correspond precisely to the partial bijections $\pi$.
The product decomposition over $\Ord$~\cite[Sec.~5.3]{VenhoekMR22} plays the
same role for sorted arrays.

\section{Proofs for \texorpdfstring{\cref{ssec:ProjectiveCoversAsStrongApproximation}}{Projective Covers as Strong Approximation}}\label{app:ProjectiveCovers}

Throughout, $\mathcal{S} = \Eq$ and we drop it from the notation.
Recall from \cref{lem:Proj} that $J'(\Inj([n], \mathbb{N})) \cong \I([n], -)$, the representable functor at $[n]$, and the action of $J$ on morphisms of $\I$ recorded in \cref{app:Proj}.
Both proofs rest on the explicit projective cover built from the following notion.

\begin{xdefinition}[new elements]\label{def:New}
  For $\mathscr{F} \in [\I, \Set]$ and $n \ge 0$, the set of \emph{new elements at level $n$} is
  \begin{equation*}
    \New_n(\mathscr{F}) = \mathscr{F}([n]) \setminus \bigcup_{\substack{i\colon [m] \hookrightarrow [n] \\ m < n}} \mathscr{F}(i)\bigl(\mathscr{F}([m])\bigr),
  \end{equation*}
  the elements of $\mathscr{F}([n])$ not in the image of any strictly smaller level.
\end{xdefinition}

Since each $\sigma \in S_n$ is a bijection of $[n]$, hence a morphism of $\I$, the group $S_n$ acts on $\mathscr{F}([n])$ via $\mathscr{F}(\sigma)$; this action preserves $\New_n(\mathscr{F})$, which therefore decomposes into a set $\Lambda_n$ of $S_n$-orbits.

To prove \cref{thm:ProjectiveCover}, we build the projective cover explicitly from the new elements of \cref{def:New}, taking one representable summand $\I([n], -)$ per $S_n$-orbit $\lambda \in \Lambda_n$, and then directly verify the two defining properties of a projective cover, namely being a superfluous epimorphism.
Superfluousness rests on the following lemma.

\begin{xlemma}[subfunctor lemma]\label{lem:Subfunctor}
  A proper subfunctor $Z \subsetneq \I(C, -)$ satisfies $Z(C) = \emptyset$.
\end{xlemma}

\begin{proof}
  Suppose $f \in Z(C)$.
  As an injection of a finite set, $f$ is a bijection.
  For any injection $g\colon C \hookrightarrow D$, the map $u = g \circ f^{-1}$ is injective and satisfies $u \circ f = g$; applying $Z(u)$ to $f$ therefore yields $Z(u)(f) = u \circ f = g$, so $g$ belongs to $Z(D)$.
  Since $g$ was an arbitrary injection, $Z(D) = \I(C, D)$ for all $D$, contradicting properness.
\end{proof}

\ProjectiveCover*

\begin{proof}
  For each $\lambda \in \Lambda_n$ choose a representative $x_\lambda \in \lambda$; by the Yoneda lemma it corresponds to a natural transformation $y_\lambda\colon \I([n], -) \to \mathscr{F}$.
  Set
  \begin{equation*}
    X' = \coprod_{n \ge 0,\; \lambda \in \Lambda_n} \Inj([n], \mathbb{N}),
    \qquad
    e = \coprod_{n, \lambda} y_\lambda \colon J'(X') \to \mathscr{F},
  \end{equation*}
  so that $J'(X') = \coprod_{n, \lambda} \I([n], -)$.
  We check $e$ is a superfluous epimorphism.
  \begin{description}
    \item[Epimorphism.]
      It suffices to show $e$ is pointwise surjective.
      Every $z \in \mathscr{F}(D)$ has the form $z = \mathscr{F}(i)(x)$ for some injection $i\colon [n] \hookrightarrow D$ and $x \in \mathscr{F}([n])$ (\eg, taking $n = |D|$ and $i$ to be any bijection $[n] \to D$, an isomorphism in $\I$); choosing $n$ minimal forces $x \in \New_n(\mathscr{F})$, and then $z$ lies in the image of $y_\lambda$ for the orbit $\lambda \ni x$.
    \item[Superfluous.]
      Assume a subobject $Z \subseteq J'(X')$ has an epimorphism $e|_Z$; we argue $Z$ cannot be proper, so $Z = J'(X')$.
      Subobjects of a coproduct in $[\I, \Set]$ decompose componentwise, so if $Z$ were proper then $Z_{\lambda_0} \subsetneq \I([n_0], -)$ for some component $(n_0, \lambda_0)$, whence $Z_{\lambda_0}([n_0]) = \emptyset$ by \cref{lem:Subfunctor}.
      Pick $x_0 \in \lambda_0 \subseteq \mathscr{F}([n_0])$; we check $x_0$ is not in the image of $(e|_Z)_{[n_0]}$.
      The image is the union over components $(n, \lambda)$ of $\{\mathscr{F}(h)(x_\lambda) \mid h \in Z_\lambda([n_0])\}$, and:
      \begin{itemize}
        \item components with $n < n_0$ miss $x_0$ since $x_0 \in \New_{n_0}(\mathscr{F})$;
        \item components with $n > n_0$ contribute nothing as $\I([n], [n_0]) = \emptyset$;
        \item components with $n = n_0$ and $\lambda \neq \lambda_0$ produce elements of the $S_{n_0}$-orbit $\lambda$, which are disjoint from $\lambda_0 \ni x_0$;
        \item and the component $(n_0, \lambda_0)$ contributes nothing because $Z_{\lambda_0}([n_0]) = \emptyset$.
      \end{itemize}
      So $x_0$ is uncovered, contradicting that $e|_Z$ is an epimorphism.
      Therefore $Z = J'(X')$, \ie, $e$ is superfluous.\qedhere
  \end{description}
\end{proof}

\OrbitCount*

\begin{proof}
  We instantiate the construction of \cref{thm:ProjectiveCover} at $\mathscr{F} = J(X)$ and read off its orbits: we identify the new elements at each level, construct a dimension-preserving bijection $\phi_d$ between the orbits of $X'$ and those of $X$, and finally check that this bijection is the one induced by $p$.

  Apply the construction from the proof of \cref{thm:ProjectiveCover} to $\mathscr{F} = J(X)$, where $J(X)(C) = \{x \in X \mid \supp(x) \subseteq C\}$, writing the dimension index as $d$.
  Since each $\Inj([d], \mathbb{N})$ is a single orbit of dimension $d$, the orbits of $X' = \coprod_{d, \lambda} \Inj([d], \mathbb{N})$ are exactly the summands $O'_{d, \lambda} \cong \Inj([d], \mathbb{N})$, one for each $\lambda \in \Lambda_d$, and $\dim(O'_{d, \lambda}) = d$.
  Unfolding $J(X)$ gives
  \begin{equation*}
    \New_d(J(X)) = \{x \in X \mid \supp(x) = [d]\},
  \end{equation*}
  the elements whose least support is exactly $[d]$.

  Let $M_d$ be the set of dimension-$d$ orbits of $X$, and define $\phi_d\colon \Lambda_d \to M_d$ by $\phi_d(S_d \cdot x) = G \cdot x$.
  We verify it is a well-defined bijection.
  \begin{description}
    \item[Well defined.]
      $\supp(x) = [d]$ gives $\dim(G \cdot x) = d$, and any $x' \in S_d \cdot x$ has $G \cdot x' = G \cdot x$ since $S_d \le G$.
    \item[Injective.]
      If $g \cdot x_1 = x_2$ with $x_1, x_2 \in \New_d(J(X))$, then $g(\supp(x_1)) = \supp(x_2)$ gives $g([d]) = [d]$, so $g|_{[d]} \in S_d$; since $x_1$ is supported by $[d]$, the action of $g$ on $x_1$ depends only on its restriction $g|_{[d]}$, so $(g|_{[d]}) \cdot x_1 = g \cdot x_1 = x_2$, hence $S_d \cdot x_1 = S_d \cdot x_2$.
    \item[Surjective.]
      Given $\mu \in M_d$ and $x \in \mu$, some $g \in G$ has $g(\supp(x)) = [d]$, so $g \cdot x \in \New_d(J(X)) \cap \mu$ and $\phi_d(S_d \cdot (g \cdot x)) = \mu$.
  \end{description}
  Thus $\phi_d$ is a bijection, and the orbits of $X'$ biject with those of $X$ dimension by dimension.

  It remains to identify $p$ with $\phi_d$ on orbits.
  The component $y_\lambda\colon \I([d], -) \to J(X)$ is Yoneda-induced by $x_\lambda \in \lambda$, so on an element $i \in \Inj([d], \mathbb{N})$ of $O'_{d, \lambda}$ it acts by $y_\lambda(i) = J(X)(i)(x_\lambda) = \hat{i} \cdot x_\lambda$, where $\hat{i} \in G$ is any permutation extending $i$.
  As $i$ ranges over $\Inj([d], \mathbb{N})$, the extension $\hat{i}$ ranges over all of $G$, so the image is the whole orbit $G \cdot x_\lambda = \phi_d(\lambda)$.
  Hence $p$ restricts to a surjection $O'_{d, \lambda} \twoheadrightarrow \phi_d(\lambda)$, so $\orb(p) = \coprod_d \phi_d$ is a dimension-preserving bijection.
\end{proof}

We conclude this appendix with the uniqueness of projective covers, used in \cref{ssec:ProjectiveCoversAsStrongApproximation} and \cref{lem:GenSystem}; the standard argument applies verbatim in $[\I, \Set]$.

\begin{xlemma}[uniqueness of projective covers]\label{lem:CoverUnique}
  Any two projective covers $(P, e)$ and $(P', e')$ of the same $\mathscr{F} \in [\I, \Set]$ are isomorphic via some $\varphi\colon P \to P'$ with $e' \circ \varphi = e$.
\end{xlemma}

\begin{proof}
  Projectivity of $P$ lifts $e$ along the epimorphism $e'$ to some $\varphi$ with $e' \circ \varphi = e$.
  The image $\operatorname{Im}(\varphi) \subseteq P'$ (computed pointwise) satisfies $e'(\operatorname{Im}(\varphi)) \supseteq \operatorname{Im}(e) = \mathscr{F}$, so superfluousness of $e'$ forces $\operatorname{Im}(\varphi) = P'$, \ie, $\varphi$ is an epimorphism.
  Projectivity of $P'$ then yields $\psi\colon P' \to P$ with $\varphi \circ \psi = \mathrm{id}_{P'}$, and $e \circ \psi = e' \circ \varphi \circ \psi = e'$, so the subobject $\operatorname{Im}(\psi) \subseteq P$ maps onto $\mathscr{F}$ via $e$; superfluousness of $e$ forces $\operatorname{Im}(\psi) = P$.
  Thus $\psi$ is both an epimorphism and a monomorphism (being split by $\varphi$), hence an isomorphism (isomorphisms in $[\I, \Set]$ are the pointwise bijections), and $\varphi = \psi^{-1}$ is an isomorphism as well.
\end{proof}

\section{Proofs for \texorpdfstring{\cref{ssec:MinimalStrongNominalAutomata}}{Strong Lifts of the Minimal Nominal Automaton}}\label{app:MinimalStrongNominalAutomata}

\StrongApproximation*

\begin{proof}
  By \cref{thm:ProjectiveCover}, $Q'$ is strong.
  Set $F' = p^{-1}(F)$.
  Since $q_0$ has empty support, its orbit is a singleton; by \cref{thm:OrbitCount}, the corresponding orbit in $Q'$ also has dimension $0$ and thus consists of a single element, which we denote $q_0'$, with $p(q_0') = q_0$ and $\supp(q_0') = \emptyset$.
  Since $\Sigma$ is strong by assumption and $Q'$ is strong, their product $Q' \times \Sigma$ is strong (applying \cref{lem:ProductOrbit} orbitwise), so $J'(Q' \times \Sigma)$ is projective in $[\I, \Set]$ by \cref{thm:Characterization}; hence the map $J(\delta \circ (p \times \mathrm{id}_\Sigma))$ lifts along the epimorphism $J(p) = e\colon J(Q') \twoheadrightarrow J(Q)$, and by full faithfulness of $J'$, this yields an equivariant $\delta'\colon Q' \times \Sigma \to Q'$ satisfying $p \circ \delta' = \delta \circ (p \times \mathrm{id}_\Sigma)$.
  Hence $p$ is an automaton homomorphism.
  A standard induction on words shows $p(\delta'^\ast(q_0', w)) = \delta^\ast(q_0, w)$; since $F' = p^{-1}(F)$, this gives $\delta'^\ast(q_0', w) \in F' \iff \delta^\ast(q_0, w) \in F$, confirming $L(\mathcal{A}') = L(\mathcal{A})$.
\end{proof}

\section{Effective Operations on the Array Representation}\label{app:Operations}

We summarize the computational costs of the effective operations on orbit-finite strong nominal sets, which were introduced in \cref{ssec:DataRepresentations}.
This analysis parallels that of~\cite[Sec.~8]{VenhoekMR22} for the total-order symmetry.

We adopt the constant-time assumption on orbit-level primitives from \cref{sec:AutomataAlgorithms}, as in \cite{VenhoekMR22}.
By representing an equivariant set as a hash set of its orbits and an equivariant map as a per-orbit index array (as detailed in \cref{ssec:DataRepresentations}), we can efficiently implement these operations.
We assume that user-supplied predicates and maps evaluate in $O(1)$ time.
The costs are then as given in \cref{tab:Operations}.

\begin{table}[t]
  \caption{Costs of the effective operations on orbit-finite strong nominal sets defined in \cref{ssec:DataRepresentations}, expressed in the orbit counts $N(\cdot)$ and dimensions $\dim(\cdot)$ of the operands; $\beta$ is the product-orbit count of \cref{prop:Product}. Orbit-level primitives are assumed constant-time as in \cref{sec:AutomataAlgorithms}; bounds relying on hashing are expected time.}\label{tab:Operations}
  \centering
  \begin{tabular}{@{}ll@{}}
    \toprule
    Operation & Cost \\
    \midrule
    Membership $x \in X$ & $O(1)$ \\
    Inclusion $Y \subseteq X$ & $O(N(Y))$ \\
    Emptiness $X = \emptyset$ & $O(1)$ \\
    Union $X \cup Y$, intersection $X \cap Y$, coproduct $X \sqcup Y$ & $O(N(X) + N(Y))$ \\
    Image $\{ f(x) \mid x \in X \}$ & $O(N(X))$ \\
    Filter $\{x \in X \mid p(x)\}$ & $O(N(X))$ \\
    Product $X \times Y$ & $O(N(X) \, N(Y) \, \beta(\dim(X), \dim(Y)))$ \\
    \bottomrule
  \end{tabular}
\end{table}

By storing orbit representations in a hash set rather than an ordered tree, we achieve expected-$O(1)$ membership tests and sorting-free image computations, in contrast to the (worst-case) $O(\log N(X))$ and $O(N(X) \log N(X))$ tree bounds of \cite{VenhoekMR22}.
The only factorial cost is the product: by \cref{prop:Product}, two single orbits of dimensions $m$ and $n$ yield $\beta(m, n)$ orbits, each built in $O(1)$ from its partial bijection.
This is the same factor $\beta$ appearing in the costs of the algorithms of \cref{sec:AutomataAlgorithms}.

The constant-time assumption on orbit-level primitives is an idealization, and its gap to actual costs is not always negligible.
Constructing the orbit of a concrete element scans its entire representation: for a word $w \in \Sigma^\ast$ of length $\ell$, whose dimension may reach $\ell \cdot \dim(\Sigma)$, the scan takes $\Omega(\ell)$ time.
Learning is particularly exposed to this gap, as the observation table is indexed by words: filling a cell computes the orbit of a word of length $O(\ell)$ for $\ell = m + O(n d)$ as in \cref{thm:Learn}, and each membership query transmits such a word, so the actual running time carries a factor of $\ell$ that the model hides and that is not easily avoided.
In our experience developing \Namae{}, orbit construction and representative selection were among the most performance-critical operations: deriving orbit construction for lists generically from their algebraic structure $\mathsf{Nil} \mid \mathsf{Cons}(x, \mathit{xs})$ computes a coincidence pattern of the pair $(x, \mathit{xs})$ at every step and is therefore quadratic in the length $\ell$, so \Namae{} instead builds the orbit of a word by a single flat scan.

\section{Proofs and Pseudocode for \texorpdfstring{\cref{ssec:LanguageEquivalence}}{Language Equivalence}}\label{app:GeneratingSystem}

\subparagraph{The generating system.}
Let $X$ be an orbit-finite strong nominal set with orbits $O_1, \dots, O_N$, where $O_i \cong \Inj([d_i], \mathbb{N})$; we identify elements of $O_i$ with injective arrays $x\colon [d_i] \to \mathbb{N}$.
A \emph{generating system} $\mathcal{G}$ represents an equivariant equivalence relation ${\sim}$ on $X$ using the following data, generalizing the register-level structures of~\cite{MurawskiRT15,MurawskiRT25,MurawskiRT18} to orbit-finite strong nominal sets:
\begin{itemize}
  \item a partition of the orbit indices $\{1, \dots, N\}$ into classes, mapping each orbit to its class representative to track ${\sim}$-related orbits;
  \item for each class representative $r$, a \emph{characteristic set} $\chi_r \subseteq [d_r]$ and a generating set $G_r$ for a subgroup $\langle G_r \rangle \le \Sym(\chi_r)$;
  \item for each orbit $O_i$ in the class of $r$, an injection (termed a \emph{ray}) $\rho_i\colon \chi_r \hookrightarrow [d_i]$ aligning the characteristic positions with those of $O_i$.
\end{itemize}
This data \emph{represents} the relation
\begin{equation}\label{eq:GenSystemRelation}
  x \sim y
  \;\iff\;
  \text{$O_i$, $O_j$ lie in the same class $r$, and $x \circ \rho_i \circ \sigma = y \circ \rho_j$ for some $\sigma \in \langle G_r \rangle$,}
\end{equation}
for $x \in O_i$ and $y \in O_j$, where $x \circ \rho_i \in \Inj(\chi_r, \mathbb{N})$ lists the \emph{characteristic atoms} of $x$.
In~\cite{MurawskiRT18}, the global freshness of fresh-register automata is additionally handled by indexing the structure with history levels and by freshness-aware bisimulation conditions~\cite[Defs.~10 and 18]{MurawskiRT18}, orthogonally to the per-level group-theoretic representation generalized here.

The operations $\textsf{create}$, $\textsf{contains}$, and $\textsf{classes}$ maintain this data directly: $\textsf{create}(X)$ makes each orbit a singleton class with $\chi_i = [d_i]$, trivial group, and identity ray; $\textsf{contains}(x, y)$ transports the coincidence pattern of $(x, y)$ to $\chi_r$ and tests membership in $\langle G_r \rangle$; and $\textsf{classes}$ reads off the classes with their characteristic tuples.
The one nontrivial operation is $\textsf{merge}$, shown in \cref{alg:Merge}.
Writing $f \mathbin{;} g$ for the partial-bijection composite that applies $f$ then $g$, $\textsf{merge}(x, y)$ forms the \emph{coincidence pattern} $s$ of $(x, y)$ (the partial bijection matching equal atoms), transports it to the characteristic set along the rays, and coarsens ${\sim}$ by one of three updates: a \emph{generator addition} when the transported pattern is a permutation of $\chi_r$, a characteristic-set \emph{shrink} when it is partial, or a class \emph{fusion} when $x$ and $y$ start in different classes.
The shrink recomputes $\chi_r$ as $\textsf{base}(\chi_r; \{\sigma\} \cup G_r)$, the largest $\Delta \subseteq \chi_r$ on which the transported pattern $\sigma$ and the generators all restrict to permutations of $\Delta$; this is the base of~\cite[Lem.~23]{MurawskiRT18}, computable by the graph reachability of~\cite[Lem.~24]{MurawskiRT18}.

\begin{algorithm}[t]
  \small
  \caption{The $\textsf{merge}$ operation of a generating system}\label{alg:Merge}
  \KwIn{elements $x \in O_i$ and $y \in O_j$}
  $s \leftarrow$ the coincidence pattern of $(x, y)$: the partial bijection with $s(a) = b \iff x(a) = y(b)$\;
  \eIf{$O_i$ and $O_j$ lie in the same class $r$}{
    $\sigma \leftarrow \rho_i \mathbin{;} s \mathbin{;} \rho_j^{-1}$\tcp*{transported pattern, a partial bijection on $\chi_r$}
    \eIf{$\dom(\sigma) = \chi_r$}{
      $G_r \leftarrow G_r \cup \{\sigma\}$\tcp*{generator addition}
    }{
      $\chi' \leftarrow \textsf{base}(\chi_r; \{\sigma\} \cup G_r)$\tcp*{shrink}
      $G_r \leftarrow \{\, g|_{\chi'} : g \in \{\sigma\} \cup G_r,\ g|_{\chi'} \in \Sym(\chi') \,\}$;\quad $\chi_r \leftarrow \chi'$\;
      restrict every ray of class $r$ to $\chi'$\;
    }
  }{
    let $r_1, r_2$ be the classes of $O_i, O_j$\tcp*{fusion}
    $\sigma_1 \leftarrow \rho_i \mathbin{;} s \mathbin{;} \rho_j^{-1}$;\quad $\sigma_2 \leftarrow \rho_j \mathbin{;} s^{-1} \mathbin{;} \rho_i^{-1}$\;
    $H \leftarrow \{\sigma_1 \mathbin{;} \sigma_2\} \cup G_{r_1} \cup \{\, \sigma_1 \mathbin{;} g \mathbin{;} \sigma_2 : g \in G_{r_2} \,\}$\;
    $\chi \leftarrow \textsf{base}(\chi_{r_1}; H)$\;
    unite the two classes with representative $r_1$\;
    $G_{r_1} \leftarrow \{\, h|_{\chi} : h \in H,\ h|_{\chi} \in \Sym(\chi) \,\}$;\quad $\chi_{r_1} \leftarrow \chi$\;
    restrict every ray of class $r_1$ to $\chi$, and replace every ray $\rho_k$ of class $r_2$ by $(\sigma_1 \mathbin{;} \rho_k)|_{\chi}$\;
  }
\end{algorithm}

The following lemma is our analogue of the invariants of \cite[Sec.~4]{MurawskiRT18}; it makes the generalization precise and identifies the $\textsf{classes}$ output with the strong approximation.
Its proof follows the forcing scheme of~\cite[Sec.~4]{MurawskiRT18}; what is new is the transport across orbits of different dimensions, via rays and fresh atoms, and the identification with the projective cover.

\begin{xlemma}[generating-system invariant]\label{lem:GenSystem}
  \leavevmode
  \begin{enumerate}
    \item\label{it:GenSysEquiv} For any generating-system data, the relation ${\sim}$ of \cref{eq:GenSystemRelation} is an equivariant equivalence relation on $X$.
    \item\label{it:GenSysFinest} After $\textsf{create}(X)$ followed by any sequence of operations $\textsf{merge}(x_1, y_1), \dots, \textsf{merge}(x_k, y_k)$, the represented relation ${\sim}$ is the finest equivariant equivalence relation on $X$ relating $x_l$ and $y_l$ for all $l \le k$.
    \item\label{it:GenSysCover} $\textsf{classes}$ returns the strong nominal set $P = \coprod_r \Inj([|\chi_r|], \mathbb{N})$, one orbit per class, together with the equivariant surjection $h\colon X \twoheadrightarrow P$ sending each $x$ to its characteristic tuple $x \circ \rho_i$ (composed with a renaming $\chi_r \cong [|\chi_r|]$ fixed once per class).
    Then $P$ corresponds to the projective cover of $J(X/{\sim})$ (unique by \cref{lem:CoverUnique}), and $h$ lifts the quotient map $q\colon X \to X/{\sim}$: $p \circ h = q$ for the equivariant surjection $p\colon P \twoheadrightarrow X/{\sim}$ corresponding to the cover.
  \end{enumerate}
\end{xlemma}

Invariants~\ref{it:GenSysEquiv} and~\ref{it:GenSysCover} are proved directly below; for the minimality half of Invariant~\ref{it:GenSysFinest}, we first extract the recurring arguments as standalone lemmas, following this strategy.
Each elementary step of \cref{alg:Merge} replaces the data $(\chi_r, G_r)$ of the affected class by $(\chi', S|_{\chi'})$, where $S$ is a set of transported patterns and $\chi' = \textsf{base}(\chi_r; S)$, so that the post-step relation relates exactly the pairs whose $\chi'$-tuples differ by a twist in $\langle S|_{\chi'} \rangle$.
Minimality therefore reduces to showing that any equivariant equivalence relation ${\approx}$ relating a witnessing pair for each pattern in $S$, in the sense of \emph{realization} defined below, must relate all these twisted pairs; this is \cref{lem:Forcing}.
Its proof moves the atoms of an element outside its $\chi'$-positions freely, which is sound because ${\approx}$ cannot separate elements by those atoms; \cref{lem:Confinement} establishes this by confining the supports of ${\approx}$-classes to $\chi'$-atoms.
Both rest on \cref{lem:TupleTransfer}, which transfers characteristic tuples freely across the orbits of a class.

The lemmas concern a fixed generating-system state representing ${\sim}$ and a class $r$.
For an element $u \in O_k$ of the class and $\Delta \subseteq \chi_r$, the \emph{$\Delta$-tuple} of $u$ is the restriction $(u \circ \rho_k)|_\Delta$ of its characteristic tuple, and the \emph{$\Delta$-atoms} of $u$ are the entries $u(\rho_k(c))$ for $c \in \Delta$; in particular, the $\chi_r$-tuple is the characteristic tuple, and the $\chi_r$-atoms are the characteristic atoms.

\begin{xlemma}[tuple transfer]\label{lem:TupleTransfer}
  Let $O_k$ be an orbit of class $r$.
  \begin{enumerate}
    \item Every $t \in \Inj(\chi_r, \mathbb{N})$ is the characteristic tuple of some $v \in O_k$, whose remaining entries can be chosen fresh for any prescribed finite set of atoms.
    \item Any $u, v$ in the class with equal characteristic tuples satisfy $u \sim v$.
  \end{enumerate}
\end{xlemma}

\begin{proof}
  For~(1), extend $t$ to an injective array of dimension $d_k$ by filling the positions outside the ray image with pairwise distinct atoms avoiding $t$ and the prescribed set; every injective array of that dimension lies in $O_k$.
  For~(2), equal characteristic tuples satisfy \cref{eq:GenSystemRelation} with the identity twist.
\end{proof}

For an equivariant equivalence relation ${\approx}$ on $X$, we say a pair $u \approx v$, with $u \in O_k$ and $v \in O_l$ in the class, \emph{realizes} a partial bijection $\tau$ on $\chi_r$ if the characteristic tuples of $u$ and $v$ agree exactly along $\tau$: $u(\rho_k(\tau(c))) = v(\rho_l(c))$ for every $c \in \dom(\tau)$, and no further $\chi_r$-atoms of $u$ and $v$ coincide.
If $(u, v)$ realizes $\tau$, then $(v, u)$ realizes $\tau^{-1}$; since $\textsf{base}$ and generated subgroups are unchanged under inverting their arguments, we use realizing pairs in either orientation.
If ${\approx}$ contains ${\sim}$, every $g \in \langle G_r \rangle$ is realized: given any $u \in O_k$, \cref{lem:TupleTransfer}(1) provides $v \in O_k$ with $v \circ \rho_k = u \circ \rho_k \circ g$ and fresh remaining entries, so $u \sim v$ and the characteristic coincidences are exactly those matched by $g$.

\begin{xlemma}[support confinement]\label{lem:Confinement}
  Let ${\approx}$ be an equivariant equivalence relation on $X$ containing ${\sim}$, let $S$ be a set of partial bijections on $\chi_r$, each realized by some ${\approx}$-pair in the class, and let $\chi' = \textsf{base}(\chi_r; S)$.
  Then $\supp([u]_\approx)$ consists of $\chi'$-atoms of $u$, for every $u$ in the class.
\end{xlemma}

\begin{proof}
  \leavevmode
  \begin{description}
    \item[Confinement to $\chi_r$.]
    A permutation fixing the $\chi_r$-atoms of $u$ fixes its characteristic tuple, so it maps $u$ to a ${\sim}$-related element by \cref{lem:TupleTransfer}(2) and therefore stabilizes $[u]_\approx$; hence $\supp([u]_\approx)$ consists of $\chi_r$-atoms of $u$.
    \item[Position constancy.]
    Let $D_u = \{c \in \chi_r : u(\rho_k(c)) \in \supp([u]_\approx)\}$ collect the positions of the support in $u$.
    Acting by $\pi \in G$ moves the support $\supp([\pi \cdot u]_\approx) = \pi \cdot \supp([u]_\approx)$ and the entries together, so $D_{\pi \cdot u} = D_u$; and equal-tuple elements share both their ${\approx}$-class and their characteristic tuple, so their position sets agree as well.
    Since an equal-tuple hop via \cref{lem:TupleTransfer}(1) followed by a $G$-move connects any two elements of the class, all position sets equal one set $D$.
    \item[Invariance under $S$.]
    Let $(u_1, v_1)$ realize $\tau \in S$.
    The atoms of $\supp([u_1]_\approx) = \supp([v_1]_\approx)$ are $\chi_r$-atoms of both $u_1$ and $v_1$ by confinement, hence matched by the pattern $\tau$; reading off their positions gives $D \subseteq \dom(\tau)$ and $\tau(D) = D$ by constancy.
    \item[Descent.]
    Every $\tau \in S$ thus restricts to a permutation of $D$, so $D$ is contained in the largest such subset, $\chi' = \textsf{base}(\chi_r; S)$.
    \qedhere
  \end{description}
\end{proof}

\begin{xlemma}[forcing]\label{lem:Forcing}
  Under the hypotheses of \cref{lem:Confinement}, with $\chi'$ as there, any $u, v$ in the class whose $\chi'$-tuples differ by a twist in $\langle S|_{\chi'} \rangle$, \ie, $(u \circ \rho_k)|_{\chi'} \circ \tau = (v \circ \rho_l)|_{\chi'}$ for some $\tau \in \langle S|_{\chi'} \rangle$, satisfy $u \approx v$.
\end{xlemma}

\begin{proof}
  \leavevmode
  \begin{description}
    \item[Equal tuples ($\tau = \mathrm{id}$).]
    Since $u$ and $v$ may lie in different orbits, and \cref{lem:TupleTransfer}(2) requires equal \emph{full} characteristic tuples, we route through fresh intermediates: by \cref{lem:TupleTransfer}(1), let $w$, in $u$'s orbit, carry a characteristic tuple agreeing with $u$'s on $\chi'$ and fresh elsewhere, and let $w'$, in $v$'s orbit, carry the same characteristic tuple as $w$ and fresh remaining entries.
    The permutation fixing the $\chi'$-atoms of $u$ and moving its remaining atoms to the fresh ones maps $u$ to $w$ and fixes $[u]_\approx$ by \cref{lem:Confinement}, so $u \approx w$; likewise $v \approx w'$; and $w \sim w'$ by \cref{lem:TupleTransfer}(2), so $u \approx v$ by transitivity.
    \item[A single generator.]
    Let the $\chi'$-tuples differ by $\tau|_{\chi'}$ for a single $\tau \in S$, realized by $(u_1, v_1)$.
    Transporting the pair by equivariance so that the $\chi'$-tuple of $u_1$ matches that of $u$, the equal-tuple case applies on both sides: $u \approx u_1 \approx v_1 \approx v$.
    \item[General twists.]
    A twist in $\langle S|_{\chi'} \rangle$ factors as a composite of generators and their inverses; interpolating elements that carry the intermediate $\chi'$-tuples, via \cref{lem:TupleTransfer}(1), and chaining the previous case, applied to the swapped realizing pair for inverse hops, gives $u \approx v$ by transitivity.
    \qedhere
  \end{description}
\end{proof}

\begin{proof}[Proof of \cref{lem:GenSystem}]
  We establish each claim in turn.

  \begin{description}
    \item[Proof of Invariant~\ref{it:GenSysEquiv}.]
      Reflexivity and symmetry hold via $\sigma = \mathrm{id}$ and $\sigma^{-1}$.
      For transitivity, $x \circ \rho_i \circ \sigma = y \circ \rho_j$ and $y \circ \rho_j \circ \sigma' = z \circ \rho_k$ give $z \circ \rho_k = x \circ \rho_i \circ (\sigma \sigma')$ with $\sigma \sigma' \in \langle G_r \rangle$.
      For equivariance, acting by $\pi \in G$ replaces the array $x$ by $\pi \circ x$, so the defining equation is preserved: $(\pi \circ x) \circ \rho_i \circ \sigma = (\pi \circ y) \circ \rho_j$.

    \item[Proof of Invariant~\ref{it:GenSysFinest}.]
      $\textsf{create}(X)$ initializes singleton classes with $\chi_i = [d_i]$, trivial groups, and identity rays; by \cref{eq:GenSystemRelation} this represents equality, the finest equivariant equivalence relation.
      $\textsf{merge}(x, y)$ of \cref{alg:Merge} coarsens the represented relation by \emph{elementary steps} until $x \sim y$ holds: a class \emph{fusion} when the two orbits lie in different classes, a \emph{generator addition} when the transported pattern $\sigma$ is a permutation of $\chi_r$, or a characteristic-set \emph{shrink} that sets $\chi_r$ to $\textsf{base}(\chi_r; \{\sigma\} \cup G_r)$ when $\sigma$ is partial.
      Each elementary step only coarsens the represented relation, and afterwards $x \sim y$ holds by construction, so the final relation is an equivariant equivalence relation (by Invariant~\ref{it:GenSysEquiv}) relating all merged pairs.

      For minimality, let ${\approx}$ be any equivariant equivalence relation with $x_l \approx y_l$ for all $l$; we show by induction over elementary steps that the represented relation stays contained in ${\approx}$.
      This holds after $\textsf{create}$.
      Consider a step triggered by $\textsf{merge}(x, y)$, where $x \approx y$ and, inductively, the pre-step relation ${\sim}$ is contained in ${\approx}$; \cref{lem:TupleTransfer,lem:Confinement,lem:Forcing} then apply to this ${\sim}$ and ${\approx}$.
      \begin{description}
        \item[Generator additions and shrinks.] These replace $(\chi_r, G_r)$ by $(\chi', (\{\sigma\} \cup G_r)|_{\chi'})$ for $\chi' = \textsf{base}(\chi_r; \{\sigma\} \cup G_r)$; when $\sigma$ is total, $\chi' = \chi_r$ and the step is a generator addition.
        The post-step relation relates exactly the pairs of the class whose $\chi'$-tuples differ by a twist in $\langle (\{\sigma\} \cup G_r)|_{\chi'} \rangle$, and each member of $S = \{\sigma\} \cup G_r$ is realized by an ${\approx}$-pair: the pair $(y, x)$ realizes $\sigma$, as $\sigma$ is exactly the transported characteristic pattern of the merged pair, and the pre-step relation realizes each $g \in G_r$, as noted after \cref{lem:TupleTransfer}.
        Hence \cref{lem:Forcing} forces every post-step pair.
        \item[Fusions.] Relating $x \approx y$ across two classes merges their ${\approx}$-connected components, so uniting the classes is forced, and the pairs represented afterwards are exactly the $\langle H|_\chi \rangle$-twists, along the new rays, of tuple-equal pairs, for $\chi = \textsf{base}(\chi_{r_1}; H)$.
        The transported rays $(\sigma_1 \mathbin{;} \rho_k)|_{\chi}$ are total and injective since $\chi \subseteq \dom(\sigma_1)$: $H$ contains $\sigma_1 \mathbin{;} \sigma_2$, the partial identity on $\dom(\sigma_1)$, and every member of $H$ restricts to a permutation of $\chi$ by the definition of $\textsf{base}$.
        Each member of $H$ is realized by an ${\approx}$-pair: each $g' \in G_{r_1}$ by the pre-step relation, as before; and a conjugate $\sigma_1 \mathbin{;} g \mathbin{;} \sigma_2$ with $g \in G_{r_2}$ by $(x, \pi \cdot x)$, where $\pi \in G$ realizes the $g$-twist of $y$ and moves the atoms of $x$ outside the characteristic atoms of $y$ to fresh ones, so that $x \approx y \approx \pi \cdot y \approx \pi \cdot x$ by the inductive hypothesis and equivariance, and the characteristic tuples of $x$ and $\pi \cdot x$ agree exactly along $\sigma_1 \mathbin{;} g \mathbin{;} \sigma_2$ (travel along $\sigma_1$, twist by $g$, return along $\sigma_2$); $g = \mathrm{id}$ covers $\sigma_1 \mathbin{;} \sigma_2$.
        The arguments of \cref{lem:Confinement,lem:Forcing} then run over the united class with $S = H$ and the new rays, with the uses of \cref{lem:TupleTransfer}(2) adjusted to cross the old class boundary as follows.
        For confinement, stage~1 runs per old class and confines supports to the old characteristic atoms on each side; invariance under the realized $\sigma_1 \mathbin{;} \sigma_2$ then places the position set $D$ inside $\dom(\sigma_1)$, so supports consist of atoms shared along the bridge, and transporting positions through $\sigma_1$ extends position constancy to the $r_2$ side.
        In the equal-tuple stage of \cref{lem:Forcing}, a cross-boundary pair $u$, on the $r_1$ side, and $v$, on the $r_2$ side, with equal $\chi$-tuples along the new rays is connected by the chain $u \approx w \sim \pi' \cdot x \approx \pi' \cdot y \sim w' \approx v$: the fresh intermediates $w$ and $w'$ are as in \cref{lem:Forcing}, $\pi'$ is chosen so that $\pi' \cdot x$ carries the full $\chi_{r_1}$-tuple of $w$ and the remaining characteristic atoms of $y$ go to fresh ones, and $w'$ carries the full $\chi_{r_2}$-tuple of $\pi' \cdot y$; the two $\sim$-hops are old-relation equal-tuple pairs, and the outer hops are the fresh-intermediate moves of \cref{lem:Forcing}.
        Hence every post-step pair is forced.
      \end{description}
      Hence the final represented relation is contained in every such ${\approx}$, and is therefore the finest one.

    \item[Proof of Invariant~\ref{it:GenSysCover}.]
      By \cref{eq:GenSystemRelation}, $q(x)$ is determined by, and determines, the class $r$ of $x$'s orbit together with the coset $[x \circ \rho_i] \in \Inj(\chi_r, \mathbb{N}) / \langle G_r \rangle$, so
      \begin{equation*}
        X/{\sim} \;\cong\; \coprod_r \Inj(\chi_r, \mathbb{N}) / \langle G_r \rangle,
      \end{equation*}
      an orbit-finite nominal set with one orbit per class, whose $r$-th orbit has support representation $(\chi_r, \langle G_r \rangle)$.
      By \cref{thm:OrbitCount}, the projective cover of $J(X/{\sim})$ has exactly one orbit per class, of matching dimension $|\chi_r|$, so its cover object is isomorphic to $J'(P)$; by \cref{lem:CoverUnique} this identification is unique up to isomorphism.
      Choosing the Yoneda representative of the $r$-th summand in the construction of \cref{thm:ProjectiveCover} to be $q(x_r)$, for a class representative $x_r$ whose characteristic tuple is the inclusion $\chi_r \subseteq \mathbb{N}$, and unwinding the construction as in the proof of \cref{thm:OrbitCount}, gives $p(h(x)) = [x \circ \rho_i] = q(x)$ for every $x$, \ie, $p \circ h = q$.
      Finally, $h$ is surjective since every tuple in $\Inj(\chi_r, \mathbb{N})$ arises as the characteristic tuple of some element of the class.
      \qedhere
  \end{description}
\end{proof}

\para{Operation costs.}
We detail the worst-case cost for a single call of each operation. The aggregate cost analysis required for the equivalence bound is deferred to the proof of \cref{thm:Equivalence}.
\begin{description}
  \item[$\textsf{create}(X)$] initializes $N$ singleton classes with full characteristic sets and identity rays in $O(N \cdot d)$ time.
  \item[$\textsf{contains}(x, y)$] derives the partial bijection from shared atoms in $O(d)$ time and performs one group-membership test in $\mu(d)$ time, for $O(d + \mu(d))$ overall.
  \item[$\textsf{merge}(x, y)$] executes one branch of \cref{alg:Merge}: a class fusion (recomputing the characteristic set via $\textsf{base}$ and all rays of the fused class in $O(N \cdot d)$, plus a Schreier--Sims rebuild in $\mu(d)$), a characteristic-set shrink (a $\textsf{base}$ computation and generator restriction, $O(d^2 + \mu(d))$; rays are restricted lazily upon access), or a generator addition (one rebuild, $\mu(d)$), for $O(N \cdot d + d \cdot \mu(d))$ worst-case per call.
  Across any run there are at most $N - 1$ fusions and at most $N \cdot d$ shrinks in total, since each fusion removes a class and each shrink strictly decreases $\sum_r |\chi_r| \le N d$.
  \item[$\textsf{classes}$] extracts the $N$ class orbits and sorts their representatives to assign contiguous identifiers, taking $O(N \cdot d + N \log N)$ time; this cost enters the quotient operation of \cref{ssec:QuotientOp} rather than the equivalence check.
\end{description}

\begin{algorithm}[t]
  \small
  \caption{Language equivalence via bisimulation with a generating system}\label{alg:Equivalence}
  \KwIn{strong nominal automata $\mathcal{A}_i = (\Sigma, Q_i, q_0^i, \delta_i, F_i)$ for $i \in \{1, 2\}$ over a common alphabet $\Sigma$}
  \KwOut{whether $L(\mathcal{A}_1) = L(\mathcal{A}_2)$}
  $\mathcal{G} \leftarrow \textsf{create}(Q_1 \sqcup Q_2)$;\quad $\mathcal{G}.\textsf{merge}(q_0^1, q_0^2)$;\quad enqueue $(q_0^1, q_0^2)$\;
  \While{the queue is nonempty}{
    dequeue $(p_1, p_2)$\;
    \lIf{$(p_1 \in F_1) \neq (p_2 \in F_2)$}{\Return $\textsf{false}$}
    \ForEach{representative input $a$ of an orbit of $(G \cdot (p_1, p_2)) \times \Sigma$}{
      $(p_1', p_2') \leftarrow (\delta_1(p_1, a), \delta_2(p_2, a))$\;
      \If{\textbf{not} $\mathcal{G}.\textsf{contains}(p_1', p_2')$}{
        $\mathcal{G}.\textsf{merge}(p_1', p_2')$\;
        enqueue $(p_1', p_2')$\;
      }
    }
  }
  \Return $\textsf{true}$\;
\end{algorithm}

\Equivalence*

\begin{proof}
  Due to determinism, language equivalence coincides with bisimilarity.
  By induction on the search, every enqueued pair is reachable from $(q_0^1, q_0^2)$ under a common word; hence if \cref{alg:Equivalence} returns $\textsf{false}$, some such pair differs in acceptance and the automata are inequivalent.
  Suppose instead it returns $\textsf{true}$, and let ${\sim}$ be the final represented relation, which by Invariant~\ref{it:GenSysFinest} of \cref{lem:GenSystem} is the finest equivariant equivalence relating every merged pair.
  Let $R \subseteq {\sim}$ consist of the pairs that agree on acceptance and whose successors on each input are again ${\sim}$-related; this $R$ is an equivariant equivalence containing all merged pairs, since every merged pair is dequeued, its acceptance verified, and its successor orbit representatives merged into ${\sim}$ (guarded by $\textsf{contains}$).
  By the minimality of ${\sim}$ we get $R = {\sim}$, so ${\sim}$ is a bisimulation up to equivariant-equivalence closure; since $(q_0^1, q_0^2) \in {\sim}$, the automata are equivalent.

  For the time complexity, note that every enqueued pair triggers an effective $\textsf{merge}$ (guarded by $\textsf{contains}$), and each effective merge performs at least one elementary step (fusion, shrink, or generator addition), so the number of visited pairs is at most the total number of elementary steps, which we now bound.
  Fusions number at most $N - 1$ and shrinks at most $N d$, as each fusion removes a class and each shrink strictly decreases $\sum_r |\chi_r| \le N d$.
  Call an \emph{epoch} of a class a maximal period during which its characteristic set stays fixed; every fusion or shrink ends at most two epochs and starts one, so there are $O(N d)$ epochs overall.
  Within one epoch the ambient group $\Sym(\chi_r)$ is fixed, so the generator additions of that epoch form a strictly increasing chain of subgroups of $\Sym(\chi_r)$, of length $O(|\chi_r|) = O(d)$ by Babai's bound on subgroup chains in symmetric groups~\cite{Babai86}.
  Hence there are $O(N d \cdot d) = O(N d^2)$ elementary steps, and as many visited pairs.
  Exploring a pair requires iterating over the representative inputs from the orbits of $(G \cdot (p_1, p_2)) \times \Sigma$.
  Since the state pair has dimension at most $2d$ and the alphabet letters have dimension $e$, \cref{prop:Product} bounds the number of branches by $N(\Sigma) \cdot \beta(2d, e)$.
  Each branch invokes $\textsf{contains}$, which takes $O(\mu(d))$ time for the group membership test. Across all $O(N \cdot d^2)$ visited pairs, this contributes $O(N \cdot d^2 \cdot N(\Sigma) \cdot \beta(2d, e) \cdot \mu(d))$ to the total time.
  The merges themselves add $O(N^2 d)$ for ray recomputations across all fusions, $O(N d \cdot (d^2 + \mu(d)))$ for the $\textsf{base}$ computations and generator restrictions of the shrinks, and one Schreier--Sims rebuild per elementary step, $O(N \cdot d^2 \cdot \mu(d))$ in total; all of these are absorbed by the $\textsf{contains}$ term and the $O(N^2 d)$ term.
  Summing these costs yields the stated time bound. Because $\beta(2d, e) = O((2d)^e e!)$, the overall time is polynomial in $N$, $N(\Sigma)$, and $d$ whenever $e$ is fixed.
\end{proof}

This generalizes the polynomial-time result for register automata~\cite{MurawskiRT18} to strong nominal automata.

\para{Comparison with fresh-register automata.}
For fresh-register automata, the $r$ registers hold pairwise distinct values and each control state fixes which registers are filled (the availability function of~\cite[Def.~1]{MurawskiRT18}), so the configurations at each control state form a single strong orbit.
Hence $N = |Q|$ and $d \le r$, where $|Q|$ is the number of control states and $r$ the number of registers.

As we do not treat global freshness, we compare against their $h = \infty$ case, where a conservative bound of $O(|Q|^2 r^2 p(r))$ is given~\cite{MurawskiRT18}, with $p$ a polynomial accounting for group-theoretic operations; substituting $N = |Q|$, $d \le r$, and $\mu$ for $p$, we write it as $O(N^2 d^2 \mu(d))$, which only strengthens their bound.
Because their analysis does not charge the cost of updating the generating system, this bound cannot be compared directly with ours; charging each of their $O(N^2 d)$ updates its worst-case cost of $O(d^2 \mu(d))$ yields $O(N^2 d^3 \mu(d))$.
At $e=1$ with $N(\Sigma)$ treated as constant, our bound $O(N d^3 \mu(d) + N^2 d)$ refines this single term into two lower-degree terms and hence is more precise.

Moreover, the comparison does not depend on this recounting, because $N$ and $d$ are not independent.
In a reachable strong nominal automaton, $\supp(q_0) = \emptyset$ and equivariance gives $\supp(\delta(q, \sigma)) \subseteq \supp(q) \cup \supp(\sigma)$, so each transition adds at most $e$ atoms to the support.
Furthermore, a shortest run reaching an orbit of maximal dimension visits each orbit at most once, as transporting its suffix along the group action would otherwise shorten it; hence $d \le e\,(N - 1)$.
At $e = 1$ with $N(\Sigma)$ treated as constant, this yields $d < N$, so our bound is no larger than even the uncharged $O(N^2 d^2 \mu(d))$: indeed $N d^3 \mu(d) \le N^2 d^2 \mu(d)$ and $N^2 d \le N^2 d^2 \mu(d)$.

\section{Proofs and Pseudocode for \texorpdfstring{\cref{ssec:MinimizationLearning}}{Minimization and Learning}}\label{app:MinimizationProofs}

This appendix provides the pseudocode and full proofs for the algorithms of \cref{ssec:MinimizationLearning}: the quotient operation in \cref{ssec:QuotientOp}, minimization via Moore's algorithm in \cref{ssec:MinApp}, and learning via $\nu L^\ast$ in \cref{ssec:NuLStar}.

\subsection{The Quotient Operation}\label{ssec:QuotientOp}

We present the quotient operation as \cref{alg:Quotient} and establish its correctness and complexity in \cref{lem:Quotient}. It computes the strong approximation of a quotient through the generating system of \cref{app:GeneratingSystem}, bypassing any non-strong intermediate structure.

\begin{algorithm}[t]
  \small
  \caption{$\textsf{Quotient}(X, {\equiv})$}\label{alg:Quotient}
  \KwIn{orbit-finite strong nominal set $X$; equivariant equivalence ${\equiv} \subseteq X \times X$}
  \KwOut{strong approximation $P$ of $X / {\equiv}$ and a lift $h\colon X \twoheadrightarrow P$}
  $\mathcal{G} \leftarrow \textsf{create}(X)$\;
  \ForEach{orbit $O_1$ of $X$}{
    \ForEach{orbit $O_2$ of $X$ not before $O_1$}{
      \ForEach{representative pair $(x, y)$ of an orbit of $O_1 \times O_2$}{
        \lIf{\textbf{not} $\mathcal{G}.\textsf{contains}(x, y)$ \textbf{and} $x \equiv y$}{$\mathcal{G}.\textsf{merge}(x, y)$}
      }
    }
  }
  \Return $\mathcal{G}.\textsf{classes}$\;
\end{algorithm}

\Quotient*

\begin{proof}
  \Cref{alg:Quotient} processes every ${\equiv}$-related representative pair through $\mathcal{G}$ and returns $\textsf{classes}$. Its correctness follows from \cref{lem:GenSystem}: $\mathcal{G}$ maintains the finest equivariant equivalence relation containing the merged pairs, which is ${\equiv}$ itself since ${\equiv}$ is equivariant and all representative pairs are processed, and $\textsf{classes}$ returns the strong approximation of $X / {\equiv}$ with its lift.
  Regarding performance, $\textsf{create}$ takes $O(N(X) d)$ time. The nested loops iterate over $O(N(X)^2)$ orbit pairs, enumerating the $\beta(d, d)$ orbits of each product as dictated by \cref{prop:Product}. Processing each such orbit requires one ${\equiv}$-test, of cost $C$, and at most one $\textsf{contains}$/$\textsf{merge}$ operation; since $\mu(d) = \Omega(d)$ absorbs the $O(d)$ for forming the representative pair and its partial bijection, the non-ray work is $O(C + \mu(d))$ per orbit.
  The remaining bookkeeping splits into three parts, bounded as in the proof of \cref{thm:Equivalence}. Ray recomputations are triggered only by class fusions (at most $N(X)-1$ of them) and cost $O(N(X)\,d)$ per fusion, hence $O(N(X)^2 d)$ in total. Characteristic-set shrinks number at most $N(X)\,d$ and cost $O(d^2 + \mu(d))$ each (a $\textsf{base}$ computation and generator restriction). Generator additions form subgroup chains of length $O(d)$ per epoch~\cite{Babai86}, for $O(N(X)\,d^2)$ elementary steps and as many Schreier--Sims rebuilds, contributing $O(N(X)\,d^2 \cdot \mu(d))$ in total. All three terms are dominated by the per-pattern work because $d^3 \le N(X)\,\beta(d, d)\,\mu(d)$ holds for all $N(X) \ge 1$ and $d \ge 1$ (as $\beta(d, d) \ge d^2$ and $\mu(d) = \Omega(d)$), and are thus absorbed into the bound.
  This yields a dominant total time of $O\!\left( N(X)^2 \, \beta(d, d) \, (C + \mu(d)) + N(X)^2 d + N(X)\, d^2 \cdot \mu(d) \right) = O\!\left( N(X)^2 \, \beta(d, d) \, (C + \mu(d)) \right)$.
  Given $\beta(d, d) = O(2^d d!)$, this bound is polynomial in $N(X)$ but factorially dependent on $d$.
\end{proof}

The generating system is not strictly required for the quotient operation.
A \emph{support-intersection} variant enumerates the $\beta(d, d)$ partial bijections per orbit pair and tests ${\equiv}$ on a representative pair of each product orbit directly.
The least support of an ${\equiv}$-class equals the intersection of the supports of its members: the inclusion into each member's support is immediate from equivariance of the quotient map, and conversely, if $a \notin \supp([x]_{\equiv})$, swapping $a$ with a fresh atom fixes the class and produces a member whose support avoids $a$.
Intersecting the supports of the equivalent representatives incrementally along this enumeration therefore yields the strong approximation together with its lift.
This replaces the $\mu(d)$ factor of \cref{lem:Quotient} by the $O(d)$ cost of the support intersections, for a total of $O(N(X)^2 \, \beta(d, d) \, (C + d))$.

The substitution extends to \cref{thm:Minimize,thm:Learn}, whose $\mu(d)$ factors are, in principle, not essential either.
For learning, each ${\rowsim}$-test reads the observation table directly, so the hypothesis construction can use the support-intersection variant as is.
For minimization, one instead iterates at the level of sets, as in Moore's algorithm for $\Ord$~\cite[Sec.~3.1]{VenhoekMR22}: maintain ${\equiv_k}$ as an equivariant subset of $Q \times Q$ and compute ${\equiv_{k+1}}$ from it by the effective operations of \cref{app:Operations} (a product ${\equiv_k} \times \Sigma$, a filter selecting the orbits whose successor pair leaves ${\equiv_k}$, an image under the projection, and a set difference), each $O(1)$ per orbit, and extract the strong approximation by a single support-intersection quotient after stabilization.
Each round then costs $O(N(Q)^2 \, \beta(d, d) \, N(\Sigma) \, \beta(2d, e))$, removing the $\mu(d)$ factor from \cref{thm:Minimize}.
In practice, however, the generating system stores the same relation far more compactly (generators instead of up to $\beta(d, d)$ patterns per orbit pair), and $\mathcal{G}.\textsf{contains}$ skips the ${\equiv}$-test for pairs already known to be equivalent, which is beneficial when $C$ dominates, as in minimization, where each ${\equiv_{k+1}}$-test iterates over $N(\Sigma) \cdot \beta(2d, e)$ representative inputs.

\subsection{Minimization}\label{ssec:MinApp}

We present minimization as \cref{alg:Minimize} and analyze its complexity in \cref{thm:Minimize}.
The input automaton is assumed reachable; the reachable part can be precomputed by a breadth-first exploration of orbits as in \cref{alg:Equivalence}.
It is Moore's algorithm: it computes the $k$-step Myhill--Nerode relations ${\equiv_k}$ until they stabilize, computing the strong approximation of each successive quotient with \cref{alg:Quotient}.
We retain the generating system $\mathcal{G}_k$ built by the $k$-th call to $\textsf{Quotient}$ across rounds, so that each query $q \equiv_k q'$ is answered by $\mathcal{G}_k.\textsf{contains}$ in $O(d + \mu(d))$ time.
Our implementation additionally restricts the enumeration in each round to pairs within a common ${\equiv_k}$-class and memoizes falsified pairs across rounds; both optimizations are sound because the relations ${\equiv_k}$ only refine, and they do not affect the worst-case bound.

\begin{algorithm}[t]
  \small
  \caption{$\textsf{Minimize}(\mathcal{A})$ via Moore's algorithm}\label{alg:Minimize}
  \KwIn{reachable strong nominal automaton $\mathcal{A} = (\Sigma, Q, q_0, \delta, F)$}
  \KwOut{strong nominal automaton $\mathcal{A}'$, a strong lift of the minimal automaton of $\mathcal{A}$}
  $(P_0, h_0) \leftarrow \textsf{Quotient}(Q, {\equiv_0})$, \textbf{where} $q \equiv_0 q' \iff (q \in F \Leftrightarrow q' \in F)$\;
  $k \leftarrow 0$\;
  \Repeat{${\equiv_k} = {\equiv_{k-1}}$}{
    $(P_{k+1}, h_{k+1}) \leftarrow \textsf{Quotient}(Q, {\equiv_{k+1}})$, \textbf{where}
      $q \equiv_{k+1} q' \iff (q \equiv_k q' \textbf{ and } \forall a \in \Sigma,\; \delta(q, a) \equiv_k \delta(q', a))$\;
    $k \leftarrow k + 1$\;
  }
  \Return the strong nominal automaton on $P_k$ induced by $h_k$ via \cref{thm:StrongApproximation}\;
\end{algorithm}

The stabilization test in the loop condition needs care: the class count $N(P_k)$ alone cannot detect a refinement that only grows the least support of a class or shrinks its local symmetry group, and stopping on the count alone may freeze a relation that is not yet a congruence.
The following lemma shows that these three quantities together witness every refinement.

\begin{xlemma}[refinement measure]\label{lem:RefinementMeasure}
  Let ${\equiv'} \subseteq {\equiv}$ be equivariant equivalence relations on an orbit-finite nominal set $X$.
  If ${\equiv'} \neq {\equiv}$, then $N(X / {\equiv'}) > N(X / {\equiv})$, or some ${\equiv'}$-class has a strictly larger least support, or a strictly smaller local symmetry group, than the ${\equiv}$-class containing it.
\end{xlemma}

\begin{proof}
  The map $X/{\equiv'} \to X/{\equiv}$, $[x]_{\equiv'} \mapsto [x]_{\equiv}$, is an equivariant surjection, so $N(X/{\equiv'}) \ge N(X/{\equiv})$; moreover, for classes $C' \subseteq C$ we have $\supp(C) \subseteq \supp(C')$, and $\Stab_G(C') \le \Stab_G(C)$, since $g \cdot C' = C'$ makes $g \cdot C$ (a class of the equivariant relation ${\equiv}$) meet $C$, forcing $g \cdot C = C$.
  Now assume all three quantities are preserved.
  First, $\supp(C') = \supp(C)$ together with equal local symmetry groups gives $\Stab_G(C') = \Stab_G(C)$: for $g \in \Stab_G(C)$, pick $g' \in \Stab_G(C')$ with $g'|_{\supp(C')} = g|_{\supp(C)}$, possible as the local symmetries coincide; then $g'^{-1} g \in \Fix_G(\supp(C')) \subseteq \Stab_G(C')$, so $g \in \Stab_G(C')$.
  Next, fix an ${\equiv}$-class $C$ and ${\equiv'}$-subclasses $C', C'' \subseteq C$.
  Since $N(X/{\equiv'}) = N(X/{\equiv})$, the orbit map induced by the surjection above is a bijection, so $C''$ and $C'$ lie in a common orbit: $C'' = g \cdot C'$ for some $g \in G$.
  Then $g \cdot C' \subseteq C \cap g \cdot C$ forces $g \cdot C = C$, so $g \in \Stab_G(C)$ and, by the first step, $g \in \Stab_G(C')$; hence $C'' = g \cdot C' = C'$.
  Thus every ${\equiv}$-class is a single ${\equiv'}$-class, i.e., ${\equiv'} = {\equiv}$.
\end{proof}

This three-part measure mirrors the convergence argument for minimizing history-dependent automata~\cite[Thm.~2]{FerrariMP02}, where each iteration either splits a block, adds a name, or shrinks a symmetry group.

\Minimize*

\begin{proof}
  By induction, ${\equiv_k}$ relates exactly the states that agree on acceptance after every word of length at most $k$; the relations refine monotonically, and once ${\equiv_k} = {\equiv_{k-1}}$ the relation is closed under transitions, hence it is the Myhill--Nerode equivalence of $\mathcal{A}$; since $\mathcal{A}$ is reachable, $Q/{\equiv_k}$ is the state space of the minimal automaton.
  The strong approximation $P_k$ of $Q/{\equiv_k}$ then serves as the state space of a strong lift of the minimal automaton per \cref{cor:MinimalStrong}; the result $\mathcal{A}'$ is this lift, with the transition function lifted onto $P_k$ by \cref{thm:StrongApproximation}.
  The lift is polynomial in the input and absorbed into the bound below.

  For the number of rounds, each round with ${\equiv_{k+1}} \neq {\equiv_k}$ advances the measure of \cref{lem:RefinementMeasure}: it increases the class-orbit count (at most $N(Q)$ times in total, since a quotient of $Q$ has at most $N(Q)$ orbits), grows the least support of some class orbit (at most $d$ times each), or strictly shrinks the local symmetry group of some class orbit (at most $O(d)$ times each, since a strictly descending chain of subgroups of $S_d$ has length $O(d)$~\cite{Babai86}).
  Hence the loop runs $O(N(Q) \cdot d)$ times.
  The generating system maintains exactly these three quantities (the classes, their characteristic sets, and their local symmetry groups), so by \cref{lem:RefinementMeasure} the stabilization test ${\equiv_k} = {\equiv_{k-1}}$ amounts to comparing them across consecutive rounds, in time absorbed by the round itself.

  Deciding ${\equiv_{k+1}}$ for a state pair, the conjunct $q \equiv_k q'$ is one $\textsf{contains}$ query on $\mathcal{G}_k$, and the universal quantification over $\Sigma$ is sound to decide on one representative input per orbit of $(G \cdot (q, q')) \times \Sigma$ by equivariance, as in \cref{alg:Equivalence}.
  The state pair has dimension at most $2d$, so \cref{prop:Product} bounds the number of branches by $N(\Sigma)\,\beta(2d, e)$, each performing one $\textsf{contains}$ query in $O(d + \mu(d)) = O(\mu(d))$; thus ${\equiv_{k+1}}$ is decidable in $C = O(N(\Sigma)\,\beta(2d, e)\,\mu(d))$.
  Each round is a call $\textsf{Quotient}(Q, {\equiv_{k+1}})$, costing $O(N(Q)^2 \, \beta(d, d)\,(C + \mu(d)))$ by \cref{lem:Quotient}; multiplying by the $O(N(Q) \cdot d)$ rounds gives the stated bound.
\end{proof}

Our result is the $\Eq$ counterpart of the $\Ord$ minimization bound of~\cite[Thm.~8.2]{VenhoekMR22}, which is $O(3^{5k} \cdot k \cdot N(Q)^3 N(\Sigma))$ where $k = \dim(Q \sqcup \Sigma)$: a product of single $\Ord$-orbits of dimensions $m$ and $n$ has $\beta_\Ord(m, n) = O(3^{m+n})$ orbits~\cite[Thm.~8.1]{VenhoekMR22}, so substituting $\beta_\Ord$ for $\beta$ in our bound, which depends on the symmetry only through this product-orbit count, yields $O(N(Q)^3 \cdot d \cdot N(\Sigma) \cdot 3^{4d + e} \cdot \mu(d))$, which recovers theirs up to the $\mu(d)$ factor, as $d \le k$ and $4d + e \le 5k$.
The remaining $\mu(d)$ factor stems from the generating system and is not essential: the set-level iteration of \cref{ssec:QuotientOp}, with a single final support-intersection quotient, avoids it entirely, yielding $O(N(Q)^3 \cdot d \cdot N(\Sigma) \cdot 3^{4d + e})$ after the substitution, which is subsumed by their bound.

\subsection{The $\nu L^\ast$ Algorithm}\label{ssec:NuLStar}

We instantiate the nominal $\nu L^\ast$ algorithm~\cite{MoermanS0KS17}, the nominal generalization of Angluin's $L^\ast$~\cite{Angluin87}, to strong nominal alphabets, replacing their implicit quotient of the prefix set by the strong approximation of \cref{lem:Quotient}.
The control flow (closedness, consistency, and counterexample handling) matches the standard $\nu L^\ast$; the only structural change is that hypothesis construction goes through $\textsf{Quotient}$, yielding a strong nominal automaton via \cref{thm:StrongApproximation} rather than a quotient in $\Nom$.

\subparagraph*{Membership and equivalence queries.}
The learner interacts with a \emph{teacher} through two oracles.
A \emph{membership query} $\textsf{MQ}(w)$ reports whether a word $w \in \Sigma^\ast$ belongs to the target language $L$, and an \emph{equivalence query} $\textsf{EQ}(\mathcal{H})$ returns $\bot$ if the hypothesis automaton $\mathcal{H}$ recognizes $L$, or otherwise a counterexample in the symmetric difference $L(\mathcal{H}) \mathbin{\triangle} L$.
Since $L$ is equivariant, both oracles are equivariant ($\textsf{MQ}(g \cdot w) = \textsf{MQ}(w)$, and $\textsf{EQ}$ may be answered up to the $G$-action), so the learner only needs to query orbit representatives, which keeps the table orbit-finite.
In our experiments, these oracles are realized black-box across a process boundary, as detailed in \cref{app:Measurement}.

\subparagraph*{Observation table.}
The basic data structure is an observation table $(S, E, T)$. Here $S$ and $E$ are equivariant orbit-finite subsets of $\Sigma^\ast$, where $S$ is prefix-closed and $E$ is suffix-closed, and $T\colon (S \cup S\!\cdot\!\Sigma)\times E \to \{ 0, 1 \}$ is the equivariant map defined by $T(s, e) = 1$ iff $s\,e \in L$. We write $\mathrm{row}(u)\colon e \mapsto T(u, e)$ for the curried counterpart of $T$, and $u \rowsim v$ for the row-equivalence relation $\mathrm{row}(u) = \mathrm{row}(v)$.
By~\cite[Lem.~1]{MoermanS0KS17}, ${\rowsim}$ is an equivariant equivalence relation refined by $\equiv_L$ (that is, $u \equiv_L v \Rightarrow u \rowsim v$).
The table admits a finite representation: each orbit of $S$, $E$, $S\!\cdot\!\Sigma$ has a canonical representative, and each cell $T_{ij}\colon G \cdot s_i \times G \cdot e_j \to \{ 0, 1 \}$ is orbit-finite, hence fillable with finitely many membership queries even though the underlying product orbit may split, as shown in \cref{prop:Product}.

\subparagraph*{Closedness and consistency.}
The table is \emph{closed} if for every $t \in S\!\cdot\!\Sigma$ there is $s \in S$ with $t \rowsim s$, and \emph{consistent} if $s_1 \rowsim s_2$ implies $s_1\,a \rowsim s_2\,a$ for every $s_1, s_2 \in S$ and $a \in \Sigma$.
Both conditions are decidable thanks to orbit-finiteness and equivariance.
For closedness, it suffices to iterate over orbit-representatives $t \in S\!\cdot\!\Sigma$ and $s \in S$ together with the finitely many support-level partial bijections $\pi$ between $\supp(s)$ and $\supp(t)$; the condition $t \rowsim s$ holds iff some such $\pi$ satisfies $\mathrm{row}(t) = \pi \cdot \mathrm{row}(s)$.
For consistency, the set of inconsistencies
\begin{equation*}
  \{(s_1, s_2, a, e) \in S\times S\times \Sigma \times E \mid s_1 \rowsim s_2 \,\wedge\, \mathrm{row}(s_1 a e) \ne \mathrm{row}(s_2 a e)\}
\end{equation*}
is an equivariant subset of an orbit-finite set and hence orbit-finite; deciding its emptiness over orbit-representatives yields, when non-empty, a witness $(s_1, s_2, a, e)$ whose separator $a\,e$ is added to $E$.

\subparagraph*{Hypothesis construction.}
When the table is closed and consistent, we extract a strong nominal automaton by quotienting $S \cup S\!\cdot\!\Sigma$ by ${\rowsim}$ via \cref{alg:Quotient}. The strong approximation $P$ of $(S \cup S\!\cdot\!\Sigma)/{\rowsim}$ returned by $\textsf{Quotient}$ serves as the state space; by \cref{thm:OrbitCount} and closedness, $P$ inherits the orbit structure of $S/{\rowsim}$.
The transition function is defined on one representative $(h(s), a)$ per orbit of $P \times \Sigma$ with $s \in S$ (closedness lets every state of $P$ be so represented) and extended equivariantly; consistency makes the quotient-level transition $[s]_{\rowsim} \mapsto [s\!\cdot\!a]_{\rowsim}$ well-defined, and our assignment lifts it along $p \circ h = q$, as in \cref{thm:StrongApproximation}.
The equivariant extension is well-defined: any $\pi \in G$ stabilizing $(h(s), a)$ fixes the atoms of $h(s)$, \ie, $\supp(\mathrm{row}(s))$, and, $\Sigma$ being strong, $\supp(a)$ pointwise; then $\pi \cdot s \rowsim s$, so consistency gives $\mathrm{row}(\pi \cdot (s\!\cdot\!a)) = \mathrm{row}(s\!\cdot\!a)$, whence $\supp(\mathrm{row}(s\!\cdot\!a)) \subseteq \supp(\mathrm{row}(s)) \cup \supp(a)$ and $\pi$ fixes the characteristic tuple $h(s\!\cdot\!a)$.

\begin{algorithm}[t]
  \small
  \caption{$\textsf{MakeHypothesis}(S, E, T)$}\label{alg:MakeHypothesis}
  \KwIn{closed and consistent observation table $(S, E, T)$}
  \KwOut{strong nominal automaton $\mathcal{H}$ consistent with the table}
  $(P, h) \leftarrow \textsf{Quotient}(S \cup S\!\cdot\!\Sigma,\ {\rowsim})$\;
  $q_0 \leftarrow h(\epsilon)$\;
  $F \leftarrow \{\, h(s) \mid s \in S,\ \mathrm{row}(s)(\epsilon) = 1 \,\}$\;
  $\delta(h(s), a) \leftarrow h(s\!\cdot\!a)$ \textbf{for one representative} $(s, a)$ \textbf{per orbit of} $h(S) \times \Sigma$, \textbf{extended equivariantly}\;
  \Return $(\Sigma, P, q_0, \delta, F)$\;
\end{algorithm}

\begin{algorithm}[t]
  \small
  \caption{$\nu L^\ast$ via distinguishing-suffix counterexamples}\label{alg:Learn}
  \KwIn{teacher with membership query $\textsf{MQ}$ and equivalence query $\textsf{EQ}$ over a strong alphabet $\Sigma$}
  \KwOut{strong nominal automaton recognizing the target language}
  $S \leftarrow \{\epsilon\}$;\ \ $E \leftarrow \{\epsilon\}$;\ \ fill $T$ over $(S \cup S\!\cdot\!\Sigma) \times E$ via $\textsf{MQ}$\;
  \While{\textbf{true}}{
    \If{the table is not closed}{
      promote a witness $t \in S\!\cdot\!\Sigma$ to $S$\;
      extend $S\!\cdot\!\Sigma$, refill, and \textbf{continue}
    }
    \If{the table is not consistent}{
      add the witness separator $a\,e$ to $E$\;
      refill, and \textbf{continue}
    }
    $\mathcal{H} \leftarrow \textsf{MakeHypothesis}(S, E, T)$\;
    $cex \leftarrow \textsf{EQ}(\mathcal{H})$\;
    \lIf{$cex = \bot$}{\Return $\mathcal{H}$}
    \ForEach{suffix $w$ of $cex$, shortest first}{
      $E \leftarrow E \cup G \cdot w$;\ \ refill $T$ over $(S \cup S\!\cdot\!\Sigma) \times G \cdot w$\;
    }
  }
\end{algorithm}

\subparagraph*{Counterexample handling.}
\Cref{alg:Learn} processes counterexamples by adding their suffixes to $E$ rather than their prefixes to $S$.
This distinguishing-suffix variant is explicitly sanctioned by~\cite[Sec.~4.1]{MoermanS0KS17}: their termination argument is insensitive to whether counterexamples are handled by row extension or column extension, so any strategy that yields a strictly refined hypothesis is admissible.
Our implementation simply adds \emph{all} suffixes of $cex$ as columns~\cite{MalerP95}, shortest first, a column-based handling also adopted for $\Ord$~\cite{VenhoekMR22}.
As~\cite{MoermanS0KS17} observe, a finer analysis of counterexamples in the style of Rivest and Schapire~\cite{RivestS93}, adding fewer columns, could be substituted as an optimization; we do not pursue it.

\Learn*

\begin{proof}
  We first note what is learned, then argue termination, then bound the query complexity, and finally discuss the per-hypothesis runtime. The argument parallels~\cite[Thm.~2]{MoermanS0KS17}, with their splitting function $f_{\mathbb{A}}$ instantiated to our $\beta$; since our variant handles counterexamples by columns rather than rows, the row and column bounds below are rebalanced accordingly.

  \begin{description}
  \item[Correctness.]
  As in~\cite[Thm.~2]{MoermanS0KS17}, at termination the closed and consistent table yields the minimal automaton: $S/{\rowsim}$ has exactly the $n$ orbits and the dimensions of the state space of $\mathcal{A}_{\min}$.
  Since \cref{alg:MakeHypothesis} constructs the final hypothesis as the strong approximation of $S/{\rowsim}$ through $\textsf{Quotient}$, as in \cref{lem:Quotient,thm:StrongApproximation}, the learned automaton is a strong lift of $\mathcal{A}_{\min}$, as stated in \cref{cor:MinimalStrong}.

  \item[Termination.]
  The three monotone parameters of~\cite[Thm.~2]{MoermanS0KS17} control the outer loop:
  \begin{itemize}
    \item \emph{Orbit count of $S/{\rowsim}$} is bounded by $n$, since $\equiv_L$ refines ${\rowsim}$~\cite[Lem.~1]{MoermanS0KS17}.
    \item \emph{Dimension of orbits of $S/{\rowsim}$} is bounded by $d$, since $\supp([u]_{\rowsim}) \subseteq \supp([u]_{\equiv_L}) \subseteq \supp(u)$~\cite[Lem.~2]{MoermanS0KS17}.
    \item \emph{Local symmetry of $[u]_{\rowsim}$} is a subgroup of $S_d$; a strictly descending chain of such subgroups has length $O(d)$~\cite{Babai86}, sharpening the $\log_2 d!$ estimate used for this chain in~\cite[Lems.~5, 6]{MoermanS0KS17}.
  \end{itemize}
  Each closedness step adds a row orbit, strictly increasing the first parameter; each consistency or counterexample step either increases the orbit count, increases a dimension, or strictly refines a local symmetry. Since all three parameters are bounded and advance monotonically, the loop terminates. The all-suffixes column extension of \cref{alg:Learn} preserves this argument because, as noted above, handling a counterexample by columns still yields a strictly refined hypothesis, advancing one of the three parameters.

  \item[Equivalence queries.]
  Each counterexample forces a strict refinement along one of the three axes, so the number of hypotheses, and hence of equivalence queries, is bounded by $n + n\,(d + O(d)) = O(n d)$ (cf.~\cite[Lem.~5]{MoermanS0KS17}).

  \item[Membership queries.]
  Since counterexamples extend only $E$, the set $S$ grows solely by closedness promotions.
  Each promotion adds a row orbit lying in a fresh ${\rowsim}$-class, and ${\rowsim}$ only refines as $E$ grows, so $N(S/{\rowsim})$ increases monotonically; as $N(S/{\rowsim}) \le n$ by~\cite[Lem.~1]{MoermanS0KS17}, there are at most $n - 1$ promotions.
  Hence $N(S) \le n$, with representatives of length at most $n - 1$, so the words of $S \cup S\!\cdot\!\Sigma$ have dimension at most $en$.
  Columns arise from two sources: a consistency defect adds one column and strictly refines the three termination parameters, so there are at most $n(d + O(d) + 1) = O(nd)$ such additions (cf.~\cite[Lem.~6]{MoermanS0KS17}); and each of the $O(n d)$ counterexamples contributes at most $m$ suffix orbits.
  Hence
  \begin{equation*}
     N(E) = O\bigl((m + 1)\, n d\bigr),
  \end{equation*}
  and every column has length at most $m + O(nd)$, since a column is either a suffix of a counterexample (of length at most $m$) or extends an existing column by one letter through a consistency step; hence $\ell = m + O(nd)$, as claimed in \cref{thm:Learn}, and every column has dimension at most $e\ell$.
  By~\cite[Lem.~7]{MoermanS0KS17}, each orbit of $S$ paired with each orbit of $\Sigma$ splits into at most $\beta(en, e)$ orbits of $S\!\cdot\!\Sigma$, so $N(S\!\cdot\!\Sigma) \le n\,N(\Sigma)\,\beta(en, e)$, and the table has
  \begin{equation*}
     C = O\bigl( n\,N(\Sigma)\,\beta(en, e) \cdot (m + 1)\, n d \bigr)
  \end{equation*}
  (symbolic) cells.
  Filling one cell (the product of a row orbit of dimension at most $en$ and a column orbit of dimension at most $e\ell$) takes one membership query per coincidence pattern, \ie, at most $\beta(en, e\ell)$ queries; this mirrors~\cite[Cor.~1]{MoermanS0KS17}, whose second argument is likewise the column dimension.
  Multiplying the cell count $C$ by the per-cell factor bounds the total by
  \begin{equation*}
     O\!\bigl( n^2 d \cdot (m + 1) \cdot N(\Sigma) \cdot \beta(en, e) \cdot \beta(en, e\ell) \bigr),
  \end{equation*}
  the membership-query bound of \cref{thm:Learn}.
  Note that only the cell count $C$ is polynomial for a fixed alphabet dimension $e$ (mirroring the corresponding remark in~\cite{MoermanS0KS17}); the per-cell splitting factor $\beta(en, e\ell)$ is factorial in the smaller of the observation dimensions $en$ and $e\ell$.

  \item[Per-hypothesis runtime.]
  \Cref{alg:MakeHypothesis} issues one call to $\textsf{Quotient}$ on $S \cup S\!\cdot\!\Sigma$. As established above, these words have length at most $n$, so their dimension is bounded by $en$. A row-equivalence test compares rows over $E$ (whose columns have dimension at most $e\ell$); each of the $N(E)$ cells is a product orbit costing at most $\beta(en, e\ell)$ coincidence-pattern inspections according to \cref{prop:Product}, so the test is an oracle of cost $C_{\rowsim} = O(N(E)\,\beta(en, e\ell))$. By \cref{lem:Quotient}, constructing the hypothesis therefore takes
  \begin{equation*}
     O\!\bigl(N(S \cup S\!\cdot\!\Sigma)^2\,\beta(en, en)\,(\mu(en) + C_{\rowsim})\bigr),
  \end{equation*}
  which, since $N(S \cup S\!\cdot\!\Sigma) = O(n\, N(\Sigma)\, \beta(en, e))$, is polynomial in $N(\Sigma)$ and $m$ but factorial in the observation dimensions $en$ and $e\ell$, as stated.
  Outside the constant-time orbit-primitive model, constructing the orbit of a length-$\ell$ word itself carries an additional $\Omega(\ell)$ factor, as discussed in \cref{app:Operations}.
  Combining the query bounds with this per-hypothesis runtime yields the complexity stated in \cref{thm:Learn}.
  \qedhere
  \end{description}
\end{proof}

Our bounds are the strong-alphabet counterparts of the query-complexity analysis of~\cite[Sec.~4.3]{MoermanS0KS17}, whose parameters $n$, $k$, $p$, $l$, and $f_{\mathbb{A}}$ correspond to our $n$, $d$, $e$, $N(\Sigma)$, and $\beta$.
Two changes shift the bounds.
First, Babai's theorem~\cite{Babai86} replaces the $k \log k$ estimate for subgroup chains in~\cite[Lems.~5, 6]{MoermanS0KS17}, sharpening the equivalence-query bound $E_{n,k} = O(n k \log k)$ to $O(n d)$ and the column count accordingly.
Second, the counterexample handling is rebalanced: they extend the rows by all prefixes of a counterexample, so their rows number $n + m E_{n,k}$ and reach dimension $p(n + m)$ while their columns stay at length $O(n k \log k)$~\cite[Lem.~6]{MoermanS0KS17}, whereas our all-suffix variant keeps $O(n)$ rows of length at most $n$ (dimension at most $e n$) and moves the counterexample length into the columns, of length at most $\ell = m + O(n d)$.
In both cases, the membership-query bound is a cell count, polynomial in every parameter except the alphabet dimension (\cite[Cor.~1]{MoermanS0KS17}; \cref{thm:Learn}), times a per-cell factor factorial in the observation dimensions, which are $p(n + m)$ and $p\,n(k + k \log k + 1)$ for them and $e n$ and $e \ell$ for us.
Finally, the learned automaton differs: their $\nu L^\ast$ returns the minimal nominal automaton itself, whereas our instantiation returns a strong lift of \cref{cor:MinimalStrong}.

\section{Full Experimental Data}\label{app:FullExperiments}

This appendix collects the full numerical results underlying \cref{sec:ImplementationAndExperiments}.  All tables are machine-generated from the benchmark logs by our scripts.

\subsection{Tools and Reproducibility}\label{app:Setup}

All experiments were conducted on the hardware and runtime versions reported in \cref{sec:ImplementationAndExperiments} (Apple M1 Pro, macOS~26.5; Java~SE~25.0.3, Scala~3.8.4, GHC~9.4.8, Apple clang~21.0.0, Z3~4.16.0).
Each comparison tool was built from its public repository at the commit listed in \cref{tab:Tools}.
The other tools (\DEQ{}, \NLambda{}, \Lois{}, \Ons{}, \OnsHs{}) were compiled as native ARM64 binaries; no memory limit was imposed on any non-JVM tool.

\begin{table}[t]
  \caption{Comparison tools and the exact versions used. Repositories are on GitHub unless noted.}\label{tab:Tools}
  \centering
  \small
  \begin{tabular}{@{}llll@{}}
    \toprule
    Tool & Language & Repository & Commit \\
    \midrule
    \Namae{}  & Scala (JVM) & this work                   & --- \\
    \DEQ{}    & Haskell     & \texttt{sjrsay/deq}         & \texttt{6234847} \\
    \NLambda{}& Haskell     & \texttt{szynwelski/nlambda} & \texttt{cf52731} \\
    \Lois{}   & C++         & \texttt{eryxcc/lois}        & \texttt{9fe3d85} \\
    \Ons{}    & C++         & \texttt{davidv1992/ONS}     & \texttt{a13823d} \\
    \OnsHs{}  & Haskell     & \texttt{Jaxan/ons-hs}       & \texttt{d96342d} \\
    \bottomrule
  \end{tabular}
\end{table}

The minimization and learning harnesses for \Ons{} and \Lois{} are taken from the \texttt{ONS\_Minimise}\footnote{\url{https://github.com/davidv1992/ONS_Minimise}} repository (commit \texttt{11b8fe3}), and the $\Ord$ learning baseline \Ons{} additionally uses \texttt{nominal-learning-ons}\footnote{\url{https://gitlab.science.ru.nl/moerman/nominal-learning-ons.git}} (commit \texttt{faffe8a}).
Under $\Ord$, we report \OnsHs{}~\cite{VenhoekMR22} as the primary baseline: it is the actively maintained Haskell successor to the C++ \Ons{} library, selected for its active maintenance rather than any performance advantage. Both \Ons{} and \OnsHs{} appear in the full tables in \cref{tab:MinimizeFull,tab:LearningFull}.
Since \Namae{} supports both symmetries, the full results in \cref{tab:MinimizeFull,tab:LearningFull} also include \Namae{} under $\Ord$, enabling a same-implementation comparison between $\Eq$ and $\Ord$.
The two symmetries share the runtime, the array representation, and the $\nu L^\ast$ loop, but dispatch to symmetry-specific traits and implementations: under $\Ord$, where local symmetries are trivial, language equivalence runs the orbit-wise breadth-first search with a visited set in place of a generating system, minimization follows the class-refinement algorithm of \OnsHs{}~\cite{VenhoekMR22}, and quotients are computed by the support-intersection variant of \cref{ssec:QuotientOp}.

\subsection{Benchmark Languages}\label{app:Benchmarks}

For the equivalence benchmarks, we follow~\cite{MurawskiRT19} and use the following register automata:
\begin{itemize}
  \item \emph{stacks}: A register automaton over the alphabet $\Sigma = \{ \mathsf{Push}(a) \mid a \in \mathbb{N} \} \cup \{ \mathsf{Pop}(a) \mid a \in \mathbb{N} \}$ that recognizes valid sequences of stack operations, where $n$ denotes the maximum capacity of the stack.
  \item \emph{CPT}: A register automaton over the alphabet $\Sigma = \{ \mathsf{populate}(a) \mid a \in \mathbb{N} \} \cup \bigcup_{1 \le i \le n} \{ \mathsf{t}_i(a) \mid a \in \mathbb{N} \}$. This benchmark is designed to generate an exceptionally large number of partial permutations during bisimulation, thereby effectively demonstrating the power of the generating system.
\end{itemize}
Each instance compares two language-equivalent encodings, the worst case for a bisimulation-based check. We omit the suite's third family, \emph{glolo}, which relies on global freshness and thus has no orbit-finite nominal counterpart, as noted in \cref{ssec:LanguageEquivalence}.

For the minimization and learning benchmarks, besides $\mathrm{Res}(n)$ and $\mathrm{Res}_{\textsc{FIFO}}(n)$ from \cref{ssec:MinimizationLearningExperiments}, the full tables also include the double-word language $ww(n)$, and, for minimization only, $\mathrm{MRes}(m)$, whose multi-dimensional alphabet would otherwise require adapting every tool's black-box learning harness.
Specifically, these languages are defined as follows:
\begin{itemize}
  \item $\mathrm{Res}(n)$: A language over the alphabet $\Sigma = \{ \mathsf{Open}(a) \mid a \in \mathbb{N} \} \cup \{ \mathsf{Close}(a) \mid a \in \mathbb{N} \}$ that models resource management. It accepts a trace if all opened resources are eventually closed. A trace is rejected if any of the following occur:
  \begin{itemize}
    \item if a $\mathsf{Close}(a)$ occurs for a resource $a$ that is not currently open,
    \item if an $\mathsf{Open}(a)$ occurs for a resource $a$ that is already open, or
    \item if the number of concurrently open resources exceeds the capacity $n$.
  \end{itemize}
  \item $\mathrm{Res}_{\textsc{FIFO}}(n)$: An order-sensitive FIFO variant of $\mathrm{Res}(n)$. It is similar to $\mathrm{Res}(n)$ but additionally requires that resources are closed in the exact order they were opened.
  \item $\mathrm{MultiRes}(m, n)$: A language that runs $m$ independent copies of $\mathrm{Res}(n)$ in parallel. Its alphabet consists of $\mathsf{Open}(\mathbf{a})$ and $\mathsf{Close}(\mathbf{a})$ where $\mathbf{a} \in (\mathbb{N} \sqcup \{\bot\})^m$ is a tuple with at least one non-$\bot$ component; for each component $a_i \neq \bot$, the corresponding $\mathsf{Open}/\mathsf{Close}$ operation is applied to the $i$-th copy of $\mathrm{Res}(n)$. The non-$\bot$ components need not be distinct: for example, $\mathsf{Open}\langle a, a \rangle$ opens $a$ in two copies at once. Since $\mathbf{a}$ carries up to $m$ distinct atoms, its alphabet has dimension $m$. We write $\mathrm{MRes}(m) = \mathrm{MultiRes}(m, 1)$ for the single-capacity case used in the tables, whose input alphabet has dimension $m$; larger capacities are supported by our harness but raise the state dimension to $mn$, which made benchmarking impractical within the timeout.
  \item $ww(n)$: The double-word language $\{ww \mid w \in \mathbb{N}^n\}$ of length $2n$ over the alphabet $\Sigma = \mathbb{N}$.
\end{itemize}

Regarding local symmetry, some states of the minimal automaton accepting $\mathrm{Res}(n)$ carry the full symmetric group $S_n$ as their local symmetry.
In contrast, for the order-sensitive languages $\mathrm{Res}_{\textsc{FIFO}}(n)$ and $ww(n)$, the local symmetries of all states in their minimal automata are trivial.
Consequently, the minimal automaton of $\mathrm{Res}_{\textsc{FIFO}}(n)$ is already strong, while a strong lift of $\mathrm{Res}(n)$'s replaces each set of open resources by an ordered tuple, so the two strong lifts have isomorphic state spaces over $\Eq$: tuples of at most $n$ pairwise distinct open resources plus a rejecting sink.
Since both transition functions are total and equivariant over the same state space and alphabet, their orbit counts coincide as well, so the two $\nu L^\ast$ runs under $\Eq$ maintain observation tables of the same shape and are expected to issue the same number of membership queries, exactly as the identical MQ and $\Delta$ columns of \cref{tab:LearningMain,tab:LearningFull} show.

\subsection{Running the Benchmarks}\label{app:Measurement}

We report the median wall-clock time of three runs subject to a 10-minute timeout as detailed in \cref{sec:ImplementationAndExperiments}.
We use wall-clock time rather than the CPU-time convention of~\cite{VenhoekMR22}, since for \Namae{} JVM-internal CPU accounting cannot be cleanly separated from JIT and garbage-collection overhead, making wall-clock the fairer common measure across the JVM, Haskell, and C++ implementations compared here.
The measurement boundary differs by task and is stated in each paragraph below.
Wherever \Namae{} is timed in-process (equivalence and minimization), JIT warm-up is excluded from the measurement.
The warm-up exercises fixed small instances distinct from the measured one (a small stack pair for equivalence and $ww(1)$ for minimization) within the same JVM as the subsequent measurement; the three measured runs use separate JVM invocations and share no JIT state.
Excluding warm-up prevents JVM start-up and JIT compilation overhead from dominating the small-$n$ measurements, which would otherwise exaggerate the gap to the natively compiled comparison tools; for large $n$ this overhead is negligible and does not affect our conclusions.
For each tool and benchmark family, we increase $n$ until the first timeout, at which point neither the remaining repetitions at that $n$ nor any larger $n$ are attempted; every \timeout{} entry in \cref{tab:MinimizeFull,tab:LearningFull} therefore reflects a single timed-out run rather than a median of three.

\para{Language equivalence.}
Following~\cite{MurawskiRT19}, \DEQ{} and \Namae{} read the two register automata to be compared from XML files and decide their equivalence, whereas \NLambda{} and \Lois{} programmatically construct the automata and perform the equivalence check internally, as in the original benchmarks.
Each tool internally measures the execution time of the equivalence check, excluding input parsing.

\para{Minimization.}
Following~\cite{VenhoekMR22}, each tool reads a serialized nominal automaton from a file, minimizes it, and serializes the result.
We designed a custom serialization format for the $\Eq$ symmetry and patched \NLambda{} to read it.
The inputs are generated by \Namae{} with a total transition function; for $\Ord$, we take the $\Eq$-generated input and split each orbit of dimension $d$ in its states and the transition function's domain into up to $d!$ orbits, realizing the orbit-splitting of \cref{rem:LocalSymmetry}.
The output formats vary across tools: \OnsHs{} and \NLambda{} emit their internal representations, \Ons{} emits only the number of states, and \Lois{} minimizes the state space only, without constructing the transition function; accordingly, the reported transition size $\Delta_{\min}$ is the number of symbolic entries for \NLambda{}, unavailable for \Lois{}, and the number of orbits in the domain of the transition function for the others (\Namae{} and \OnsHs{}).
We check the correctness of each tool's output by confirming that its reported minimized state count matches the expected $N_{\min}$.
We measure each external tool's process time; \Namae{}'s in-process measurement still includes input parsing and output serialization.
For reference, with JVM start-up and warm-up included, \Namae{} takes $0.44$\,s instead of $0.06$\,s when minimizing $\mathrm{Res}(3)$, but $11.6$\,s instead of $11.13$\,s on $\mathrm{Res}(7)$; this in-process/external-process asymmetry favors \Namae{}, though the baselines' process start-up remains small against the multi-second runtimes at the scales where we draw our conclusions.

\para{Learning.}
Following~\cite{VenhoekMR22}, the teacher and the learner run as separate processes that communicate over standard input/output, ensuring a genuinely black-box learning setup; we reuse their communication protocol.
Our teacher is implemented in Scala on top of \Namae{}'s automaton representation; the stdin/stdout round-trip latency between teacher and learner is negligible against the measured runtimes.
We measure the wall-clock time of the learner process uniformly for every learner, including \Namae{}, unlike the in-process exception used for equivalence and minimization.
The equivalence query returns a pre-computed counterexample, again as in~\cite{VenhoekMR22}.
As in minimization, the transition size $\Delta$ is the number of symbolic entries for \NLambda{} and the number of orbits in the domain of the transition function for the others.

\newpage

\subsection{Full Results}\label{app:FullResults}

\begin{figure}[t]
  \centering
  \includegraphics[width=0.85\linewidth]{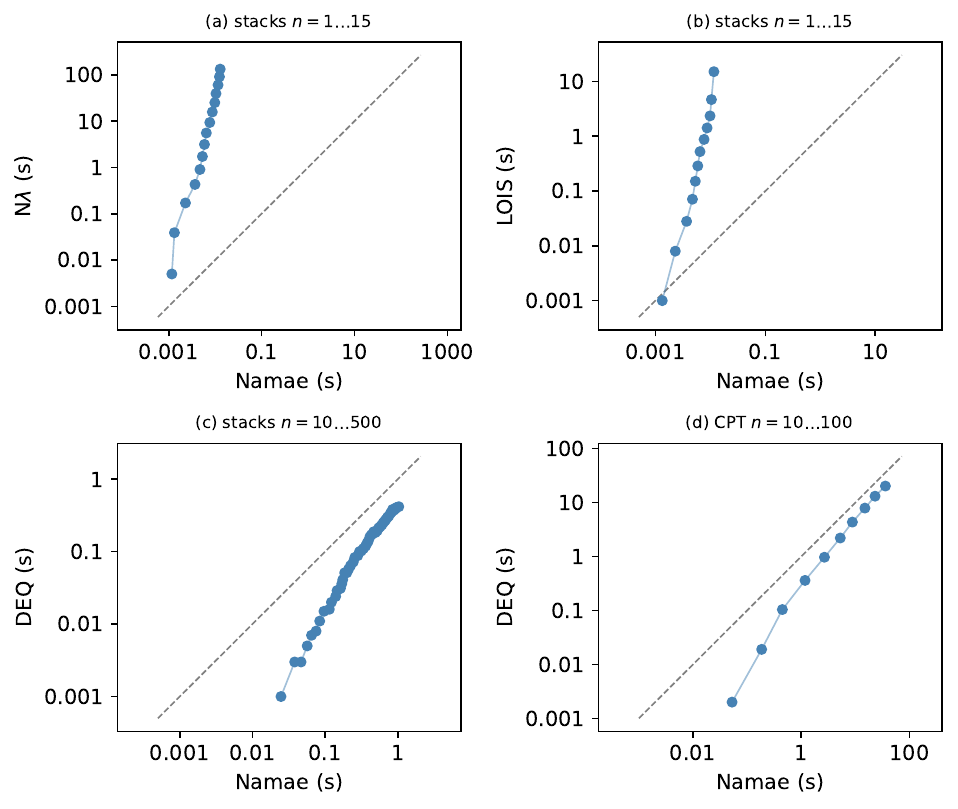}
  \caption{Language equivalence runtimes on stacks $n=1\ldots15$, all four tools (seconds, log-log). (a) \Namae{} vs.\ \NLambda{}. (b) \Namae{} vs.\ \Lois{}. (c) \Namae{} vs.\ \DEQ{} on stacks $n=10,\dots,500$. (d) \Namae{} vs.\ \DEQ{} on CPT $n=10,\dots,100$.}\label{fig:EquivFull}
\end{figure}

For language equivalence, \cref{fig:EquivFull} shows all four-tool comparisons on stacks and CPT, and \cref{tab:EquivStacksSmall,tab:EquivStacksLarge,tab:EquivCPT} report the corresponding raw runtimes.
For minimization and learning, \cref{tab:MinimizeFull,tab:LearningFull} give the full results across all benchmark languages and tools: runtimes, orbit counts, and transition sizes (with membership-query counts for learning).
On the shared $\Ord$ inputs, the differences between \Namae{} and \OnsHs{} are implementation-level constants, in both directions: \OnsHs{} minimizes $\mathrm{Res}(4)$ and $\mathrm{Res}(5)$ about $6\times$ faster, whereas \Namae{} learns $\mathrm{Res}(3)$ about $10\times$ faster and completes $\mathrm{Res}_{\textsc{FIFO}}(3)$ learning where \OnsHs{} times out.
Both swings are dwarfed by the symmetry effect visible within the same tables: on $\mathrm{Res}(5)$ minimization, the same \Namae{} code takes $0.26$\,s under $\Eq$ and $141.49$\,s under $\Ord$.

\begin{table}[p]
  \centering
  \begin{minipage}[t]{0.5\linewidth}
    \centering
    \caption{Language-equivalence runtimes on the \emph{stacks} benchmark for $n=1,\dots,15$, comparing \Namae{} with \DEQ{}, \NLambda{}, and \Lois{} (median of 3 runs, seconds; entries marked \timeout{} hit the timeout, `---' denotes an instance not attempted after a smaller instance timed out, and $0.000$ denotes a time below the $1$\,ms measurement resolution). These data underlie \cref{fig:EquivMain}(a) and \cref{fig:EquivFull}.}\label{tab:EquivStacksSmall}
    \small
    \begin{tabular}{rrrrr}
  \toprule
  $n$ & Namae & DEQ & NLambda & LOIS \\
  \midrule
  1 & 0.001 & 0.000 & 0.005 & 0.000 \\
  2 & 0.001 & 0.000 & 0.039 & 0.001 \\
  3 & 0.002 & 0.000 & 0.171 & 0.008 \\
  4 & 0.004 & 0.000 & 0.430 & 0.028 \\
  5 & 0.005 & 0.000 & 0.903 & 0.071 \\
  6 & 0.005 & 0.000 & 1.723 & 0.151 \\
  7 & 0.006 & 0.000 & 3.142 & 0.288 \\
  8 & 0.006 & 0.000 & 5.552 & 0.525 \\
  9 & 0.008 & 0.000 & 9.344 & 0.876 \\
  10 & 0.009 & 0.000 & 15.766 & 1.421 \\
  11 & 0.010 & 0.000 & 25.134 & 2.357 \\
  12 & 0.010 & 0.001 & 39.463 & 4.678 \\
  13 & 0.012 & 0.000 & 60.392 & 15.119 \\
  14 & 0.012 & 0.000 & 90.933 & \timeout{} \\
  15 & 0.013 & 0.001 & 132.899 & --- \\
  \bottomrule
\end{tabular}

    \vspace{2\baselineskip}

    \caption{Language-equivalence runtimes on the \emph{CPT} benchmark for $n=10,20,\dots,100$, comparing \Namae{} with \DEQ{} (median of 3 runs, seconds). These data underlie \cref{fig:EquivMain}(c).}\label{tab:EquivCPT}
    \small
    \begin{tabular}{rrr}
  \toprule
  $n$ & Namae & DEQ \\
  \midrule
  10 & 0.053 & 0.002 \\
  20 & 0.185 & 0.019 \\
  30 & 0.448 & 0.104 \\
  40 & 1.174 & 0.361 \\
  50 & 2.680 & 0.970 \\
  60 & 5.262 & 2.205 \\
  70 & 8.798 & 4.345 \\
  80 & 14.986 & 7.897 \\
  90 & 23.095 & 13.144 \\
  100 & 35.857 & 20.255 \\
  \bottomrule
\end{tabular}

  \end{minipage}\hfill
  \begin{minipage}[t]{0.4\linewidth}
    \centering
    \caption{Language-equivalence runtimes on the \emph{stacks} benchmark for $n=1,\dots,15$ and $n=20,30,\dots,500$, comparing \Namae{} with \DEQ{} (median of 3 runs, seconds). These data underlie \cref{fig:EquivMain}(b).}\label{tab:EquivStacksLarge}
    \scriptsize
    \begin{tabular}{rrr}
  \toprule
  $n$ & Namae & DEQ \\
  \midrule
  1 & 0.001 & 0.000 \\
  2 & 0.001 & 0.000 \\
  3 & 0.002 & 0.000 \\
  4 & 0.004 & 0.000 \\
  5 & 0.005 & 0.000 \\
  6 & 0.005 & 0.000 \\
  7 & 0.006 & 0.000 \\
  8 & 0.006 & 0.000 \\
  9 & 0.008 & 0.000 \\
  10 & 0.009 & 0.000 \\
  11 & 0.010 & 0.000 \\
  12 & 0.010 & 0.001 \\
  13 & 0.012 & 0.000 \\
  14 & 0.012 & 0.000 \\
  15 & 0.013 & 0.001 \\
  20 & 0.016 & 0.000 \\
  30 & 0.025 & 0.001 \\
  40 & 0.038 & 0.003 \\
  50 & 0.047 & 0.003 \\
  60 & 0.057 & 0.005 \\
  70 & 0.065 & 0.007 \\
  80 & 0.075 & 0.008 \\
  90 & 0.084 & 0.011 \\
  100 & 0.096 & 0.015 \\
  110 & 0.114 & 0.016 \\
  120 & 0.122 & 0.020 \\
  130 & 0.139 & 0.024 \\
  140 & 0.146 & 0.029 \\
  150 & 0.162 & 0.031 \\
  160 & 0.169 & 0.036 \\
  170 & 0.175 & 0.041 \\
  180 & 0.185 & 0.051 \\
  190 & 0.198 & 0.051 \\
  200 & 0.210 & 0.057 \\
  210 & 0.223 & 0.064 \\
  220 & 0.244 & 0.072 \\
  230 & 0.256 & 0.083 \\
  240 & 0.281 & 0.088 \\
  250 & 0.297 & 0.100 \\
  260 & 0.313 & 0.102 \\
  270 & 0.336 & 0.110 \\
  280 & 0.359 & 0.119 \\
  290 & 0.376 & 0.130 \\
  300 & 0.391 & 0.139 \\
  310 & 0.412 & 0.160 \\
  320 & 0.433 & 0.171 \\
  330 & 0.466 & 0.188 \\
  340 & 0.490 & 0.183 \\
  350 & 0.520 & 0.194 \\
  360 & 0.544 & 0.212 \\
  370 & 0.569 & 0.220 \\
  380 & 0.601 & 0.233 \\
  390 & 0.629 & 0.250 \\
  400 & 0.662 & 0.265 \\
  410 & 0.686 & 0.278 \\
  420 & 0.721 & 0.298 \\
  430 & 0.749 & 0.309 \\
  440 & 0.790 & 0.337 \\
  450 & 0.821 & 0.353 \\
  460 & 0.847 & 0.382 \\
  470 & 0.880 & 0.372 \\
  480 & 0.921 & 0.389 \\
  490 & 0.960 & 0.407 \\
  500 & 1.029 & 0.417 \\
  \bottomrule
\end{tabular}

  \end{minipage}
\end{table}

\begin{sidewaystable}[p]
  \caption{Full minimization results across all benchmark languages ($ww$, $\mathrm{Res}$, $\mathrm{Res}_{\textsc{FIFO}}$, $\mathrm{MRes}$) and tools, under both the equality symmetry $\Eq$ (\Namae{}, \NLambda{}) and the total-order symmetry $\Ord$ (\Namae{}, \Ons{}, \OnsHs{}, \NLambda{}, \Lois{}); median of 3 runs. $N_{\mathrm{in}}$/$N_{\min}$ are the state orbit counts before/after minimization, $\dim$ the state dimension, and $\Delta_{\min}$ the size of the minimized transition function; times are in seconds, entries marked \timeout{} hit the timeout, `---' denotes values not measured or unavailable, and \textbf{bold} marks the fastest tool overall. \Cref{tab:MinimizeMain} excerpts representative rows.}\label{tab:MinimizeFull}
  \centering
  \scriptsize
  \scalebox{0.8}{  \begin{tabular}{lrrrrrrrrrrrrrrrrrrrrr}
    \toprule
    & \multicolumn{7}{c}{Equality Symmetry} & \multicolumn{13}{c}{Total Order Symmetry} \\
    \cmidrule(lr){2-8}\cmidrule(lr){9-21}
    & \multicolumn{3}{c}{size} & \multicolumn{2}{c}{\textsc{Namae}} & \multicolumn{2}{c}{\textsc{N}$\lambda$}
    & \multicolumn{3}{c}{size} & \multicolumn{2}{c}{\textsc{Namae}} & \multicolumn{2}{c}{\textsc{Ons}} & \multicolumn{2}{c}{\textsc{Ons-hs}} & \multicolumn{2}{c}{\textsc{N}$\lambda$} & \multicolumn{2}{c}{\textsc{LOIS}} \\
    \cmidrule(lr){2-4}\cmidrule(lr){5-6}\cmidrule(lr){7-8}
    \cmidrule(lr){9-11}\cmidrule(lr){12-13}\cmidrule(lr){14-15}\cmidrule(lr){16-17}\cmidrule(lr){18-19}\cmidrule(lr){20-21}
    Target & $N_\mathrm{in}$ & $N_{\min}$ & $\dim$ & time (s) & $\Delta_{\min}$ & time (s) & $\Delta_{\min}$
           & $N_\mathrm{in}$ & $N_{\min}$ & $\dim$ & time (s) & $\Delta_{\min}$ & time (s) & $\Delta_{\min}$ & time (s) & $\Delta_{\min}$ & time (s) & $\Delta_{\min}$ & time (s) & $\Delta_{\min}$ \\
    \midrule
    \rowcolor{gray!20}$ww(1)$ & 4 & 4 & 1 & 0.03 & 5 & 0.03 & 5 & 4 & 4 & 1 & 0.03 & 6 & \textbf{0.01} & --- & 0.03 & 6 & 0.03 & 5 & 0.08 & --- \\
    $ww(2)$ & 8 & 7 & 2 & 0.05 & 12 & 2.06 & 15 & 10 & 8 & 2 & 0.06 & 22 & \textbf{0.01} & --- & 0.03 & 22 & 7.32 & 24 & \timeout{} & --- \\
    \rowcolor{gray!20}$ww(3)$ & 21 & 14 & 3 & 0.08 & 32 & 306.75 & 68 & 46 & 24 & 3 & 0.23 & 110 & 0.34 & --- & \textbf{0.06} & 110 & \timeout{} & --- & --- & --- \\
    $ww(4)$ & 71 & 34 & 4 & \textbf{0.27} & 99 & \timeout{} & --- & 320 & 112 & 4 & 6.13 & 726 & 58.84 & --- & 4.72 & 726 & --- & --- & --- & --- \\
    \rowcolor{gray!20}$ww(5)$ & 286 & 101 & 5 & \textbf{1.90} & 354 & --- & --- & 2880 & --- & 5 & \timeout{} & --- & \timeout{} & --- & \timeout{} & --- & --- & --- & --- & --- \\
    $ww(6)$ & 1296 & 356 & 6 & \textbf{35.40} & 1434 & --- & --- & 28734 & --- & 6 & --- & --- & --- & --- & --- & --- & --- & --- & --- & --- \\
    \rowcolor{gray!20}$ww(7)$ & 6420 & --- & 7 & \timeout{} & --- & --- & --- & 336370 & --- & 7 & --- & --- & --- & --- & --- & --- & --- & --- & --- & --- \\
    \midrule
    $\mathrm{Res}(1)$ & 3 & 3 & 1 & 0.03 & 8 & 0.02 & 7 & 3 & 3 & 1 & 0.03 & 10 & \textbf{0.01} & --- & 0.02 & 10 & 0.02 & 7 & 0.18 & --- \\
    \rowcolor{gray!20}$\mathrm{Res}(2)$ & 4 & 4 & 2 & 0.04 & 14 & 0.50 & 11 & 5 & 4 & 2 & 0.05 & 20 & \textbf{0.01} & --- & 0.02 & 20 & 2.21 & 15 & 594.85 & --- \\
    $\mathrm{Res}(3)$ & 5 & 5 & 3 & 0.06 & 22 & 6.99 & 15 & 11 & 5 & 3 & 0.19 & 34 & 0.06 & --- & \textbf{0.04} & 34 & \timeout{} & --- & \timeout{} & --- \\
    \rowcolor{gray!20}$\mathrm{Res}(4)$ & 6 & 6 & 4 & \textbf{0.10} & 32 & 461.26 & 19 & 35 & 6 & 4 & 3.79 & 52 & 1.53 & --- & 0.62 & 52 & --- & --- & --- & --- \\
    $\mathrm{Res}(5)$ & 7 & 7 & 5 & \textbf{0.26} & 44 & \timeout{} & --- & 155 & 7 & 5 & 141.49 & 74 & 67.76 & --- & 21.59 & 74 & --- & --- & --- & --- \\
    \rowcolor{gray!20}$\mathrm{Res}(6)$ & 8 & 8 & 6 & \textbf{1.22} & 58 & --- & --- & 875 & --- & 6 & \timeout{} & --- & \timeout{} & --- & \timeout{} & --- & --- & --- & --- & --- \\
    $\mathrm{Res}(7)$ & 9 & 9 & 7 & \textbf{11.13} & 74 & --- & --- & 5915 & --- & 7 & --- & --- & --- & --- & --- & --- & --- & --- & --- & --- \\
    \midrule
    \rowcolor{gray!20}$\mathrm{Res}_\textsc{FIFO}(1)$ & 4 & 3 & 1 & 0.03 & 8 & 0.03 & 7 & 4 & 3 & 1 & 0.03 & 10 & \textbf{0.01} & --- & 0.03 & 10 & 0.03 & 7 & 1.38 & --- \\
    $\mathrm{Res}_\textsc{FIFO}(2)$ & 7 & 4 & 2 & 0.05 & 14 & 4.48 & 19 & 10 & 5 & 2 & 0.08 & 30 & \textbf{0.02} & --- & 0.03 & 30 & 13.24 & 31 & \timeout{} & --- \\
    \rowcolor{gray!20}$\mathrm{Res}_\textsc{FIFO}(3)$ & 11 & 5 & 3 & \textbf{0.07} & 22 & 80.53 & 51 & 34 & 11 & 3 & 0.58 & 114 & 0.77 & --- & 0.38 & 114 & \timeout{} & --- & --- & --- \\
    $\mathrm{Res}_\textsc{FIFO}(4)$ & 16 & 6 & 4 & \textbf{0.20} & 32 & \timeout{} & --- & 154 & 35 & 4 & 42.43 & 546 & 98.97 & --- & 52.10 & 546 & --- & --- & --- & --- \\
    \rowcolor{gray!20}$\mathrm{Res}_\textsc{FIFO}(5)$ & 22 & 7 & 5 & \textbf{0.84} & 44 & --- & --- & 874 & --- & 5 & \timeout{} & --- & \timeout{} & --- & \timeout{} & --- & --- & --- & --- & --- \\
    $\mathrm{Res}_\textsc{FIFO}(6)$ & 29 & 8 & 6 & \textbf{8.82} & 58 & --- & --- & 5914 & --- & 6 & --- & --- & --- & --- & --- & --- & --- & --- & --- & --- \\
    \rowcolor{gray!20}$\mathrm{Res}_\textsc{FIFO}(7)$ & 37 & 9 & 7 & \textbf{99.98} & 74 & --- & --- & 46234 & --- & 7 & --- & --- & --- & --- & --- & --- & --- & --- & --- & --- \\
    \midrule
    $\mathrm{MRes}(1)$ & 3 & 3 & 1 & 0.04 & 8 & 0.04 & 7 & 3 & 3 & 1 & 0.03 & 10 & \textbf{0.01} & --- & 0.02 & 10 & 0.03 & 7 & 0.18 & --- \\
    \rowcolor{gray!20}$\mathrm{MRes}(2)$ & 6 & 6 & 2 & 0.05 & 102 & 7.62 & 58 & 7 & 7 & 2 & 0.09 & 298 & 0.06 & --- & \textbf{0.04} & 298 & 22.32 & 85 & \timeout{} & --- \\
    $\mathrm{MRes}(3)$ & 16 & 16 & 3 & \textbf{0.26} & 1752 & \timeout{} & --- & 27 & 27 & 3 & 17.09 & 18730 & 62.44 & --- & 19.77 & 18730 & \timeout{} & --- & --- & --- \\
    \rowcolor{gray!20}$\mathrm{MRes}(4)$ & 53 & 53 & 4 & \textbf{38.37} & 42292 & --- & --- & 151 & --- & 4 & \timeout{} & --- & \timeout{} & --- & \timeout{} & --- & --- & --- & --- & --- \\
    \bottomrule
  \end{tabular}
}
\end{sidewaystable}

\begin{sidewaystable}[p]
  \caption{Full $\nu L^\ast$ learning results across all benchmark languages ($ww$, $\mathrm{Res}$, $\mathrm{Res}_{\textsc{FIFO}}$) and tools, under both the equality symmetry $\Eq$ (\Namae{}, \NLambda{}) and the total-order symmetry $\Ord$ (\Namae{}, \Ons{}, \OnsHs{}, \NLambda{}); median of 3 runs. $N$ is the state orbit count of the target language, $\dim$ the state dimension, MQs the number of membership queries, and $\Delta$ the size of the learned transition function; times are in seconds, entries marked \timeout{} hit the timeout, `---' denotes values not measured or unavailable, and \textbf{bold} marks the fastest tool overall. \Cref{tab:LearningMain} excerpts representative rows.}\label{tab:LearningFull}
  \centering
  \scriptsize
  \scalebox{0.8}{  \begin{tabular}{lrrrrrrrrrrrrrrrrrrrrrrr}
    \toprule
    & \multicolumn{8}{c}{Equality Symmetry} & \multicolumn{14}{c}{Total Order Symmetry} \\
    \cmidrule(lr){2-9}\cmidrule(lr){10-23}
    & \multicolumn{2}{c}{size} & \multicolumn{3}{c}{\textsc{Namae}} & \multicolumn{3}{c}{\textsc{N}$\lambda$}
    & \multicolumn{2}{c}{size} & \multicolumn{3}{c}{\textsc{Namae}} & \multicolumn{3}{c}{\textsc{Ons}} & \multicolumn{3}{c}{\textsc{Ons-hs}} & \multicolumn{3}{c}{\textsc{N}$\lambda$} \\
    \cmidrule(lr){2-3}\cmidrule(lr){4-6}\cmidrule(lr){7-9}\cmidrule(lr){10-11}\cmidrule(lr){12-14}\cmidrule(lr){15-17}\cmidrule(lr){18-20}\cmidrule(lr){21-23}
    Target & $N$ & $\dim$ & time (s) & MQs & $\Delta$ & time (s) & MQs & $\Delta$ & $N$ & $\dim$ & time (s) & MQs & $\Delta$ & time (s) & MQs & $\Delta$ & time (s) & MQs & $\Delta$ & time (s) & MQs & $\Delta$ \\
    \midrule
    \rowcolor{gray!20}$ww(1)$ & 4 & 1 & 0.37 & \textbf{39} & 5 & 2.20 & \textbf{39} & 3 & 4 & 1 & 0.41 & 111 & 6 & 0.45 & 134 & 6 & \textbf{0.21} & 111 & 6 & 6.32 & 168 & 3 \\
    $ww(2)$ & 7 & 2 & \textbf{0.47} & \textbf{199} & 12 & \timeout{} & --- & --- & 8 & 2 & 0.78 & 1259 & 22 & 95.14 & 2559 & 22 & 1.07 & 1259 & 22 & \timeout{} & --- & --- \\
    \rowcolor{gray!20}$ww(3)$ & 14 & 3 & \textbf{1.20} & \textbf{3111} & 32 & --- & --- & --- & 24 & 3 & 195.47 & 130427 & 110 & \timeout{} & --- & --- & \timeout{} & --- & --- & --- & --- & --- \\
    $ww(4)$ & 34 & 4 & \textbf{85.85} & \textbf{94852} & 99 & --- & --- & --- & 112 & 4 & \timeout{} & --- & --- & --- & --- & --- & --- & --- & --- & --- & --- & --- \\
    \rowcolor{gray!20}$ww(5)$ & 101 & 5 & \timeout{} & --- & --- & --- & --- & --- & 728 & 5 & --- & --- & --- & --- & --- & --- & --- & --- & --- & --- & --- & --- \\
    \midrule
    $\mathrm{Res}(1)$ & 3 & 1 & 0.37 & \textbf{31} & 8 & 1.11 & \textbf{31} & 6 & 3 & 1 & 0.38 & 67 & 10 & 0.21 & 119 & 10 & \textbf{0.20} & 67 & 10 & 1.84 & 67 & 6 \\
    \rowcolor{gray!20}$\mathrm{Res}(2)$ & 4 & 2 & \textbf{0.50} & \textbf{315} & 14 & 26.62 & 322 & 8 & 4 & 2 & 0.61 & 1074 & 20 & 0.62 & 1183 & 20 & 0.51 & 1074 & 20 & 126.89 & 1720 & 8 \\
    $\mathrm{Res}(3)$ & 5 & 3 & \textbf{0.94} & \textbf{2638} & 22 & 525.50 & 2699 & 10 & 5 & 3 & 2.90 & 10671 & 34 & 9.68 & 10887 & 34 & 29.67 & 10671 & 34 & \timeout{} & --- & --- \\
    \rowcolor{gray!20}$\mathrm{Res}(4)$ & 6 & 4 & \textbf{11.09} & \textbf{24259} & 32 & \timeout{} & --- & --- & 6 & 4 & 105.23 & 94083 & 52 & 299.18 & 94472 & 52 & \timeout{} & --- & --- & --- & --- & --- \\
    $\mathrm{Res}(5)$ & 7 & 5 & \timeout{} & --- & --- & --- & --- & --- & 7 & 5 & \timeout{} & --- & --- & \timeout{} & --- & --- & --- & --- & --- & --- & --- & --- \\
    \midrule
    \rowcolor{gray!20}$\mathrm{Res}_\textsc{FIFO}(1)$ & 3 & 1 & 0.39 & \textbf{31} & 8 & 1.09 & \textbf{31} & 6 & 3 & 1 & 0.37 & 67 & 10 & 0.21 & 119 & 10 & \textbf{0.20} & 67 & 10 & 1.79 & 67 & 6 \\
    $\mathrm{Res}_\textsc{FIFO}(2)$ & 4 & 2 & \textbf{0.48} & \textbf{315} & 14 & 25.15 & 322 & 8 & 5 & 2 & 0.95 & 3213 & 30 & 1.68 & 3379 & 30 & 1.84 & 3213 & 30 & \timeout{} & --- & --- \\
    \rowcolor{gray!20}$\mathrm{Res}_\textsc{FIFO}(3)$ & 5 & 3 & \textbf{0.91} & \textbf{2638} & 22 & 498.29 & 2698 & 10 & 11 & 3 & 209.69 & 241431 & 114 & \timeout{} & --- & --- & \timeout{} & --- & --- & --- & --- & --- \\
    $\mathrm{Res}_\textsc{FIFO}(4)$ & 6 & 4 & \textbf{11.40} & \textbf{24259} & 32 & \timeout{} & --- & --- & 35 & 4 & \timeout{} & --- & --- & --- & --- & --- & --- & --- & --- & --- & --- & --- \\
    \rowcolor{gray!20}$\mathrm{Res}_\textsc{FIFO}(5)$ & 7 & 5 & \timeout{} & --- & --- & --- & --- & --- & 155 & 5 & --- & --- & --- & --- & --- & --- & --- & --- & --- & --- & --- & --- \\
    \bottomrule
  \end{tabular}
}
\end{sidewaystable}

\end{document}